\documentclass[11pt,a4paper]{article}
\pdfoutput=1

\usepackage{jheppub}
\usepackage{latexsym}
\usepackage{multirow}
\usepackage{color}
\usepackage[usenames,dvipsnames,table]{xcolor}
\usepackage{graphicx}
\usepackage{epsfig} 
\usepackage{epsf}   
\usepackage{dcolumn}
\usepackage{bm}
\usepackage{dcolumn}
\usepackage{textcomp}
\usepackage{float}
\usepackage{subfig}
\usepackage{hypcap}
\usepackage[]{hyperref}
\usepackage{makecell}
\usepackage{color}
\usepackage{pifont}
\usepackage{appendix}
\usepackage{amsmath}
\usepackage{multirow,bigdelim}  
\usepackage{lineno}
\usepackage[normalem]{ulem}
\graphicspath{{fig/}}

\newcommand{\comment}[1]{}
\usepackage{amssymb}
\hypersetup{
	bookmarks=true,         
	unicode=false,          
	pdftoolbar=true,        
	pdfmenubar=true,        
	pdffitwindow=true,   
	pdfstartview={FitH},    
	pdfsubject={scalar NSI},   
	pdfnewwindow=true,      
	pdfcreator={RevTeX},
	colorlinks=true,     
	linkcolor=red,         
	citecolor=blue,        
	filecolor=black,     
	urlcolor=blue,           
}

\preprint{}
         
\title{Impact of scalar NSI on the neutrino mass ordering sensitivity at DUNE, HK and KNO}

\author[a,1]{Arnab Sarker,} 
\author[a,2]{Abinash Medhi,}
\author[a,3]{Dharitree Bezboruah,}
\author[a,4]{Moon Moon Devi,}
\author[b,5]{and Debajyoti Dutta}

\affiliation[a]{Department of Physics, Tezpur University, Napaam, Sonitpur, Assam-784028, India}
\affiliation[b]{Department of Physics, Assam Don Bosco University, Kamarkuchi, Sonapur, Assam-782402, India}

\emailAdd{$^{1}$arnabs@tezu.ernet.in}
\emailAdd{$^{2}$amedhi@tezu.ernet.in}
\emailAdd{$^{3}$dbbphy1@tezu.ernet.in}
\emailAdd{$^{4}$devimm@tezu.ernet.in}
\emailAdd{$^{5}$debajyoti.dutta@dbuniversity.ac.in}

\date{\today}

	\abstract{The study of neutrino non-standard interactions (NSI) is a well-motivated phenomenological scenario to explore new physics beyond the Standard Model. The possible scalar coupling of neutrinos ($\nu$) with matter is one of such new physics scenarios that appears as a sub-dominant effect that can impact the $\nu$-oscillations in matter. The presence of scalar NSI introduces an additional contribution directly to the $\nu$-mass matrix in the interaction Hamiltonian and subsequently to the $\nu$-oscillations. This indicates that scalar NSI may have a significant impact on measurements related to $\nu$-oscillations e.g. leptonic CP phase $(\delta_{CP})$, $\theta_{23}$ octant and neutrino mass ordering (MO). The linear scaling of the effects of scalar NSI with matter density also motivates its exploration in long-baseline (LBL) experiments. In this paper, we study the impact of a scalar-mediated NSI on the MO sensitivity of DUNE, HK and HK+KNO, which are upcoming LBL experiments. We study the impact on MO sensitivities at these experiments assuming that scalar NSI parameters are present in nature and is known from other non-LBL experiments. We observe that the presence of diagonal scalar NSI elements can significantly affect the $\nu$-mass ordering sensitivities. We then also combine the data from DUNE with HK and HK+KNO to explore possible synergy among these experiments in a wider parameter space. We also observe a significant enhancement in the MO sensitivities for the combined analysis.
}

\begin{document}
	\maketitle
\section{Introduction} \label{sec:introduction}
The Standard Model (SM) of particle physics stands as one of the most successful theories to explain the fundamental nature of our universe. Still, there exist various phenomena that cannot be explained by theories within the SM, such as the phenomenon of $\nu$-oscillations, the presence of dark matter and dark energy, baryon asymmetry etc. This compels us to explore physics beyond the Standard Model (BSM). Neutrino oscillations decisively affirm the existence of non-zero masses for neutrinos, marking the initial experimental indication of BSM physics \cite{SNO:2002tuh, Super-Kamiokande:1998kpq}. The current \& upcoming $\nu$-experiments are primarily focused on determining the three major unknowns in the neutrino sector viz. neutrino mass ordering (MO), octant of $\theta_{23}$ \& leptonic CP phase. The various sub-dominant non-standard effects, like non-standard interactions (NSI) \cite{Liao:2016orc,Friedland:2012tq,Coelho:2012bp,Rahman:2015vqa,Coloma:2015kiu,deGouvea:2015ndi,Liao:2016hsa,Forero:2016cmb,Huitu:2016bmb,Bakhti:2016prn,Kumar:2021lrn,Agarwalla:2015cta,Agarwalla:2014bsa,Agarwalla:2012wf,Blennow:2016etl,Blennow:2015nxa,Deepthi:2016erc,Masud:2021ves,Soumya:2019kto,Masud:2018pig,Masud:2017kdi,Masud:2015xva,Ge:2016dlx,Fukasawa:2016lew,Chatterjee:2021wac,Medhi:2023ebi,Chaves:2021kxe,Brahma:2023wlf,Davoudiasl:2023uiq,Chatterjee:2020kkm,Choubey:2014iia,Singha:2021jkn,Denton:2018xmq,Denton:2020uda,Farzan:2015hkd}, Lorentz invariance violation (LIV)  \cite{Kostelecky:2003cr, SNO:2018mge,Mewes:2019dhj, Huang:2019etr, ARIAS2007401,LSND:2005oop,MINOS:2008fnv,MINOS:2010kat,IceCube:2010fyu,MiniBooNE:2011pix,DoubleChooz:2012eiq, Sarker:2023mlz,Sarkar:2022ujy,Majhi:2022fed}, neutrino decay \cite{Berryman:2014qha,PICORETI201670,SNO:2018pvg,GOMES2015345,Coloma:2017zpg,Abrahao:2015rba}, neutrino decoherence \cite{PhysRevD.56.6648,Benatti:2000ph,PhysRevD.100.055023,PhysRevD.95.113005,Lisi:2000zt} etc., are also being explored through these experiments. The precise determination of the unknowns is of immense significance for the complete understanding of the standard 3$\nu$ paradigm. The presence of NSI may affect the physics sensitivities of these experiments. In this study, we have mainly focused on exploring the impact of a scalar-mediated NSI on the MO sensitivities in the LBL sector, in a model-independent way. In our study, we include the upcoming LBL experiments namely DUNE, HK and HK+KNO. We have also performed a combined analysis of the MO sensitivities for DUNE with HK/HK+KNO.

The neutrino mass eigenstates may be defined in different ways, as described in the references \cite{Denton:2020exu,Denton:2021vtf,deGouvea:2008nm}. In this work, we adopt the formalism such that the condition $|U_{e1}|>|U_{e2}|>|U_{e3}|$ holds true. The ordering of the absolute neutrino masses $(m_{1}, m_{2}, m_{3})$ are not yet fully known, where $m_{1}$, $m_{2}$ and $m_{3}$ are the mass eigenstates. From the solar sector, we know that $m_{2}>m_{1}$ \cite{SNO:2001kpb,SNO:2002tuh, SNO:2002hgz, Harrison:2002er}, however, the relative ordering of $m_{3}$ is yet to be determined. We have two possible ordering of $\nu$-masses i.e. Normal Ordering (NO) $(m_{3}>m_{2}>m_{1})$ and Inverted Ordering (IO) $(m_{2}>m_{1}>m_{3})$ \cite{Esteban:2020cvm}. Various current and upcoming $\nu$-experiments from the solar \cite{Alivisatos:1998it,Cleveland:1998nv,SNO:2001kpb,SAGE:1999nng,Aleksan:1988qh,Bellini:2011rx}, atmospheric \cite{ICAL:2015stm,IceCube-PINGU:2014okk,Super-Kamiokande:2002weg,KM3Net:2016zxf,IceCube-Gen2:2020qha}, reactor \cite{Ardellier:2004ui,RENO:2010vlj,DayaBay:2007fgu,JUNO:2021vlw} and LBL accelerator \cite{NOvA:2004blv,T2K:2001wmr,Abi:2020loh,Hyper-KamiokandeProto-:2015xww,Hyper-Kamiokande:2016srs, ESSnuSB:2013dql,Akindinov:2019flp} sectors are equipped with advanced technologies and systematics which are expected to be highly sensitive to MO determination using different oscillation channels. The global analysis of neutrino oscillation data from LBL (T2K + NO$\nu$A), reactor (Daya Bay, RENO, Double Chooz) and atmospheric (Super-K) sectors prefers NO over IO at 2.7$\sigma$ Confidence Level (CL) as shown in reference \cite{Esteban:2020cvm}. In the current precision era, the future experiments are designed to have much higher sensitivities towards MO determination. All these experiments can individually achieve 3$\sigma$ CL in MO determination \cite{JUNO:2015zny, Devi:2014yaa,DUNE:2020jqi,Hyper-Kamiokande:2016srs, Hyper-KamiokandeProto-:2015xww,ESSnuSB:2021azq}. In addition, combining experiments from different sectors can further enhance the sensitivities by leveraging the synergy and a wider parameter space. In the reference \cite{Nunokawa:2005nx}, the determination of neutrino mass ordering using disappearance channel independent of matter effects and via a combination of different experiments are discussed. The authors have shown that highly precise measurements of atmospheric mass splitting in $\nu_{e}$ and $\nu_{\mu}$ disappearance channels can determine the mass ordering from the comparison of the location of the minima. They have compared for different choices of atmospheric $\Delta m^{2}$. In \cite{Fukasawa:2016lew}, the MO sensitivity up to 15$\sigma$ is obtained for synergy among T2HK, HK and DUNE. Combining current experiments T2K and NO$\nu$A with DUNE and ESS$\nu$SB, a 5$\sigma$ sensitivity for ordering determination is obtained in \cite{ Chakraborty:2019jlv}. A combination of T2K-II, NO$\nu$A and JUNO can also provide a similar sensitivity, as studied in reference \cite{Cao:2020ans}. In \cite{Ghosh:2012px}, the authors have explored the MO sensitivity at INO-ICAL and have also performed a combined analysis with T2K, NO$\nu$A and reactor experiments. Combining JUNO with IceCube Upgrade, and PINGU can also enhance the ordering sensitivity  \cite{IceCube-Gen2:2019fet}. In reference \cite{Choubey:2022gzv}, the MO sensitivity by combining JUNO and T2HK is studied. Though, JUNO alone has the potential of discovering MO at $3\sigma$ CL with $3\%$ energy resolution, a synergy between JUNO and T2HK increases it up to $9\sigma$. A synergy among T2HK, JUNO and atmospheric $\nu$-experiment INO-ICAL also has the potential of determining the MO with up to $7\sigma$ CL  with $5\%$ energy resolution of JUNO \cite{Raikwal:2023txr}.

Moreover, the MO sensitivities may be substantially affected by various BSM physics scenarios which makes it crucial to explore and understand such subdominant effects. In reference \cite{Capozzi:2019iqn, Esteban:2019lfo}, the data from the ongoing LBL experiments T2K and NO$\nu$A are fitted in the presence of vector NSI. The inclusion of vector NSI has completely washed out the preference for NO as expected in the standard scenario. In reference \cite{Masud:2016nuj}, the MO sensitivity and discovery potential, in the presence of vector NSI, are studied for the accelerator experiments T2K and NO$\nu$A along with DUNE. The presence of vector NSI has a significant impact on the MO sensitivity. The sensitivity to mass ordering at 5$\sigma$ CL is observed in DUNE for most values of $\delta_{CP}$ except for the case when $\epsilon_{ee}$ is negative. In presence of vector NSI, oscillation experiments alone can not determine the neutrino MO and $\theta_{12}$ octant due to the existence of exact degeneracy known as the LMA-Dark \cite{Coloma:2016gei, Bakhti:2014pva, deGouvea:2000pqg}. Scattering experiments, primarily coherent elastic neutrino-nucleus scattering (CE$\nu$NS) and CE$\nu$NS data from reactor experiments can break this degeneracy for mediator mass up to a few MeVs \cite{Denton:2018xmq,Denton:2022nol}. The generalized MO degeneracy in the presence of vector NSI \cite{Liao:2016hsa} deteriorates the ordering sensitivities of all three LBL experiments T2HK, T2HKK and DUNE as shown in the study \cite{Liao:2016orc}. In \cite{Deepthi:2016erc}, the authors have studied the effect of vector NSI on the MO sensitivities, considering the diagonal NSI parameter $\epsilon_{ee}$ to be equal to -1 so that it can nullify the standard matter effect. In this scenario, if additionally $\delta_{CP}$ is $\pm \pi/2$, there will be an intrinsic degeneracy between MO and NSI that cannot be lifted even if we can precisely measure the values of $\epsilon_{ee}$ and $\delta_{CP}$. Apart from vector NSI, other BSM phenomena can also affect the MO determination in LBL experiments. In reference \cite{Dutta:2016czj}, the authors have discussed how the non-unitarity in the $\nu$-mixing matrix can affect MO determination in the accelerator experiments DUNE, NO$\nu$A and T2K. The non-unitary mixing is found to decrease the MO sensitivity of these experiments. Though, DUNE will be sensitive to mass ordering discrimination even in the presence of non-unitarity for some values of $\delta_{CP}$; its sensitivity mostly decreases. In the case of NO$\nu$A and T2K, the scenario worsens as a result of their shorter baseline and less matter interaction contribution. The MO sensitivity of DUNE, T2HK and T2HKK in a $(3+1)$ scenario, with one sterile neutrino is studied in \cite{Choubey:2017cba}. The MO sensitivity deteriorates for small mixing with the sterile neutrino, but with increasing sterile mixing the MO sensitivity improves for all the experiments. The impact of various BSM physics scenarios on the MO sensitivities has been well explored for various LBL $\nu$-experiments. Long-baseline accelerator experiments will provide the best measurement of $\Delta m_{31}^{2}$. Note that, the presence of new physics can impact the measurement of $|\Delta m_{31}^{2}|$ as well as the determination of neutrino MO. The presence of SNSI in nature would be no exception. In this paper, we particularly probe the impact of diagonal scalar NSI elements on the MO sensitivities in the context of upcoming LBL experiments. We also perform a combined analysis for a chosen combination of LBL experiments as it can lead to increased statistics and broader parameter space. It may also help in better constraining the oscillation parameters.

The study of scalar NSI is a well-motivated phenomenological scenario to probe new physics searches in the leptonic sector. The possible scalar coupling of neutrinos with environmental fermions can modify the $\nu$-oscillation probabilities and also affect the physics sensitivities of various $\nu$-oscillation experiments \cite{Ge:2018uhz, Yang:2018yvk, Khan:2019jvr,Gupta:2023wct}. This kind of unique coupling can manifest as medium-dependent corrections directly to the $\nu$-mass matrix. Scalar NSI can be a unique probe to explore new physics phenomena in the neutrino sector. The effects of scalar NSI show a linear scaling relationship with environmental matter density, which makes LBL experiments a promising candidate for probing its effects. 

At present, there are no stringent experimental bounds on the scalar NSI $\eta_{\alpha\beta}$ parameters. The reference \cite{Ge:2018uhz} has shown a preliminary attempt at estimating scalar NSI parameter using data from Borexino. It suggests a preference for the scalar NSI parameter $\eta_{ee}$ to be -0.16, specifically within the context of the solar sector, which of course needs to be translated by the density of the ratio. Considering the wide range of the Sun's density from $10^{2}$ to $10^{-2}$ over the Sun's radius, we may approximately translate the $\eta$ to obtain a range of $\eta_{ee}$ from $\sim$[-0.004, -0.464], if the Earth's density is fixed at $\rho=2.9g/cc$. It may be noted that, the fit shown in \cite{Ge:2018uhz} is rather naive, and further investigation using data from various experiments would be required for a clearer picture. 

The values of $\eta_{\alpha\beta}$ that we use in our study serve solely as illustrative examples, allowing us to examine how scalar NSI influences the sensitivities of neutrino mass ordering across various long-baseline experiments. As the scalar NSI would be a sub-dominant effect, the modified Hamiltonian is expected to be a small perturbation to the original Hamiltonian. Some studies like \cite{Babu:2019iml,Venzor:2020ova} have put bounds on the scalar NSI parameters under astrophysical and cosmological limits, although they are not so stringent. The authors in reference \cite{Medhi:2021wxj,Medhi:2022qmu} have explored the impact of scalar NSI on the CP-measurement sensitivities at DUNE and have also performed a combined analysis with T2HK and T2HKK. In reference \cite{Denton:2022pxt}, the authors have investigated various new physics scenarios in LBL experiments (NOvA, T2K and DUNE) and have found that in many cases DUNE will have better model discrimination. The authors in reference \cite{Singha:2023set} have studied the effects of scalar NSI on the determination of CP-violation, $\theta_{23}$ octant and MO in the context of P2SO and DUNE. In reference \cite{Medhi:2023ebi}, we have constrained the lightest $\nu$-mass in the presence of scalar NSI as it has a direct dependence on the absolute $\nu$-mass. 

In this paper, we have explored the impact of scalar-mediated NSI on the neutrino MO sensitivity at the LBL experiments i.e. DUNE, HK (a.k.a. T2HK) and HK+KNO (a.k.a. T2HKK). We study the impact on mass ordering sensitivities at these experiments assuming that scalar NSI parameters are present in nature and are known from other non-LBL experiments. We have considered flavor conserving texture of scalar NSI and probed the impact of diagonal scalar NSI elements $(\eta_{ee},\eta_{\mu\mu},\eta_{\tau\tau})$ on the MO sensitivities for the chosen LBL experiments by taking only one element at a time. We find that the presence of scalar NSI can significantly improve/deteriorate the MO sensitivity of LBL experiments. A positive $\eta_{ee}$, mostly enhances the MO sensitivity as compared to the standard case for all three experiments. In this study, for the first time, we have performed an MO analysis in the presence of scalar NSI with the synergy of experiments viz. DUNE+HK and DUNE+HK+KNO. For the combined analysis of experiments, the overall MO sensitivities have significantly improved. It can also help in resolving underlying degeneracy in MO determination in the presence of scalar NSI. This is expected as the synergy of the experiments improves the statistics as well as widens the parameter region of the analysis. Currently, most of the $\nu$-experiments aim at determining the neutrino MO unambiguously. The determination of neutrino MO is crucial as it would help us to understand the nature of neutrinos and could also provide insights into the origin of $\nu$-mass. 

The paper is organized into different sections as follows. We first discuss the formalism of scalar NSI in section \ref{sec:scalar_NSI}. The methodology and experimental details of the analysis have been described in section \ref{sec:methdology}. We present our findings, along with associated discussions, in section \ref{sec:results}. We then summarise the study in section \ref{sec:summary}.

\section{Formalism of Scalar NSI} \label{sec:scalar_NSI}

In SM, neutrinos interact with the environmental fermions by weak interactions while passing through matter. These interactions can either be charged-current (CC) type, mediated by a W$^\pm$ or neutral-current (NC) type, mediated by a Z$^0$ boson \cite{Linder:2005fc}. The CC interactions can affect $\nu$-oscillations by introducing a flavor-dependent matter potential, whereas NC interactions are flavor-independent and do not affect the $\nu$-oscillations. Therefore, the effective Lagrangian for standard matter effect in $\nu$-oscillations can be written as \cite{Wolfenstein:1977ue, Nieves:2003in, Nishi:2004st, Maki:1962mu, Bilenky:1987ty}, 

\begin{equation}\label{eq:CC_Lag}
\mathcal{L}_{CC}^{eff} = - \frac{4 G_F}{\sqrt{2}} [\bar{\nu_e}(p_3) \gamma_\mu P_L \nu_e (p_2)][\bar e (p_1) \gamma^\mu P_L e(p_4)].
\end{equation}

\noindent In equation \ref{eq:CC_Lag}, $G_F$ is the Fermi coupling constant,  $P_L = (1-\gamma_5)/2$ is the chiral projection operator for left-handed neutrinos and $p_i$'s are momenta of incoming and outgoing fermions as well as neutrino states. Therefore, the effective Hamiltonian for matter-effect in $\nu$-oscillations expressed in flavor basis is,
\begin{equation} \label{eq:Hamiltonian_SI}
\mathcal{H}_{matter}= E_\nu + \frac{M M^\dagger}{2 E_\nu} \pm V_{SI}.
\end{equation}
\noindent In equation \ref{eq:Hamiltonian_SI}, $V_{SI}$ is the matter potential with positive and negative signs indicating neutrinos and anti-neutrinos respectively. $E_{\nu}$ is the energy of neutrinos. M is the $\nu$-mass matrix in flavor basis, which is given by $\mathcal{U} D_\nu \mathcal{U}^\dagger$, where $D_\nu = diag(m_1, m_2, m_3)$ i.e. the diagonal mass matrix of neutrinos. $\mathcal{U}$ is the PMNS mixing matrix \cite{Pontecorvo:1957cp,Pontecorvo:1957qd,Pontecorvo:1967fh,ParticleDataGroup:2020ssz}. The effective Hamiltonian can be simplified to the following form in terms of the mass-squared differences ($\Delta m_{ij}^2 = m_i^2 - m_j^2$) as


\begin{equation} \label{eq:Hamiltonian_SI_eff}
\mathcal{H}_{eff}= E_\nu + \frac{1}{2 E_\nu} \mathcal{U}.diag( 0, \Delta m_{21}^2, \Delta m_{31}^2).\mathcal{U}^\dagger + diag(V_{CC} ,0,0).
\end{equation}

\noindent In BSM scenarios, the vector-mediated NSI has been extensively explored in the past few years. However, the possibility of NSIs mediated by a scalar particle is another intriguing scenario, which has very compelling phenomenological consequences. The effective Lagrangian for $\nu$-fermion coupling mediated by a scalar particle, $\phi$ having mass $m_\phi$ is,

\begin{equation} \label{eq:scalar_lag}
  \mathcal{L}_{eff}^s = \frac{y_f y_{\alpha \beta}}{m_\phi^2} [\bar \nu_{\alpha} (p_3) \nu_\beta (p_2)] [\bar f(p_1) f(p_4)],
\end{equation}

where,
\begin{itemize}
    \item $\alpha, \beta = (e, \mu, \tau)$ refers to the three active $\nu$-flavors,
    \item $f$($\bar f$) represents matter fermions (anti-fermions),
    \item $y_f$ is the Yukawa coupling of scalar $\phi$ with matter fermions $f$, and,
    \item $y_{\alpha \beta}$ is the Yukawa coupling of $\phi$ with the active $\nu$-flavors.
\end{itemize}

\noindent The presence of Yukawa terms in equation \ref{eq:scalar_lag} prevents us from converting it to vector currents. Consequently, it does not manifest itself as an additional matter potential. In contrast, the contribution of scalar NSI is a perturbation to the mass term in the $\nu$-Hamiltonian. Thus, the presence of scalar NSI modifies the Dirac equation as follows. 

\begin{equation} \label{eq:Dirac_snsi}
 \bar \nu_\beta \left[ i \partial_\mu \gamma^\mu + \left( M_{\beta \alpha}   + \frac{ \sum_f n_f y_f y_{\alpha \beta}}{m_\phi^2} \right) \right] \nu_\alpha =0,
\end{equation}

\noindent where $n_f$ represents the number density of matter fermions. The effective form of matter Hamiltonian with scalar NSI correction can be framed as, 
\begin{equation}\label{eq:Heff_SNSI}
   \mathcal{H}_{SNSI} \equiv E_\nu + \frac{M_{eff} M_{eff}^\dagger}{2 E_\nu} \pm V_{SI}.
\end{equation}

\noindent In equation \ref{eq:Heff_SNSI}, the effective $\nu$-mass matrix ($M_{eff}$) has contributions from the genuine $\nu$-mass matrix (M) as well as the correction due to scalar NSI i.e. $M_{SNSI} \equiv \sum_f n_f y_f y_{\alpha \beta}/ m_{\phi}^2$. The effective mass matrix can be expressed as $M_{eff} = M + M_{SNSI}$. The $\nu$-mass matrix can be diagonalized by the mixing matrix $\mathcal{U}^\prime$. $\mathcal{U}^\prime$ has the form $P \mathcal{U} Q^\dagger$, with P and Q representing an unphysical rephasing matrix and the Majorana rephasing matrix respectively. We can rotate away the Majorana rephasing matrix by $QD_{\nu} Q^\dagger = D_\nu$, but the diagonal rephasing matrix P can not be rotated away. Therefore, rotating the unphysical rephasing matrix P into scalar NSI correction and denoting it as $\delta M$, we have $M_{eff}$ as,
\begin{equation} \label{eq:Meff_final}
 M_{eff}   \equiv \mathcal{U} D_\nu \mathcal{U}^\dagger + P^\dagger M_{SNSI}P \equiv M+ \delta M,
\end{equation}
where,
\begin{equation} \label{eq:deltaM_para}
    \delta M \equiv S_{m} \begin{pmatrix} \eta_{ee} & \eta_{e \mu} & \eta_{e \tau}\\
\eta_{e \mu}^{*} & \eta_{\mu \mu} & \eta_{\mu \tau} \\
\eta_{e \tau}^{*} & \eta_{\mu\tau}^{*} & \eta_{\tau \tau}
    \end{pmatrix}.
\end{equation}

\noindent We use a similar parameterization of $\delta M$ as shown in \cite{Ge:2018uhz,Medhi:2021wxj,Medhi:2022qmu}. A rescaling factor of $S_{m}$ is introduced in order to match the dimension with the $\nu$-mass matrix. In this work, $S_{m}$ is fixed to be $\sqrt{2.55\times 10^{-3}eV^{2}}$ which corresponds to the magnitude of atmospheric mass-squared splitting $\sqrt{|\Delta m_{31}^{2}|}$. The factor $S_{m}$ introduced is merely a rescaling of the matrix such that $\eta_{\alpha\beta}$ are dimensionless parameters that signify only the strength of scalar NSI. The hermiticity of the Hamiltonian allows the diagonal parameters to be real and the off-diagonal parameters to be complex. In this work, the effects of diagonal scalar NSI parameters are explored by taking one element at once. The forms of $M_{eff}$ for the three cases are shown in table \ref{tab:M_eff}. Note that, using these forms of $M_{eff}$ in equation \ref{eq:Heff_SNSI}, we can calculate $\mathcal{H}_{SNSI}$. Due to the presence of $\delta M$ term, we will see a dependence on the absolute $\nu$-mass in addition to the dependence on $\Delta m_{ij}^{2}$.

\begin{table}[h]
    \centering
    \begin{tabular}{|c|c|c|}
    \hline 
    Case &  Scalar NSI parameter & ${M_{eff}}$\tabularnewline
    \hline
    I & $\eta_{ee}$ & $\mathcal{U} \boldsymbol{\cdot} D_{\nu} \boldsymbol{\cdot}  \mathcal{U}^\dagger + S_{m}diag(\eta_{ee}, 0, 0)$\tabularnewline
    \hline 
    II & $\eta_{\mu \mu}$ & $\mathcal{U} \boldsymbol{\cdot} D_{\nu} \boldsymbol{\cdot} \mathcal{U}^\dagger + S_{m} diag(0,\eta_{\mu \mu}, 0)$\tabularnewline
    \hline 
    III & $\eta_{\tau \tau}$ & $\mathcal{U} \boldsymbol{\cdot} D_{\nu} \boldsymbol{\cdot} \mathcal{U}^\dagger + S_{m} diag(0,0,\eta_{\tau \tau})$\tabularnewline
    \hline 
    \end{tabular}
\caption{Cases of $M_{eff}$ for the diagonal scalar NSI parameters.}
\label{tab:M_eff}
\end{table}

\noindent An intriguing implication of scalar NSI is its dependence on the absolute $\nu$-mass as explored in reference \cite{Medhi:2023ebi}. All the values of scalar NSI parameters $\eta_{\alpha\beta}$ are defined at matter density $\rho$=2.9 g/cc. We have scaled the $\eta_{\alpha\beta}$ values for the densities of the experiments i.e. DUNE, HK and HK+KNO. However, the impact of rescaling $\eta_{\alpha\beta}$ values is nominal for the chosen LBL experiments.

\section{Methodology}\label{sec:methdology}
We describe the methodology of our work here. We first give the details of the simulation of the experiments in section \ref{sec:sim_details}. We then mention the details and specifications of the three chosen LBL experiments in section \ref{sec:expt_details}.

\subsection{Simulation Details}\label{sec:sim_details}
In this work, we have mainly focused on exploring the impact of diagonal scalar NSI parameters on the MO sensitivities at LBL experiments. We have used the simulation package i.e. General Long Baseline Experiment Simulator (GLoBES) \cite{Huber:2004ka, Kopp:2006wp, Huber:2007ji} for our numerical simulations of the experiments. The values of oscillation parameters used in our analysis are listed in table \ref{tab:mixing_parameters}. 
\begin{table}[!h]
    \centering
    \begin{tabular}{|c|c|c|c|}
    \hline 
    \multicolumn{2}{|c|}{Parameters} & True Values & Marginalization range\tabularnewline
    \hline 
    \multicolumn{2}{|c|}{$\theta_{12}$ $[^{\circ}]$} & $34.51$ & fixed \\
    \multicolumn{2}{|c|}{$\theta_{23}$ $[^{\circ}]$} & $47$ & 40 $\rightarrow$ 50\\
    \multicolumn{2}{|c|}{$\theta_{13}$ $[^{\circ}]$} & $8.44$ & fixed\\
    \multicolumn{2}{|c|}{$\delta_{CP}$} & $-\pi/2$ & $-\pi$ $\rightarrow$ $\pi$\\
    \multicolumn{2}{|c|}{$\Delta m_{21}^{2}$ [$10^{-5}eV^{2}$]} & $7.56$ & 6.82 $\rightarrow$ 8.04\\
    \multicolumn{2}{|c|}{$\Delta m_{31}^{2}$ (NO) [$10^{-3}eV^{2}$]} & $2.55$ & (2.25 - 2.65)\\
    \multicolumn{2}{|c|}{$\Delta m_{31}^{2}$ (IO) [$10^{-3}eV^{2}$]} & $-2.49$ & - (2.25 - 2.65)\\
    \hline
    \end{tabular}
    \caption{Benchmark values of $\nu$-oscillation parameters used in GLoBES simulation \cite{NuFIT5.0}.}
\label{tab:mixing_parameters}
\end{table}

\noindent In order to study the sensitivity of experiments towards the determination of neutrino MO, we define a statistical $\chi^2$ as
\begin{equation}\label{eq:chisq}
\chi^2 \equiv  \min_{\eta}  \sum_{i} \sum_{j}
\frac{\left[N_{true}^{i,j} - N_{test}^{i,j} \right]^2 }{N_{true}^{i,j}},
\end{equation}
where, $N_{true}^{i,j}$ ($N_{test}^{i,j}$) are the number of true (test) events in the $\{i,j\}$-th bin. The measure of statistical significance is obtained via a minimization over all the systematic uncertainties which are incorporated in the $\chi^{2}$ using pull method \cite{Fogli:2002pt,Huber:2004ka}. We have also incorporated the uncertainties on the oscillation parameters in the GLoBES simulation for the calculation of $\chi^{2}$. We have marginalized over the oscillation parameters $\theta_{23}$, $\delta_{CP}$ and the mass-squared splitting in the range as shown in the third column of table \ref{tab:mixing_parameters}. Throughout this work, we have fixed the lightest neutrino mass $m_{1}$ $(m_{3})$ at $10^{-5}$ eV for normal (inverted) mass ordering and the other neutrino masses are calculated accordingly from the best-fit values of $\Delta m_{21}^2$ and $\Delta m_{31}^2$ respectively. However, a prior scan of probability and sensitivity with respect to the various choices of the absolute neutrino masses is shown in Appendix \ref{app:abs_mass_dep}. This also accounts for the cosmological bound on the sum of neutrino masses i.e. $\sum_{i}m_{i}<0.12$ eV \cite{Planck:2018vyg}. In our analysis, we have considered the higher octant of $\theta_{23}$ as the true octant. In order to study the impact of scalar NSI on the MO sensitivities, we have considered three upcoming LBL experiments viz. DUNE, HK and HK+KNO. The detailed experimental setup and configuration of the detectors are discussed in the section \ref{sec:expt_details}.

\subsection{Overview of the three LBL experiments}\label{sec:expt_details}
The experimental systematics and background information are taken from the Technical Design Reports (TDR) of the corresponding experiments. The details of the experiments are summarised in table \ref{tab:experimental_setup}. In the subsequent sections, we briefly discuss the technical details of the three future LBL $\nu$-experiments viz. DUNE, HK, and HK+KNO.

\subsubsection{DUNE}
The Deep Underground Neutrino Experiment (DUNE) \cite{DUNE:2016hlj, DUNE:2015lol, DUNE:2016rla, DUNE:2020ypp, DUNE:2021tad} is a flagship accelerator-based $\nu$-experiment to be hosted by Fermilab. The Long-Baseline Neutrino Facility (LBNF) of Fermilab which can deliver $1.1 \times 10 ^{21}$ proton on target (POT) per year with its 1.2 MW proton beam will produce the neutrinos for the experiment. It will have two detectors, the near detector at Fermilab will be situated 60m underground and will have a baseline of 574 meters. The far detector with a 1300 km baseline will be situated 1.5 km below the earth's surface at the Sanford Underground Research Facility, South Dakota. The far detector module will consist of four Liquid Argon Time Projection Chambers (LArTPC) each having a fiducial mass of 10 kt. The excellent energy reconstruction, 3D image tracking and particle identification capabilities of LArTPC technology will provide exceptional precision to DUNE in probing neutrino properties.

\begin{table}[!t]
    \label{tab:detector_details}
    \centering
    \begin{tabular}{|c|c|c|c|}
    \hline
       \multirow{2}{*}{Experiment details} &\multirow{2}{*}{Channels} & \multicolumn{2}{|c|}{Normalization error} \\\cline{3-4}
        & & Signal & Background \\
        \hline 
        \hline
         \textbf{DUNE} &  & & \\
         Baseline = 1300 km &  $\nu_e (\bar \nu_e)$ appearance & 2 \% (2\%) & 5 \% (5 \%) \\
         L/E = 1543 km/GeV & & & \\
         Fiducial mass = 40 kt (LArTPC) & $\nu_\mu (\bar \nu_\mu)$ disappearance & 5 \% (5 \%) & 5 \% (5 \%) \\ 
         Runtime = 3.5 yr $\nu + 3.5$ yr $\bar \nu$ & & & \\
    \hline
        \textbf{HK} &  & & \\
         Baseline = 295 km &  $\nu_e (\bar \nu_e)$ appearance & 3.2 \% (3.9 \%) & 10 \% (10 \%) \\
         L/E = 527 km/GeV & & & \\
         Fiducial mass = 187 kt (WC) & $\nu_\mu (\bar \nu_\mu)$ disappearance & 3.6 \% (3.6 \%) & 10 \% (10 \%) \\ 
         Runtime = 2.5 yr $\nu + 7.5$ yr $\bar \nu$ & & & \\
    \hline
        \textbf{HK+KNO} &  & & \\
         Baseline = 295, 1100km &  $\nu_e (\bar \nu_e)$ appearance & 3.2 \% (3.9 \%) & 10 \% (10 \%) \\
         L/E = 527, 1964 km/GeV & & & \\
         Fiducial mass = 187, 187 kt(WC) & $\nu_\mu (\bar \nu_\mu)$ disappearance & 3.6 \% (3.6 \%) & 10 \% (10 \%) \\ 
         Runtime = 2.5 yr $\nu + 7.5$ yr $\bar \nu$ & & & \\
    \hline
    \end{tabular}
    \caption{Detector details and systematic uncertainties for DUNE, HK and HK+KNO.}
    \label{tab:experimental_setup}
\end{table}

\subsubsection{HK}
In our analysis, we have referred the Tokai to Hyper-Kamiokande (T2HK) \cite{Hyper-KamiokandeProto-:2015xww} detector as HK. This will be a long-baseline counterpart of the Hyper-Kamiokande (Hyper-K) experiment which is an upgradation of the successful Super-Kamiokande (Super-K) experiment. It will have a water Cherenkov (WC) detector of fiducial mass of 187 kt which will be 8.4 times larger than its predecessor Super-K. It will use neutrinos produced by the 1.2 MW beam at the J-PARC beam facility in Japan. It can produce 27$\times 10^{21}$ POT per year. The neutrinos will be detected by the Hyper-K detector situated 2.5$^\circ$ off-axis from the beam facility at the baseline of 295 km.

\subsubsection{HK+KNO}
Tokai to Hyper-Kamiokande to Korea \cite{Hyper-Kamiokande:2016srs} a.k.a. T2HKK is referred to as HK+KNO in our analysis. Korean Neutrino Observatory (KNO) is the second water Cherenkov detector which is planned to be placed at Korea at a distance of 1100 km from the J-PARC facility. It is an extension of Hyper-K experiment which will consist of two WC detectors, one 187 kt detector at a baseline of 295 km in Japan (HK), and another identical detector with a baseline of 1100 km in Korea (KNO). The 2.5$^\circ$ off-axis beam at a baseline of HK+KNO will enable us to study the oscillations at second oscillations maxima at an energy $\sim 0.6 eV$.

\section{Results and Discussion} \label{sec:results}
We discuss the impact of diagonal scalar NSI elements ($\eta_{ee}$, $\eta_{\mu \mu}$, $\eta_{\tau \tau}$) on the appearance $\nu$-oscillation probability ($P_{\mu e}$), for NO and IO, and how it affects the MO sensitivities in LBL experiments (DUNE, HK, HK+KNO). The presence of scalar NSI leads to a correction in the $\nu$-mass matrix which can modify values of $P_{\mu e}$. In section \ref{sec:oscillation_probabilities}, we first explore the impact of diagonal scalar NSI parameters on $P_{\mu e}$ for varied neutrino energies. We also study its effects on the MO-asymmetry parameter to further investigate its impact on MO determination at the probability level in section \ref{sec:MH_asymm}. The impact of scalar NSI on the MO sensitivities is presented in section \ref{sec:MH_sensitiivty}. The MO sensitivities for the synergy of DUNE with HK and HK+KNO in the presence of scalar NSI are shown in section \ref{sec:Combined_study}. In section \ref{sec:prec}, we explore the $\Delta m_{31}^{2}$ constraining capability of the experiments in presence of scalar NSI.

\subsection{Effects on the appearance channel $P_{\mu e}$}\label{sec:oscillation_probabilities}
\begin{figure}[!h]
	\centering
	\includegraphics[width=0.45\linewidth, height = 5cm]{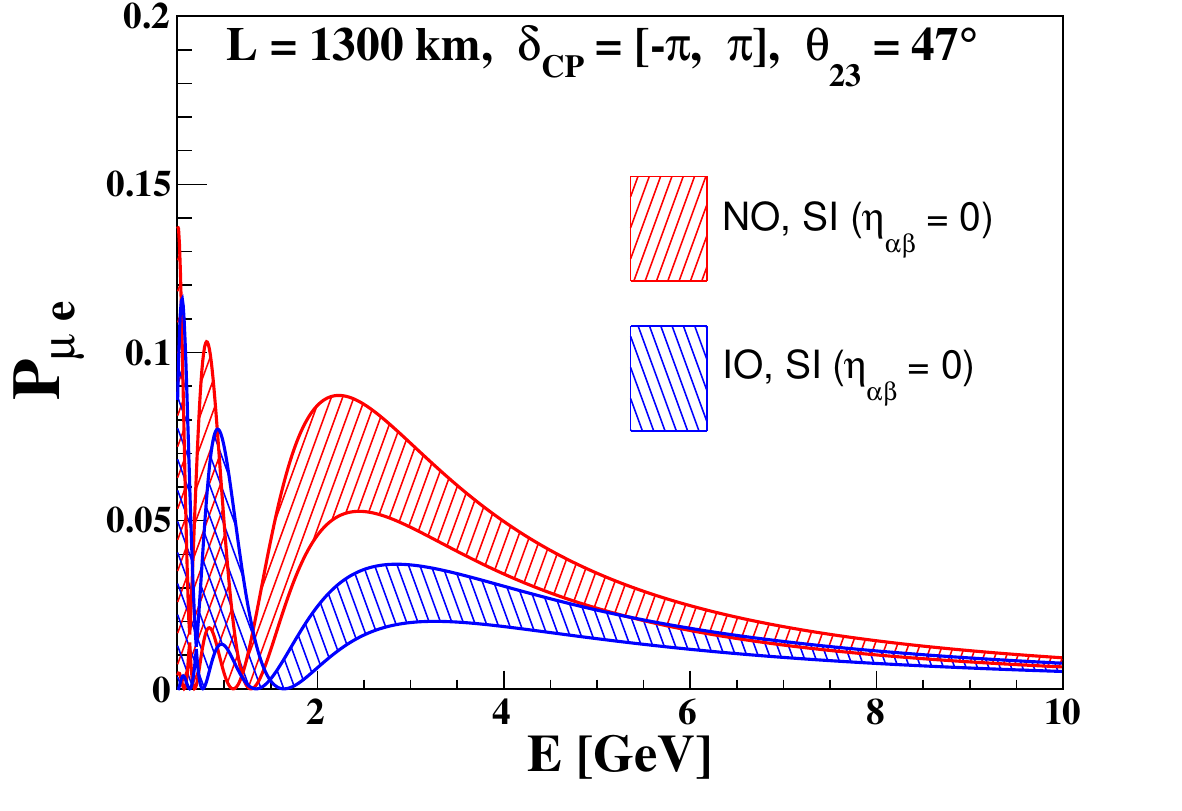} 
	\includegraphics[width=0.45\linewidth, height = 5cm]{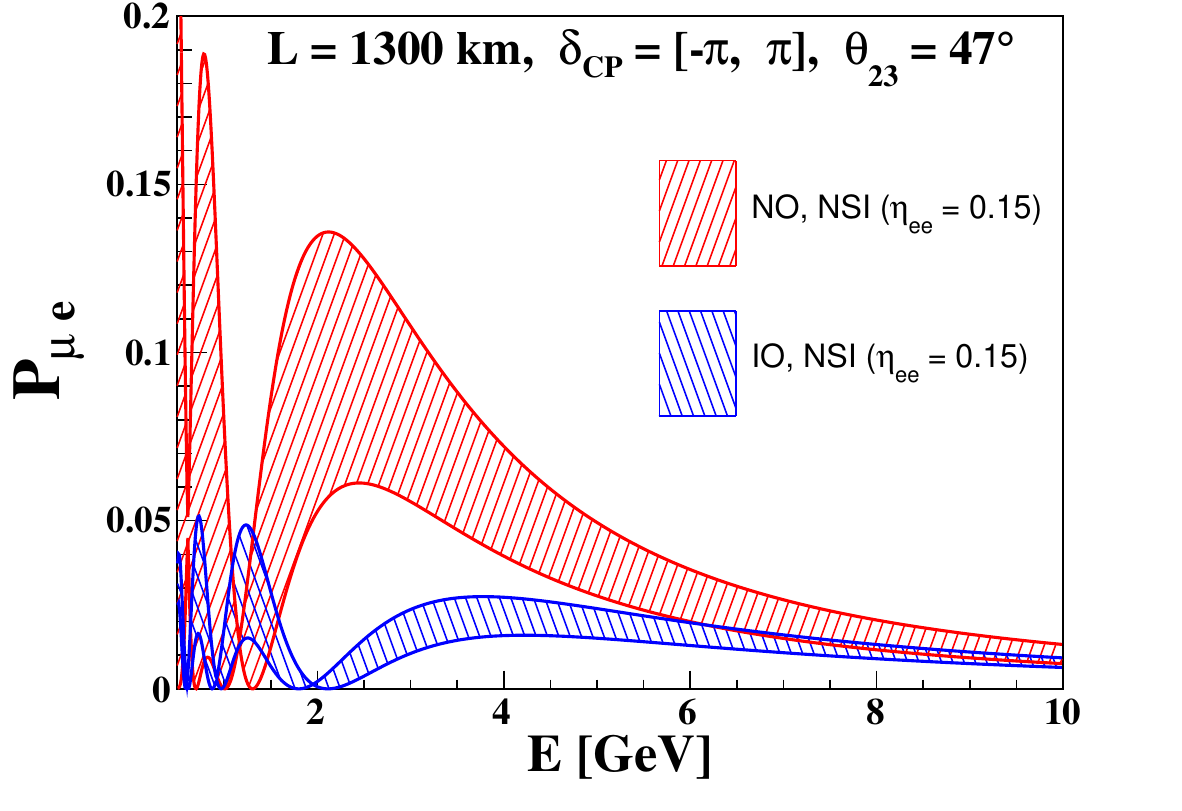} 
	\includegraphics[width=0.45\linewidth, height = 5cm]{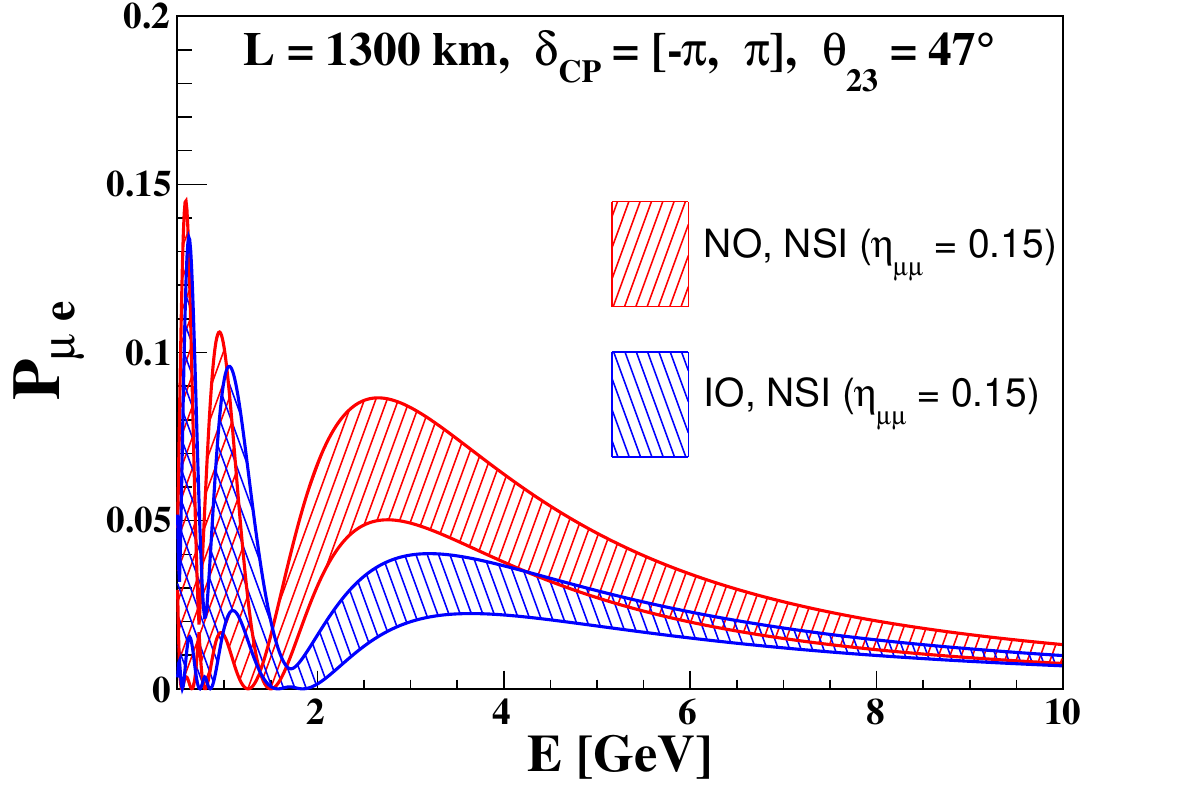} 
 	\includegraphics[width=0.45\linewidth, height = 5cm]{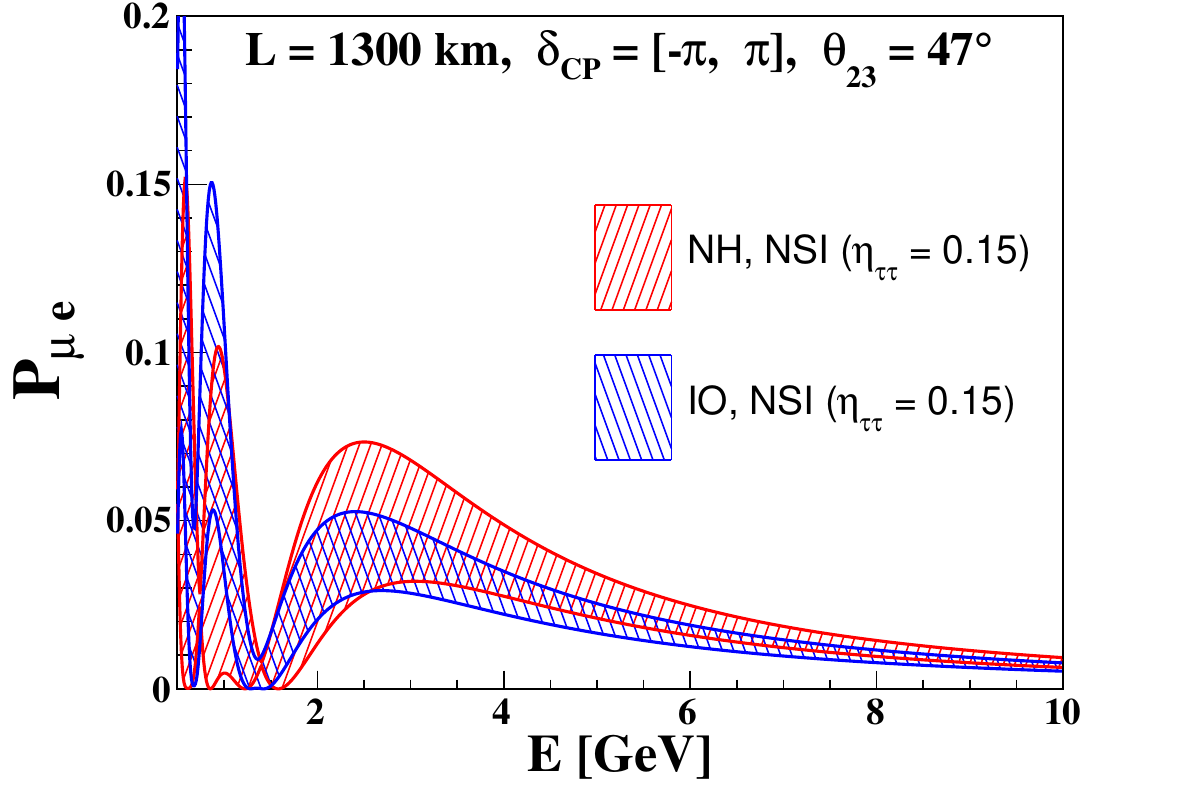} 
	\caption{$P_{\mu e}$ vs E at the DUNE-baseline of 1300km in the presence of scalar NSI, for NO (red band) and IO (blue band). Both the bands correspond to $\delta_{CP}$ values $\in[-\pi,\pi]$. The panels correspond to no scalar NSI (top-left), $\eta_{ee}$ (top-right), $\eta_{\mu \mu}$ (bottom-left) and $\eta_{\tau \tau}$ (bottom-right).}
	\label{fig:probability_1}
\end{figure}

The impact of diagonal scalar NSI elements ($\eta_{ee}$, $\eta_{\mu \mu}$, $\eta_{\tau \tau}$) on $P_{\mu e}$ for varying neutrino energy (E) at baselines corresponding to DUNE, HK and HK+KNO are shown in figure \ref{fig:probability_1}, \ref{fig:probability_2} and \ref{fig:probability_3} respectively. We observe that, 
\begin{itemize}
    \item For no scalar NSI case, the DUNE baseline can clearly distinguish between the NO and IO in the energy range $\sim$ [1.5, 4] GeV. The presence of $\eta_{ee}$ widens the energy range as well as the separation between the mass orderings. The $\eta_{\mu\mu}$ element marginally reduces the separation between the mass orderings and the energy window of the ordering separation. For $\eta_{\tau\tau}$ element, a significant overlapping of bands can be seen.
    
    \item For HK baseline, we observe an overlap between the NO and IO bands for the no scalar NSI case. The presence of $\eta_{ee}$ leads to a larger separation between the bands in the energy range $\sim$ [0.4, 0.7] GeV. For instance, if $\eta_{ee}$ is known to be exactly 0.15 from other experiments, a larger separation between the two bands would indicate a better measurement of the mass ordering. The chosen value of the element $\eta_{\mu\mu}$ shows a nominal change compared to the no scalar NSI case. For $\eta_{\tau\tau}$, the IO band lies above the NO band without any distinct separation between the mass orderings. 

\begin{figure}[!h]
	\centering
	\includegraphics[width=0.45\linewidth, height = 5cm]{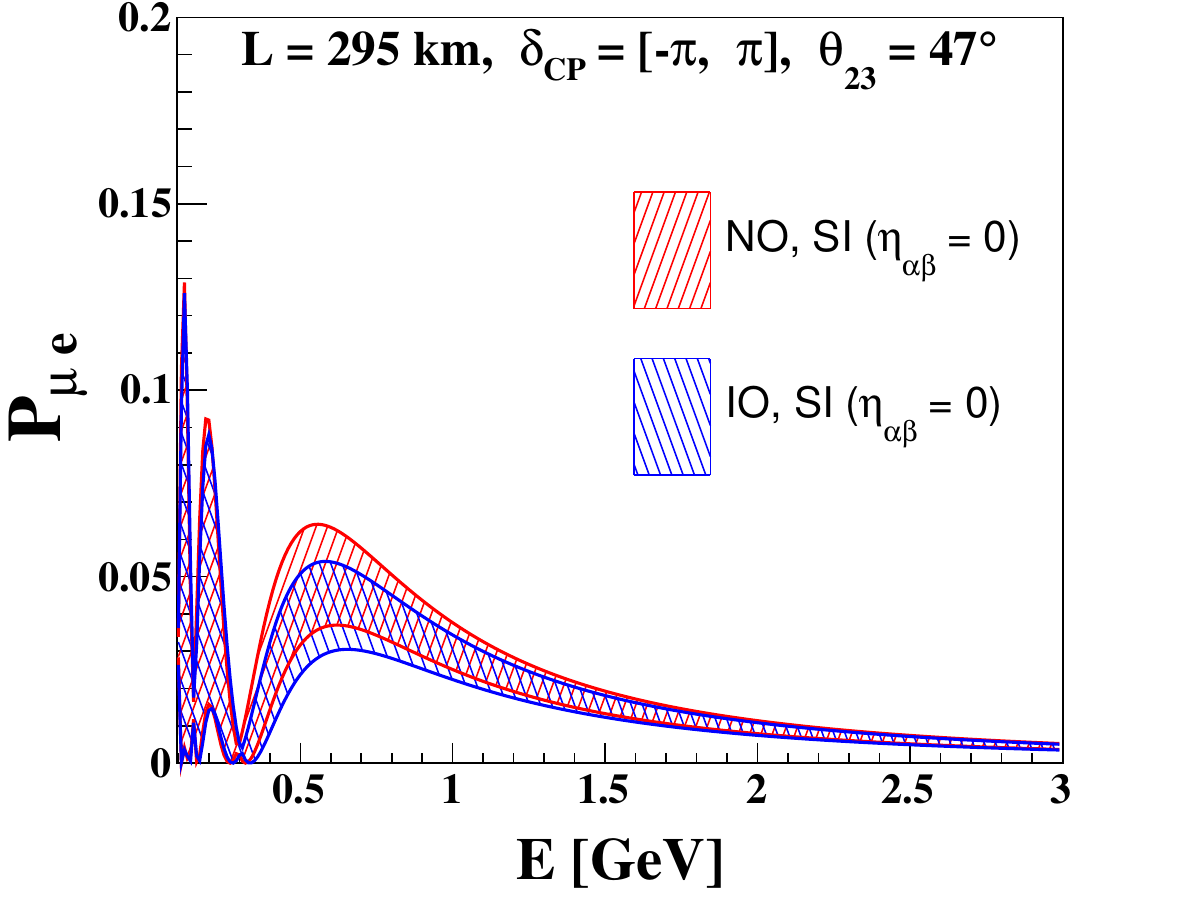} 
	\includegraphics[width=0.45\linewidth, height = 5cm]{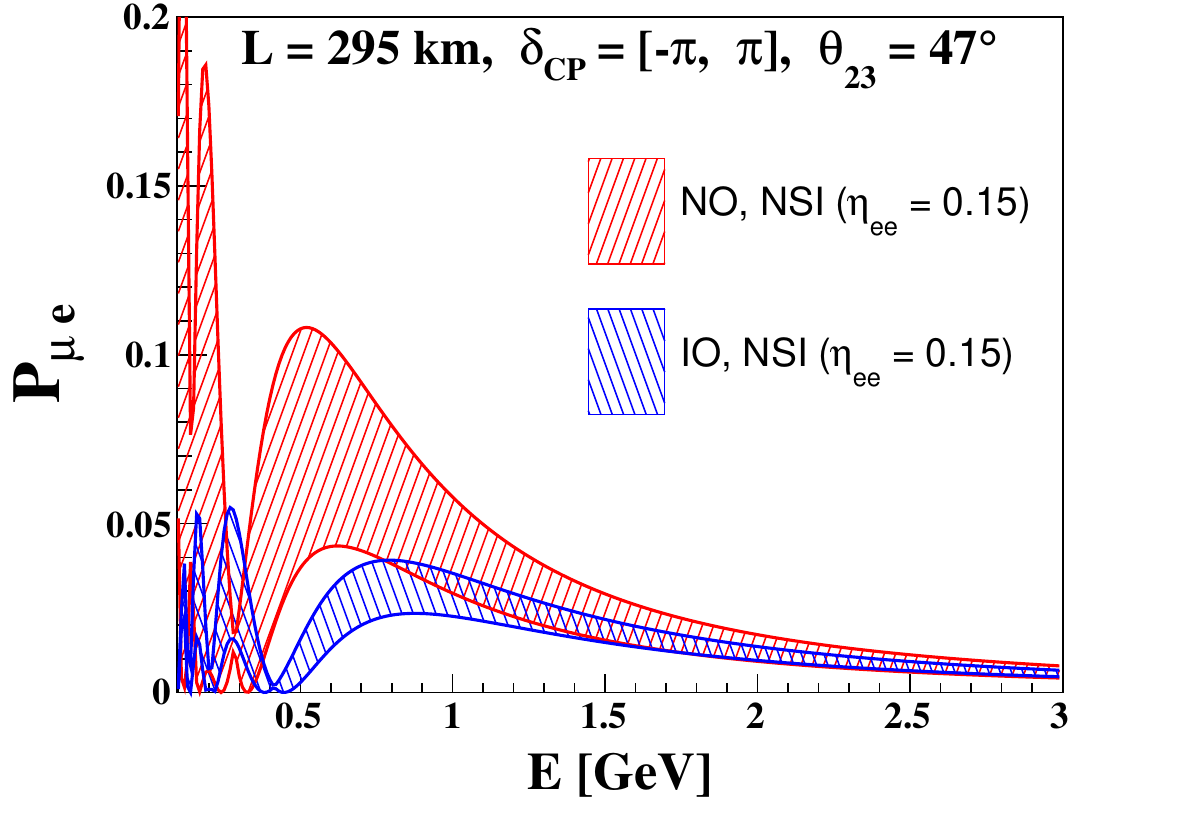} 
	\includegraphics[width=0.45\linewidth, height = 5cm]{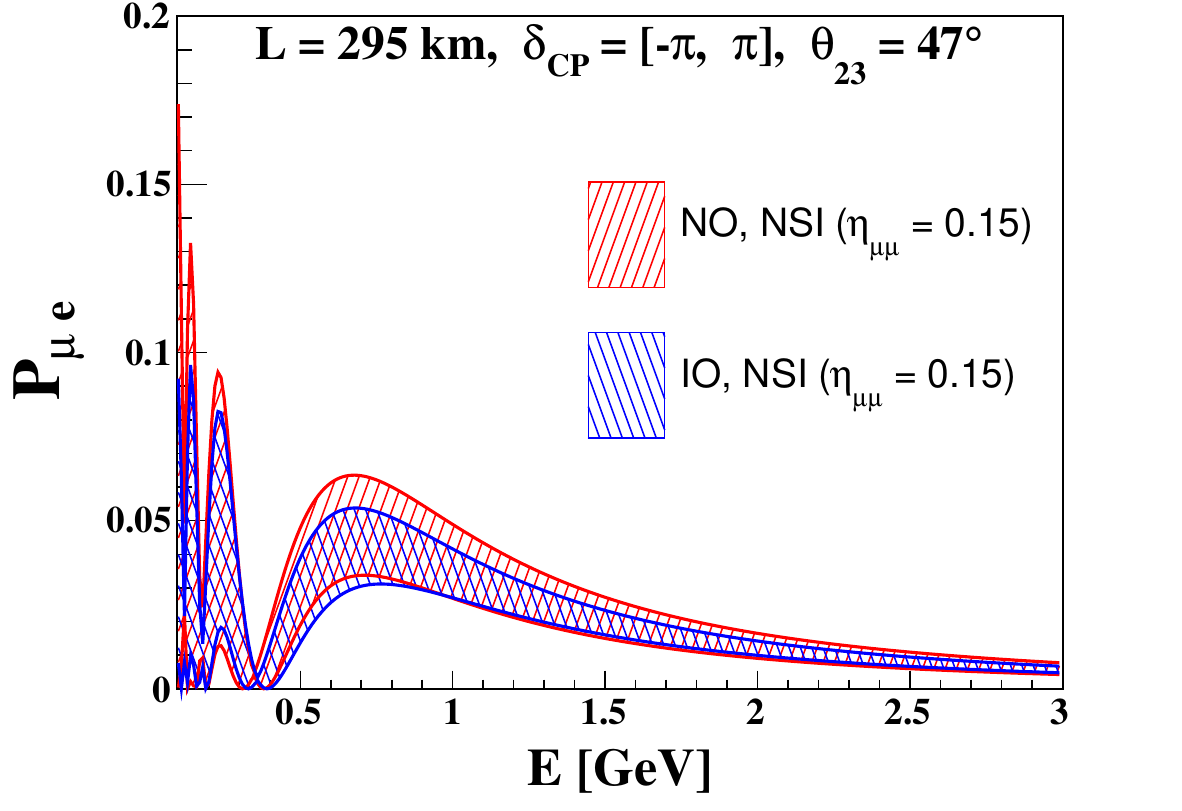} 
 	\includegraphics[width=0.45\linewidth, height = 5cm]{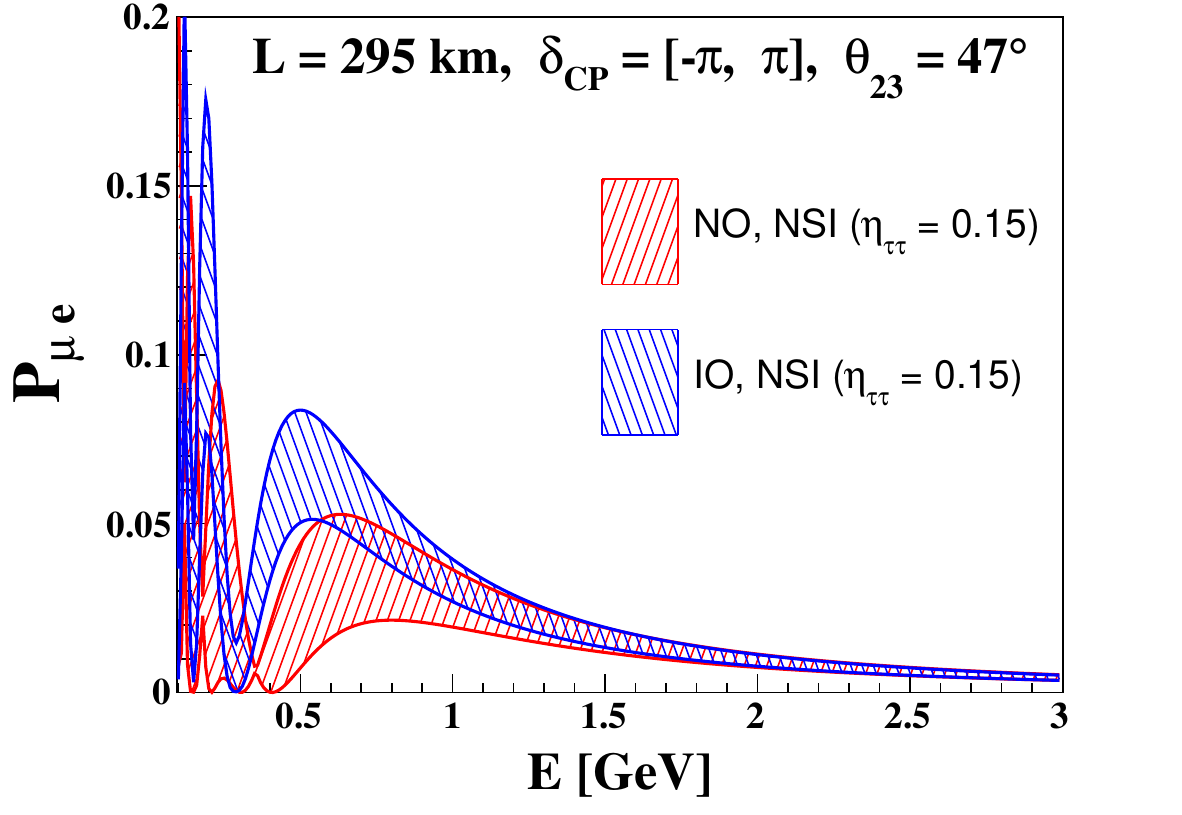} 
  \caption{$P_{\mu e}$ vs E at the HK-baseline of 295km in the presence of scalar NSI, for NO (red band) and IO (blue band). Both the bands correspond to $\delta_{CP}$ values $\in[-\pi,\pi]$. The panels correspond to no scalar NSI (top-left), $\eta_{ee}$ (top-right), $\eta_{\mu \mu}$ (bottom-left) and $\eta_{\tau \tau}$ (bottom-right).}
	\label{fig:probability_2}
\end{figure}

    \item For HK+KNO baseline, we observe clear discrimination between the mass orderings at the probability level in the energy window $\sim$ [1.5, 3] GeV in the absence of scalar NSI. The presence of $\eta_{ee}$ element widens the band separation as well as the energy window. The element $\eta_{\mu\mu}$ reduces both the amplitude as well as the discriminating power of the neutrino MO. For $\eta_{\tau\tau}$, both the NO and IO band are seen to completely overlap. 
\end{itemize}

\begin{figure}[!h]
	\centering
	\includegraphics[width=0.45\linewidth, height = 5cm]{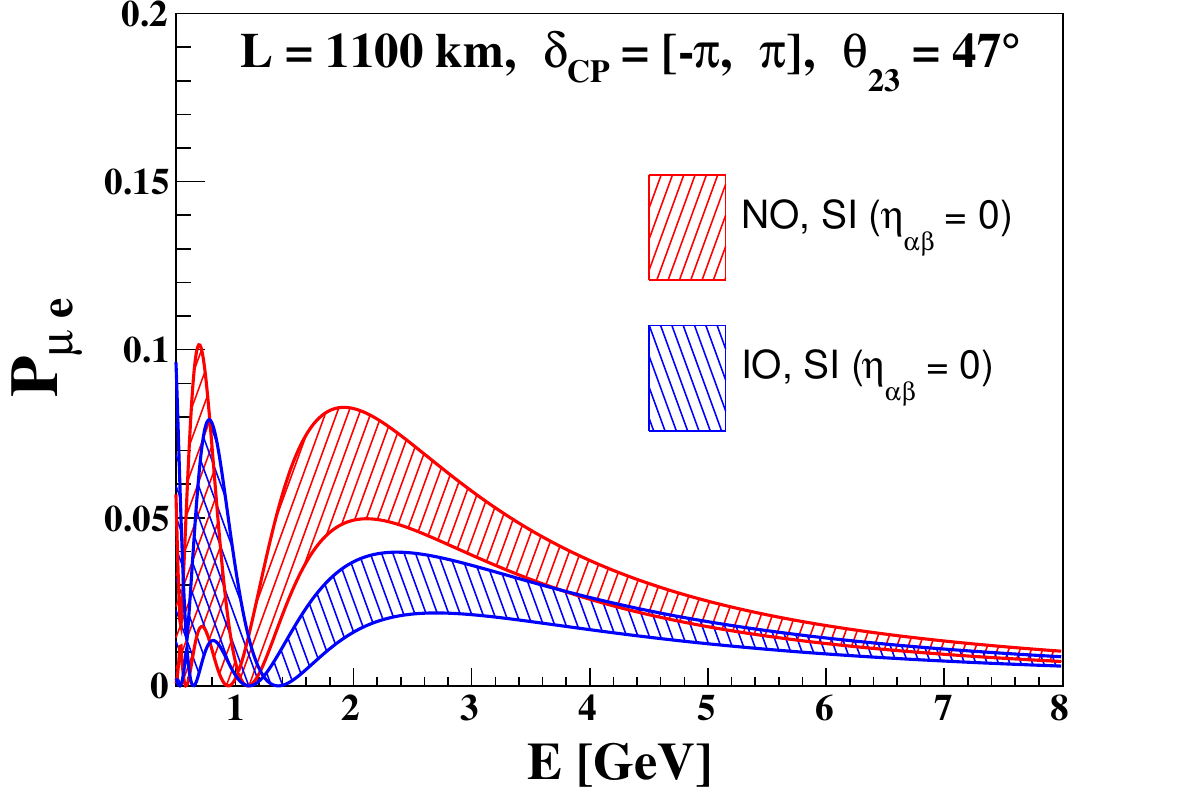} 
	\includegraphics[width=0.45\linewidth, height = 5cm]{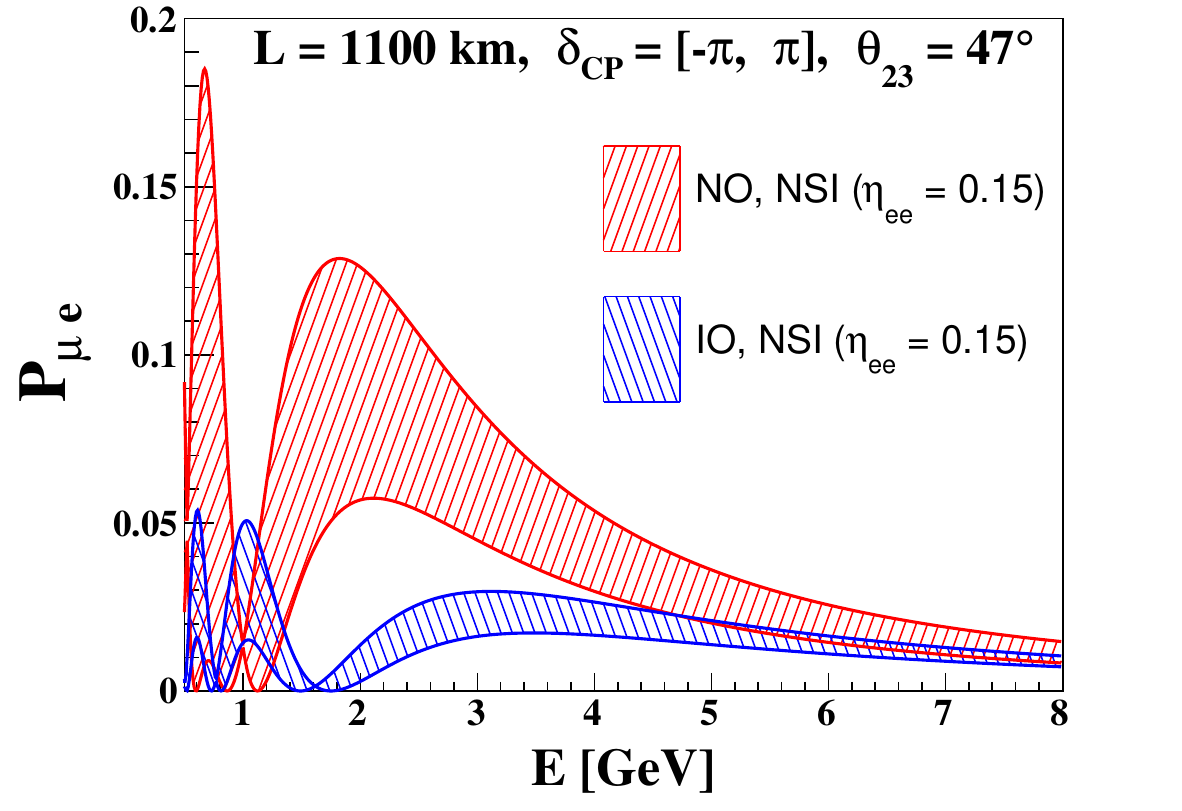} 
	\includegraphics[width=0.45\linewidth, height = 5cm]{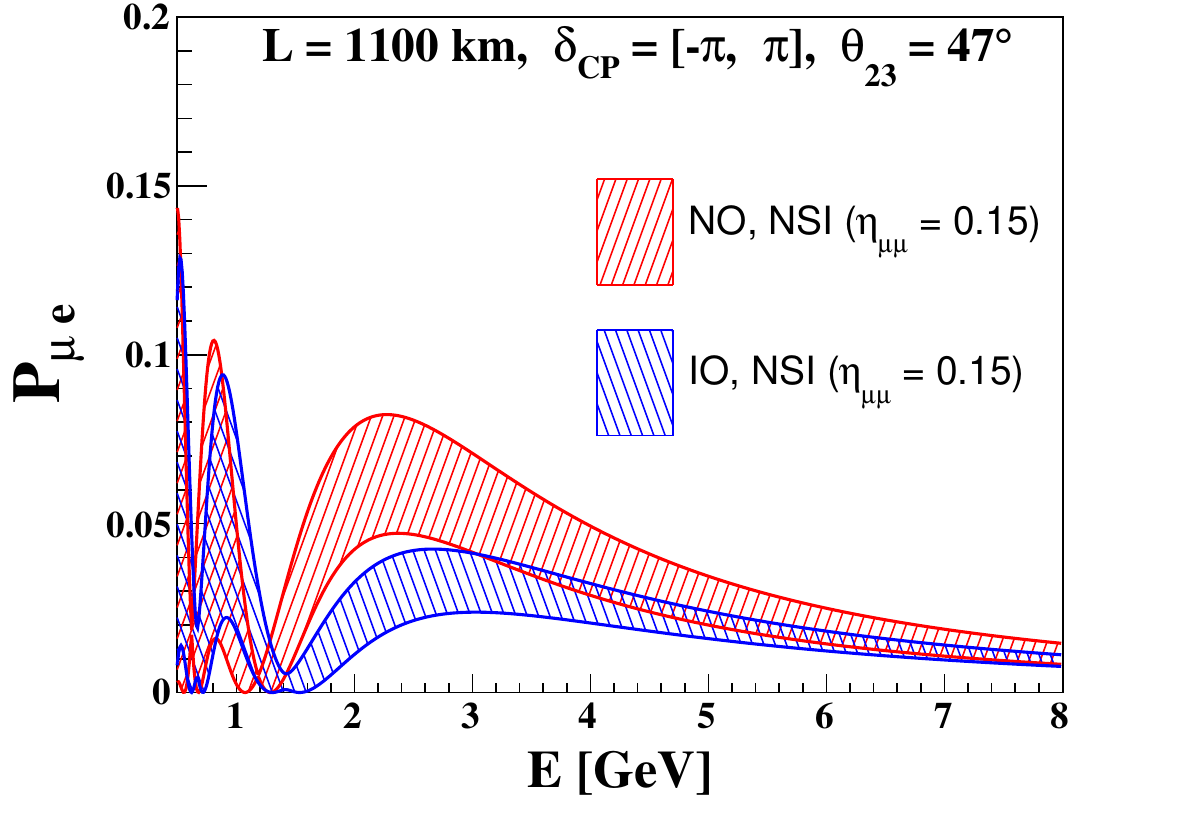} 
 	\includegraphics[width=0.45\linewidth, height = 5cm]{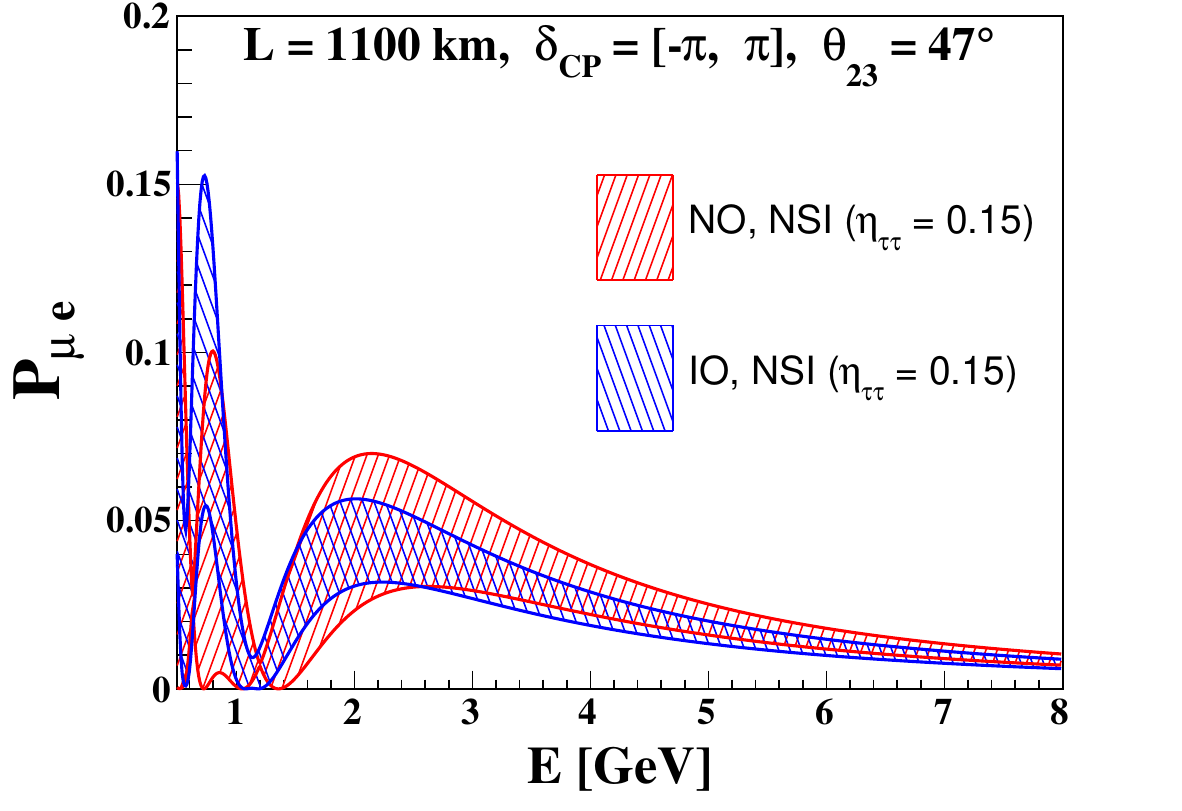}
    \caption{$P_{\mu e}$ vs E at the HK+KNO-baseline of 1100km in the presence of scalar NSI, for NO (red band) and IO (blue band). Both the bands correspond to $\delta_{CP}$ values $\in[-\pi,\pi]$. The panels correspond to no scalar NSI (top-left), $\eta_{ee}$ (top-right), $\eta_{\mu \mu}$ (bottom-left) and $\eta_{\tau \tau}$ (bottom-right).}
	\label{fig:probability_3}
\end{figure}
We have also further explored the impact of scalar NSI as a function of $\Delta P_{\mu e} (= P_{\mu e}^{NSI}- P_{\mu e}^{SI})$ in the $\Delta m_{31}^{2}$-E parameter space for all the three experiments which are shown in appendix \ref{app:deg}. 

\subsection{Imprints on the MO asymmetry} \label{sec:MH_asymm}
In order to quantify the effects of scalar NSI on the MO determination at the probability level, we define an mass ordering asymmetry parameter as
\begin{equation}\label{def_asym}
A_{MO} = \frac{P_{\mu e}^{NO}-{P}_{\mu e}^{IO}}{P_{\mu e}^{NO}+{P}_{\mu e}^{IO}}\,,
\end{equation}
where, $P_{\mu e}^{NO}$ and ${P}_{\mu e}^{IO}$ are the appearance probabilities for NO and IO respectively. This asymmetry parameter may quantify an experiment's capability towards discriminating between the two $\nu$-mass orderings. 

\begin{figure}[!t]
	\centering
	\includegraphics[width=0.33\linewidth, height = 5.5cm]{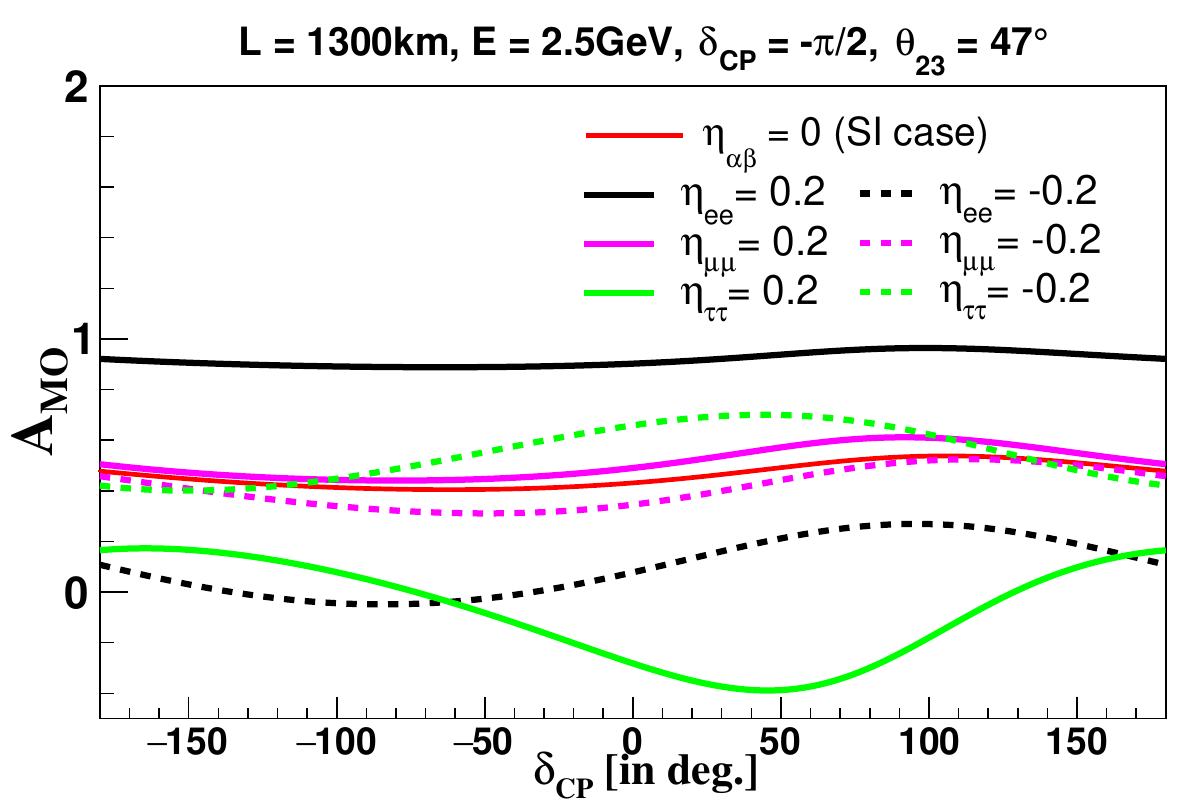} 
	\includegraphics[width=0.33\linewidth, height = 5.5cm]{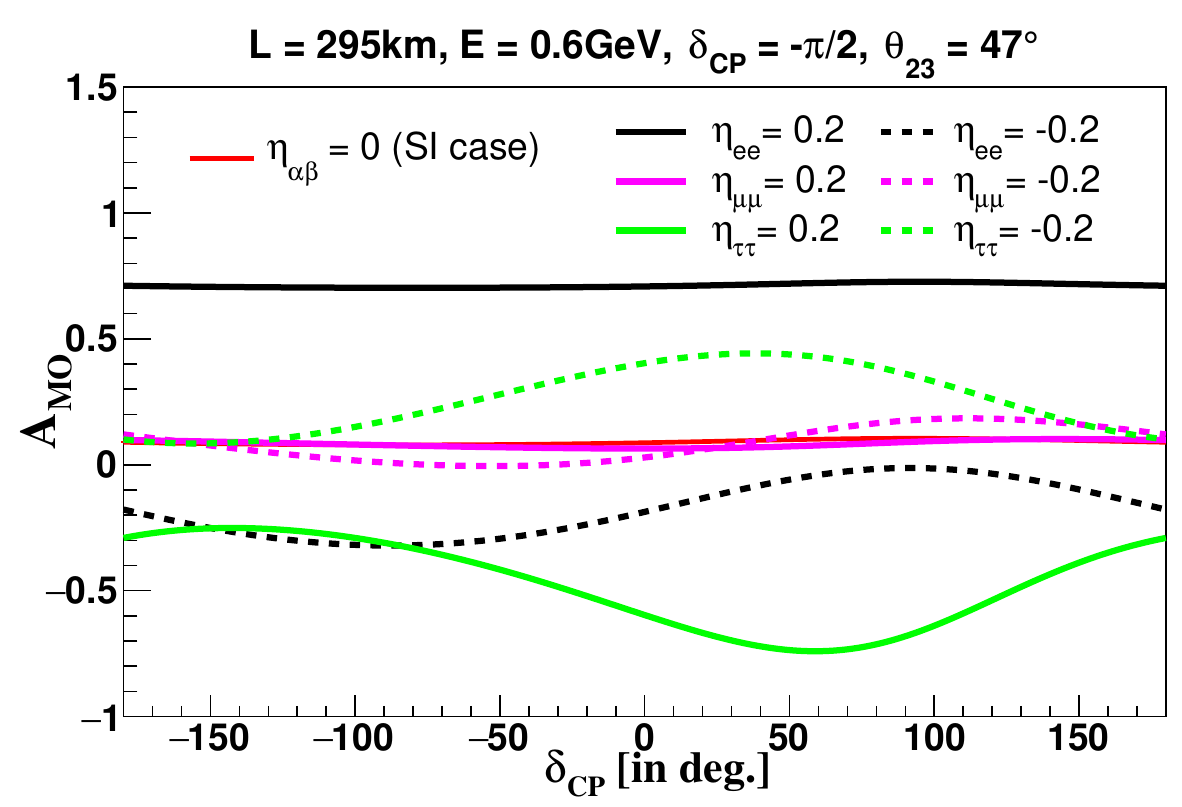} 
	\includegraphics[width=0.32\linewidth, height = 5.5cm]{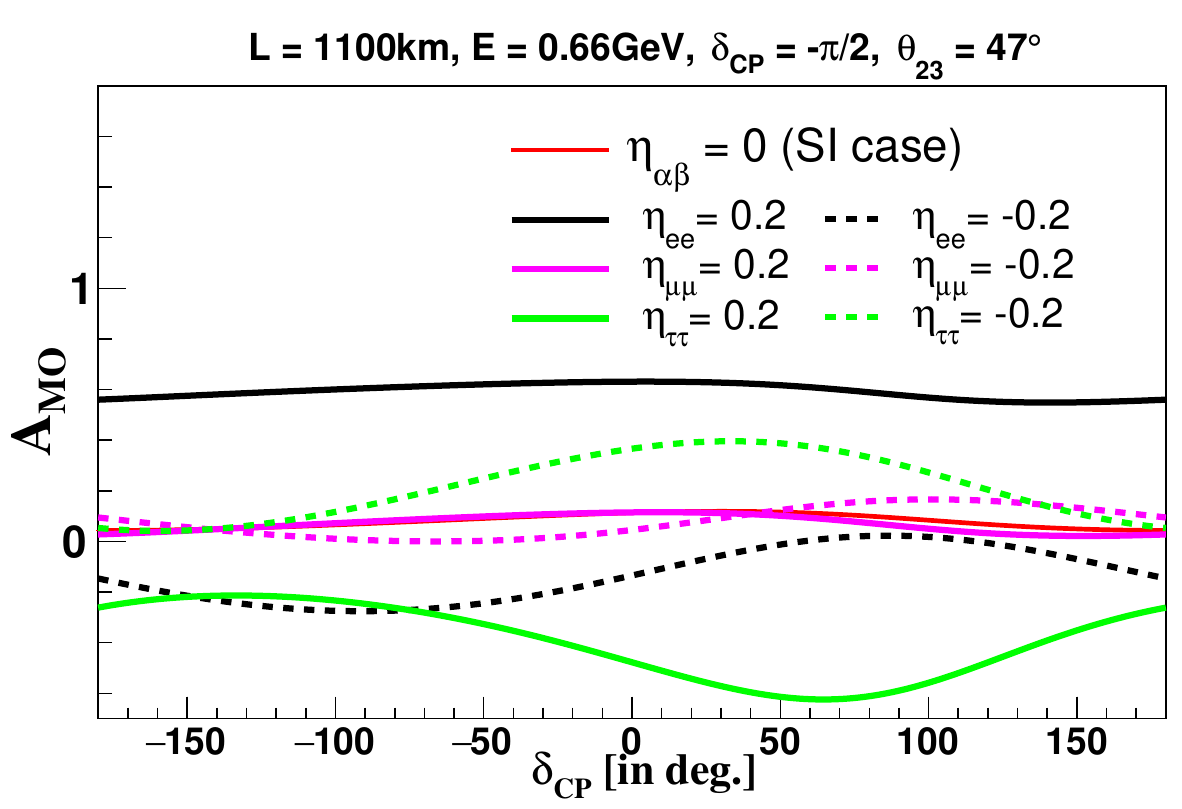} 
	\caption{Impact of scalar NSI on $A_{MO}$ at DUNE (left--panel), HK (middle--panel) and HK+KNO (right--panel) baselines. The red solid line corresponds to no scalar NSI case. The solid (dashed) lines correspond to positive (negative) diagonal scalar NSI parameters as indicated in the legends.}
    \label{fig:MH_assymetry}
\end{figure}

In figure \ref{fig:MH_assymetry}, we show the deviation of $A_{MO}$ as a function of $\delta_{CP}$ for the baselines of DUNE (left--panel), HK (middle--panel) and HK+KNO (right--panel) respectively. We have chosen the neutrino energies as their respective peak energies i.e. 2.5 GeV (DUNE), 0.6 GeV (HK) and 0.66 GeV (HK+KNO). We note the following.
\begin{itemize}
    \item  At DUNE, the presence of positive (negative) $\eta_{ee}$ can enhance (suppress) $A_{MO}$ as compared to no scalar NSI case. The effect of $\eta_{\mu\mu}$ is nominal on $A_{MO}$. A positive value of the $\eta_{\tau\tau}$ suppresses $A_{MO}$, while a negative $\eta_{\tau\tau}$ nominally enhances $A_{MO}$.

    \item For no scalar NSI, HK has nominal sensitivity towards the neutrino MO as compared to DUNE. The presence of positive (negative) $\eta_{ee}$ can enhance (suppress) $A_{MO}$. The $\eta_{\mu\mu}$ element shows a similar effect on $A_{MO}$ as compared to no scalar NSI case. However, a negative (positive) $\eta_{\tau\tau}$ enhances (suppresses) $A_{MO}$.

    \item For HK+KNO, we see a suppression in $A_{MO}$ for positive values of $\eta_{\tau\tau}$ in the range $\delta_{CP}$ $\sim$ [$40^\circ$, $180^\circ$]. A negative $\eta_{ee}$, mostly suppresses $A_{MO}$ except in the range $\delta_{CP}$ $\sim$ [$0^\circ$, $90^\circ$]. A negative $\eta_{\mu\mu}$ mostly suppresses $A_{MO}$, while a positive $\eta_{\mu\mu}$ enhances $A_{MO}$ in the range $\delta_{CP}$ $\sim$ [$-70^\circ$, $60^\circ$]. The positive $\eta_{\tau\tau}$ reduces the experiment's sensitivity towards MO for the complete $\delta_{CP}$ parameter space.
\end{itemize}
We observe that the presence of diagonal scalar NSI elements leads to a significant modification in the values of $P_{\mu e}$. As a result, it can substantially enhance/hamper the MO sensitivities of these LBL experiments.

\subsection{Impact on the MO sensitivities}\label{sec:MH_sensitiivty}
\begin{figure}[!h]
    \centering
    \includegraphics[width=0.33\linewidth, height = 5cm]{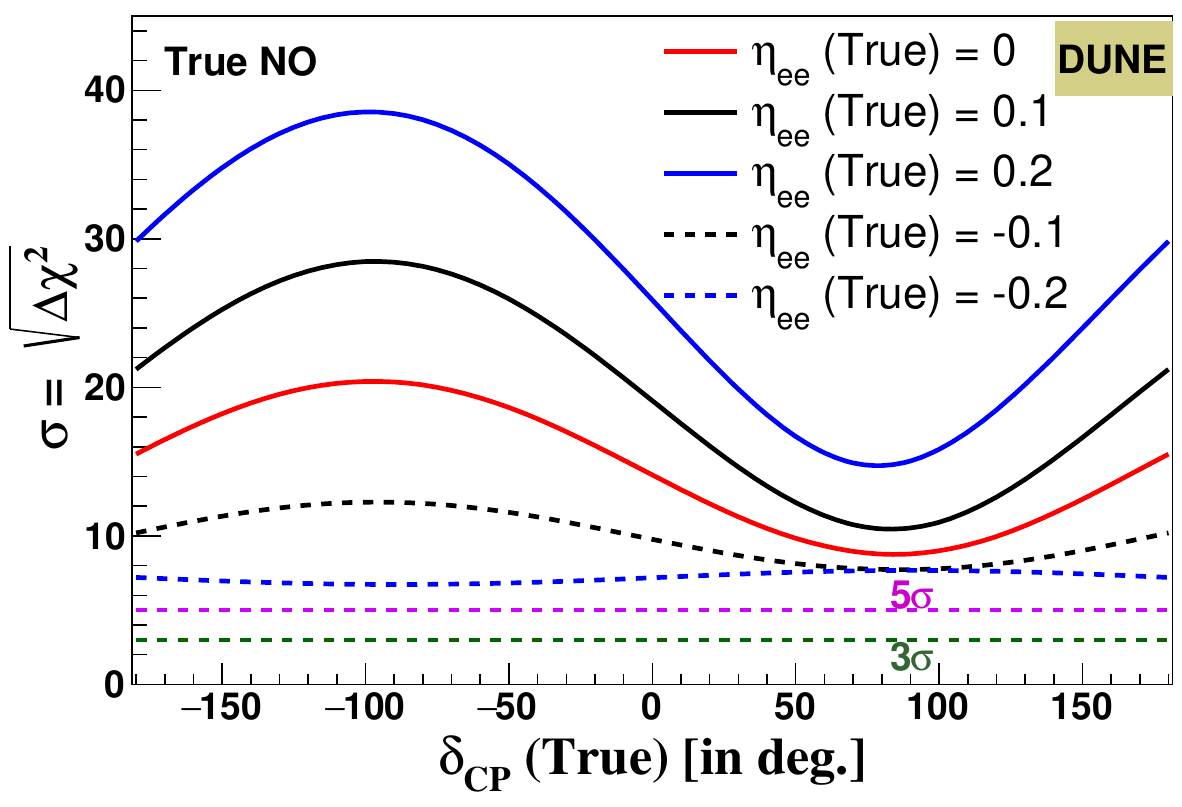} 
    \includegraphics[width=0.33\linewidth, height = 5cm]{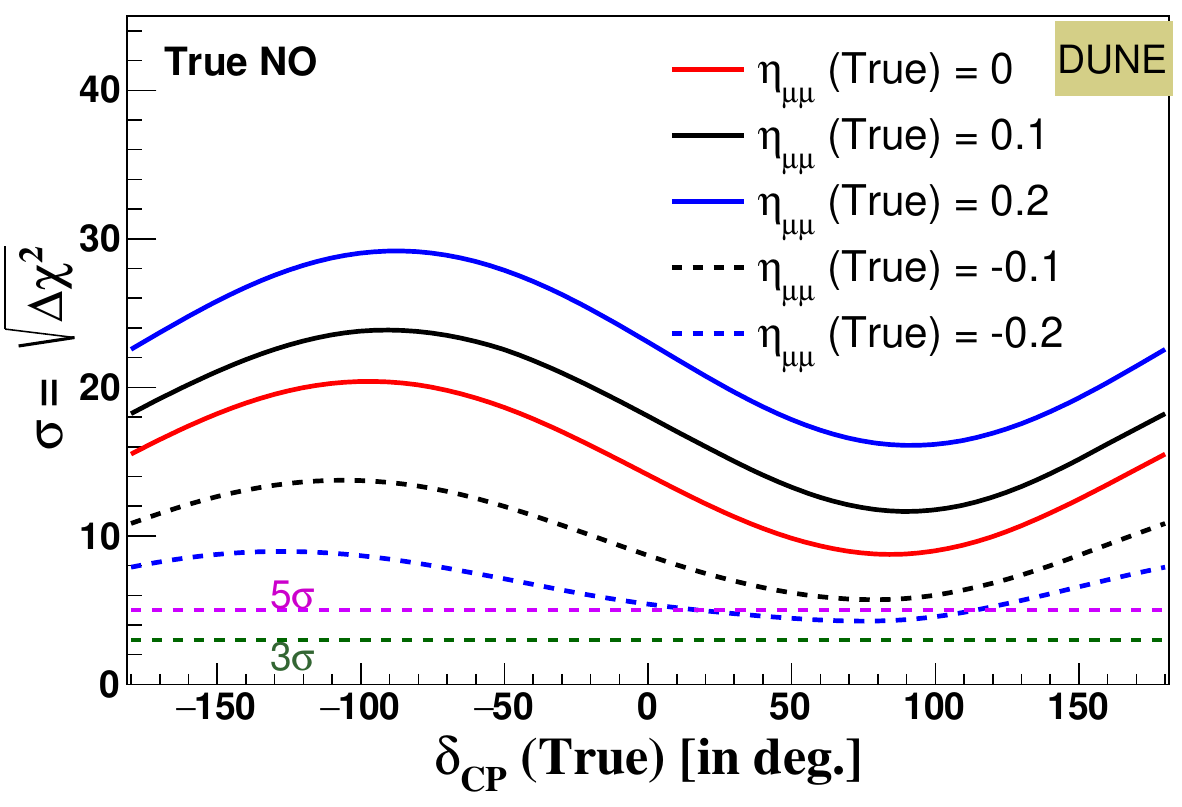} 
    \includegraphics[width=0.32\linewidth, height = 5cm]{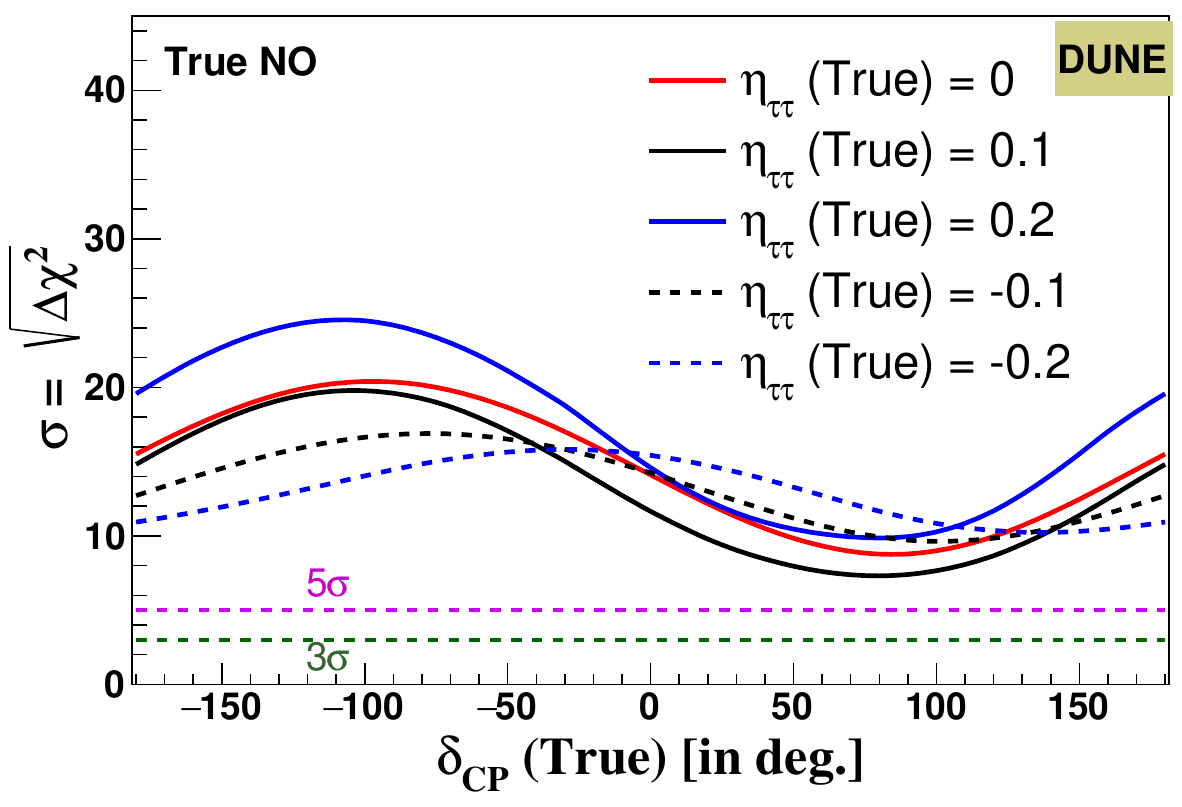} 
    \includegraphics[width=0.33\linewidth, height = 5cm]{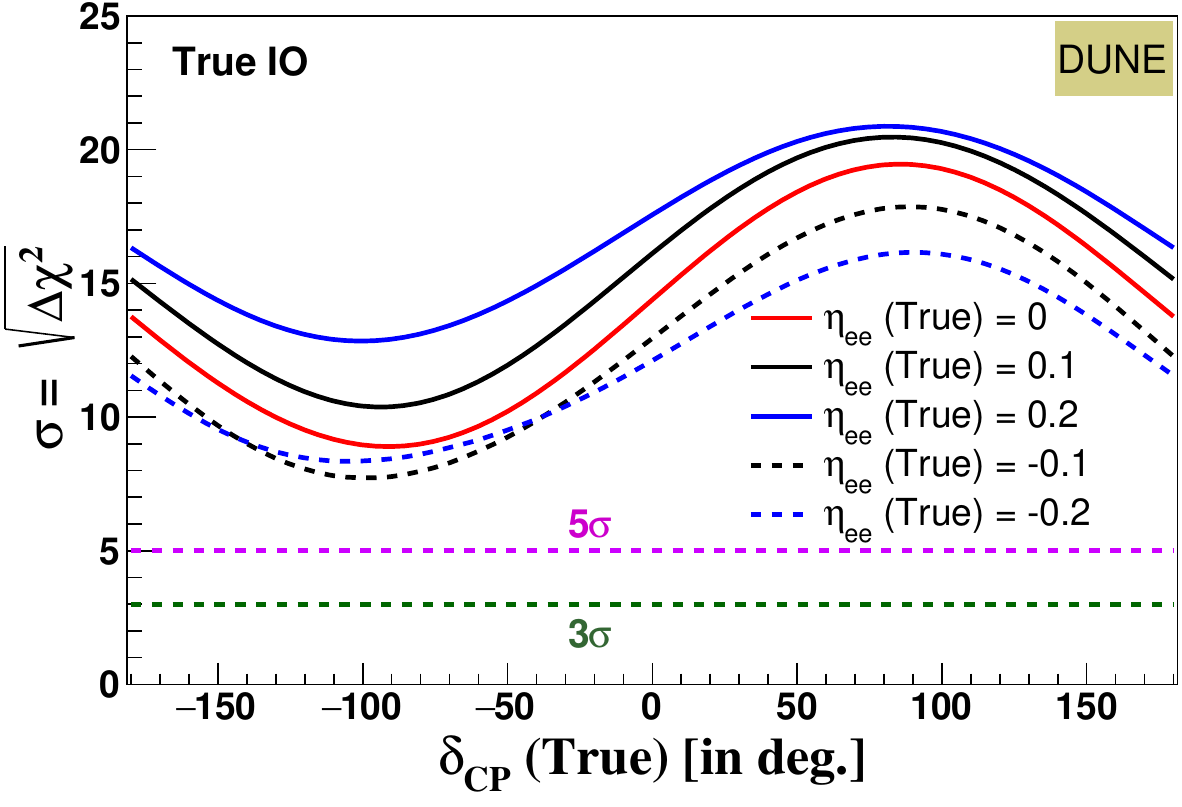}  
    \includegraphics[width=0.33\linewidth, height = 5cm]{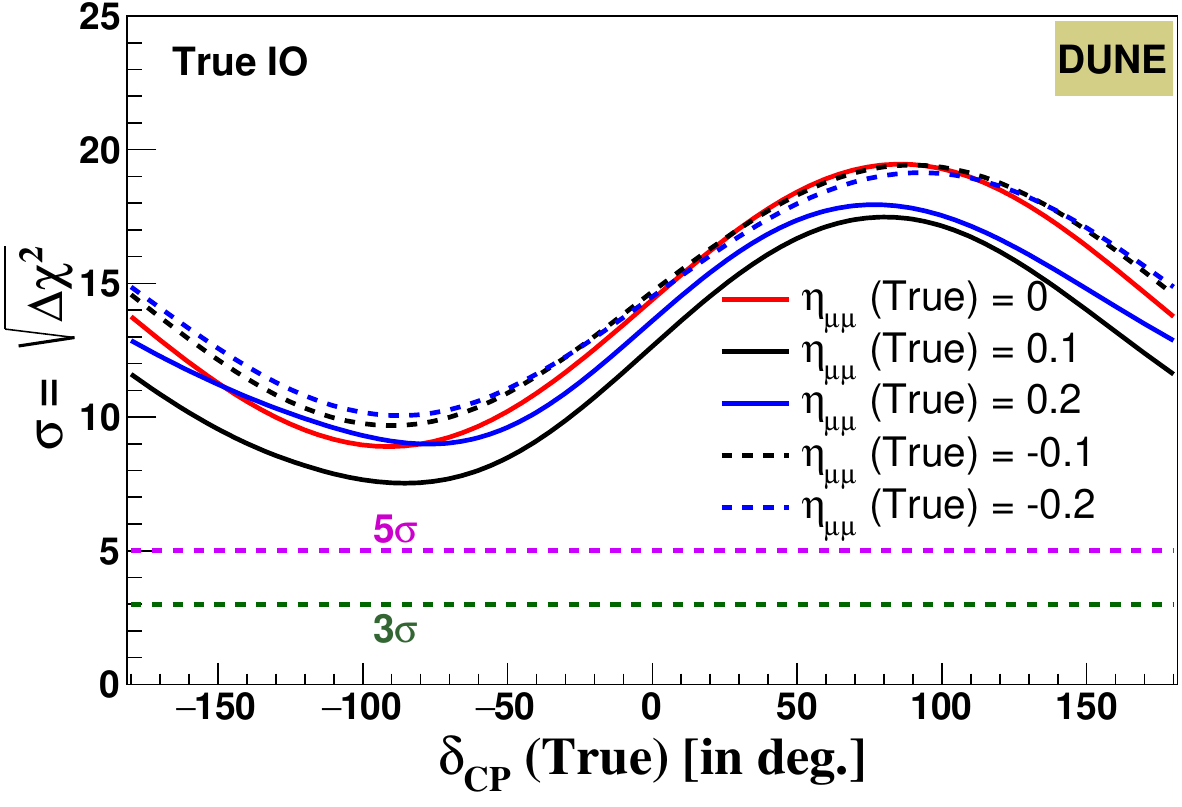}  
    \includegraphics[width=0.32\linewidth, height = 5cm]{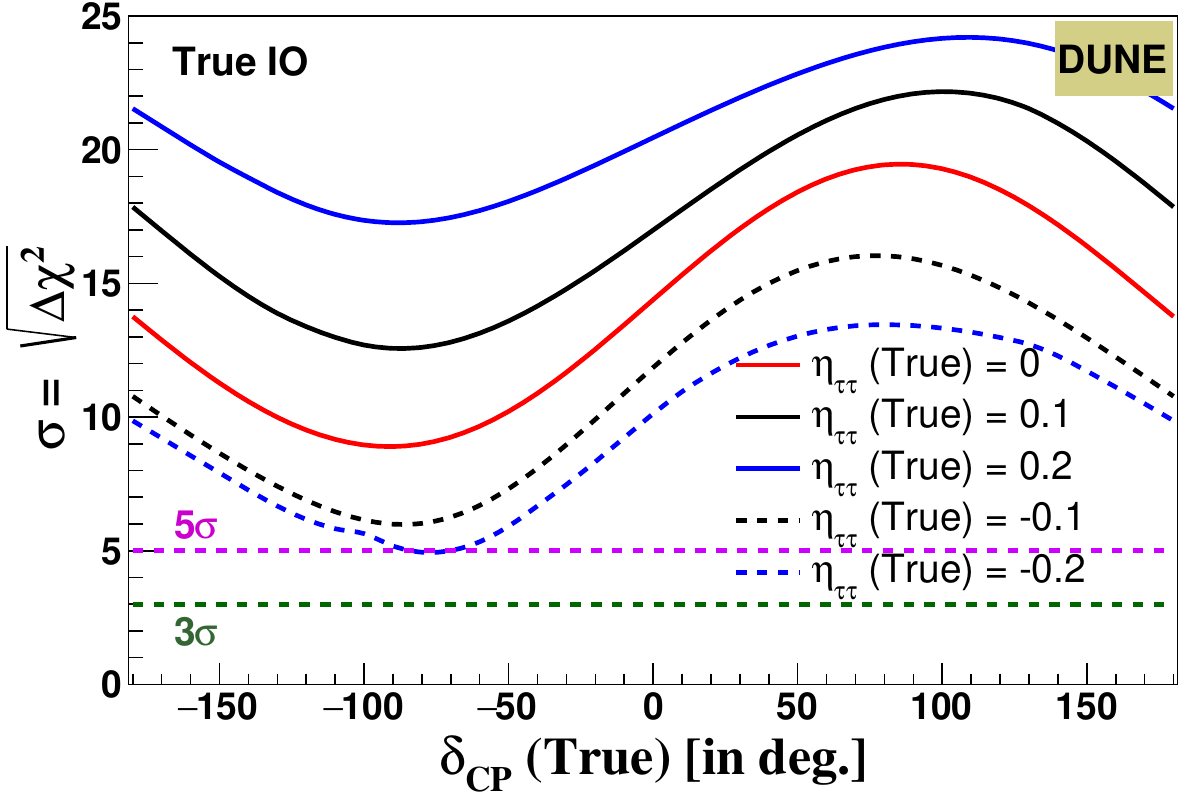}  
    \caption{The MO sensitivity of DUNE, in the presence of scalar NSI, for true NO (top--panel) and true IO (bottom--panel). The effects of diagonal scalar NSI parameters $\eta_{ee}$, $\eta_{\mu \mu}$ and $\eta_{\tau \tau}$ are shown in left, middle and right panels respectively. The solid red line is for no--scalar NSI case. The solid (dashed) lines are for non-zero positive (negative) scalar NSI parameters.}
    \label{fig:MH_1}
\end{figure}

\begin{figure}[!h]
    \centering
    \includegraphics[width=0.33\linewidth, height = 5cm]{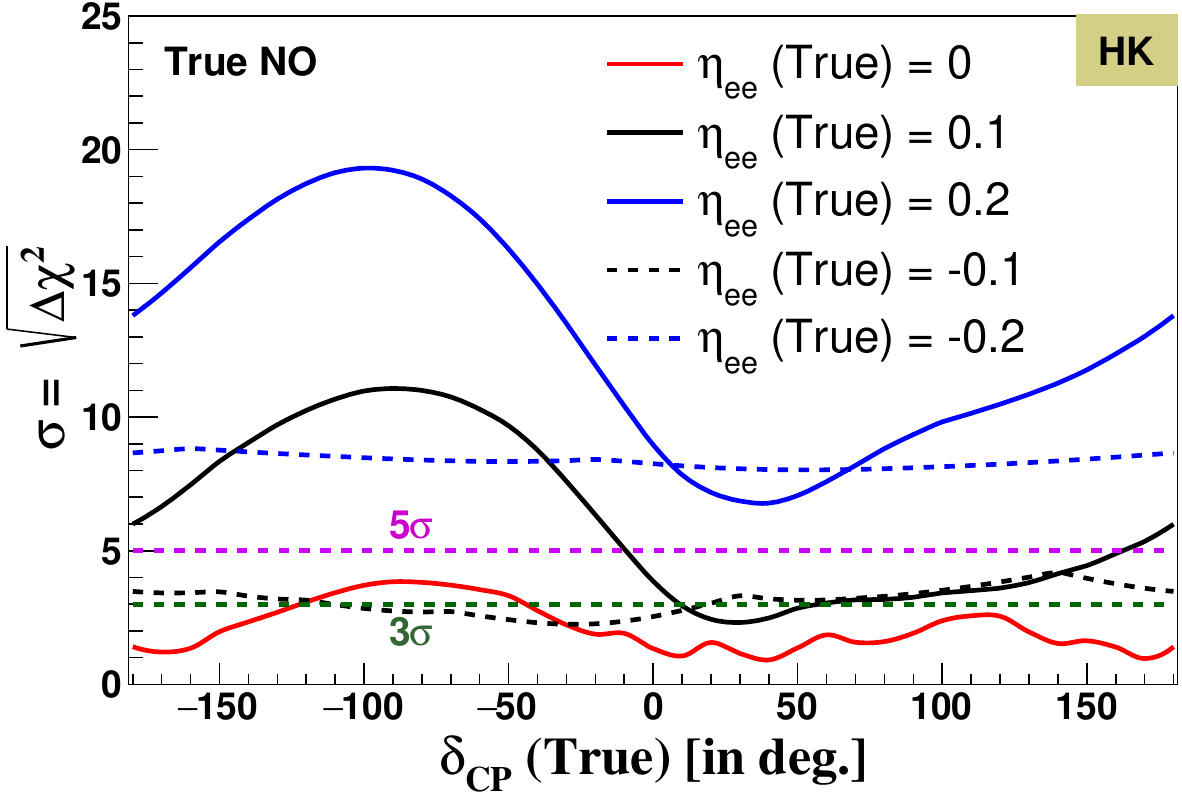} 
    \includegraphics[width=0.33\linewidth, height = 5cm]{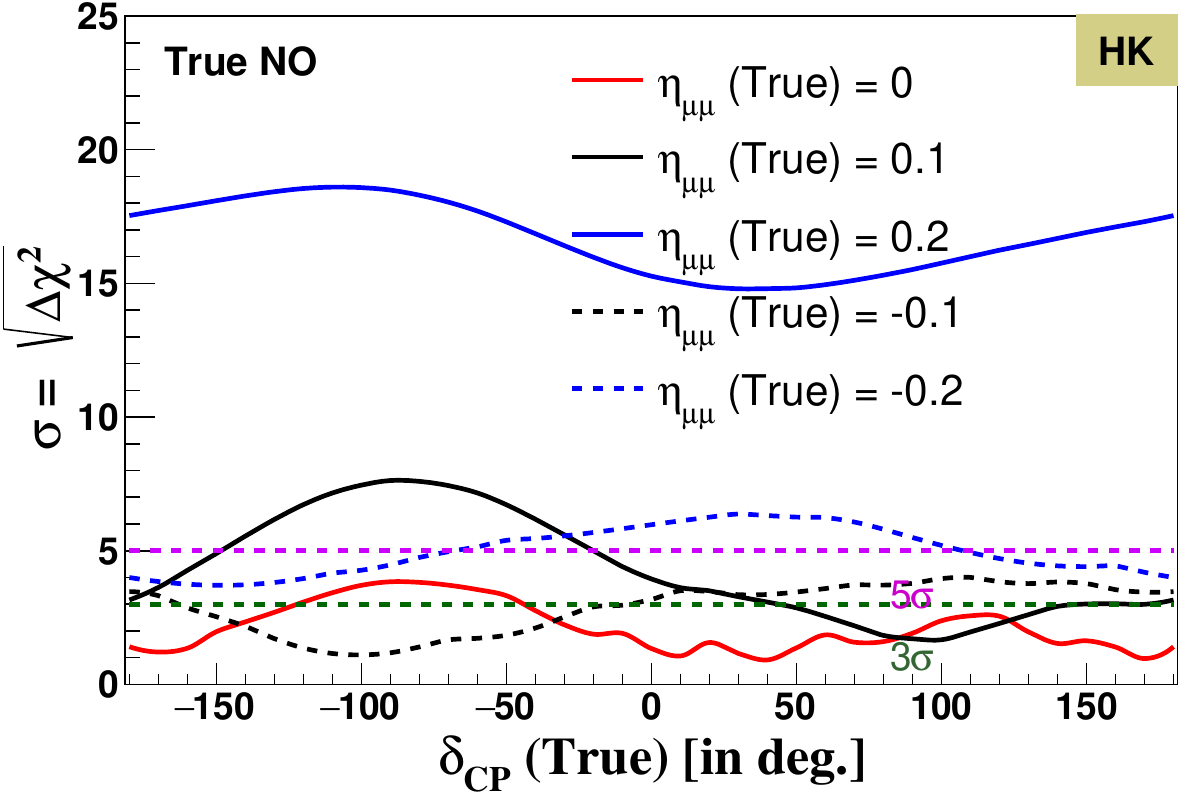}
    \includegraphics[width=0.32\linewidth, height = 5cm]{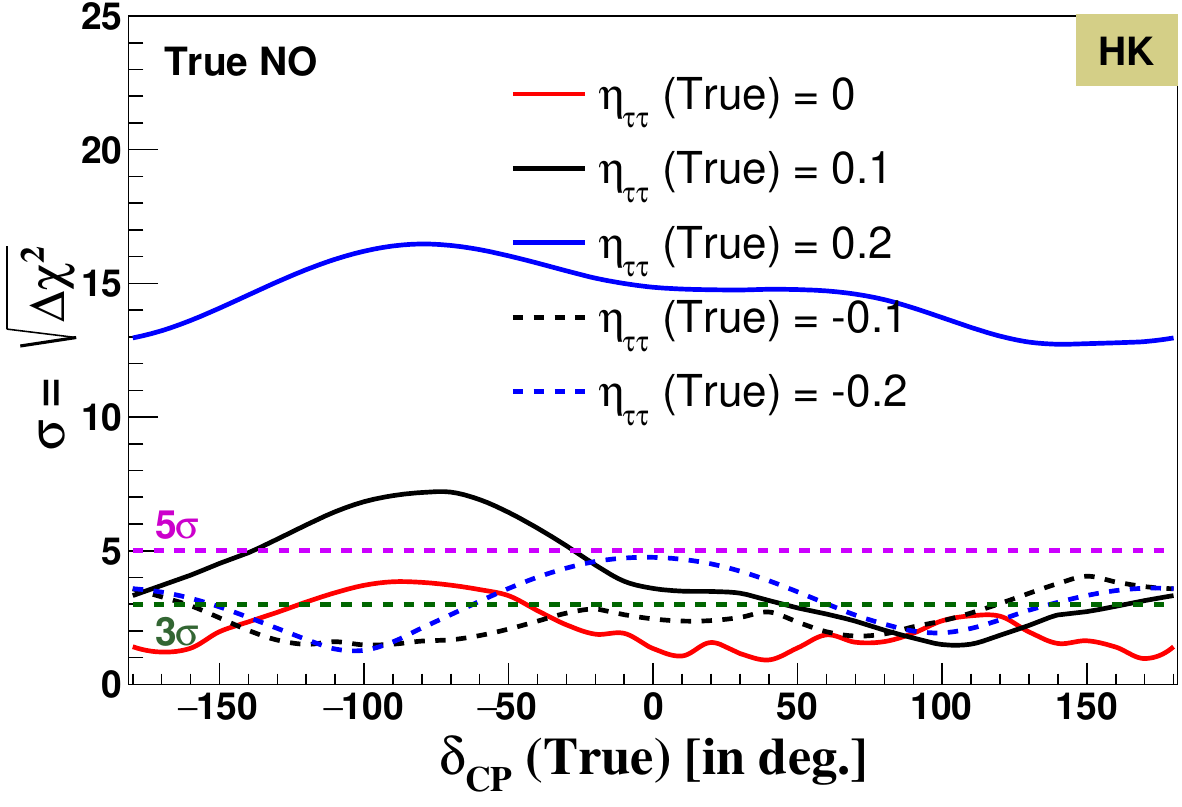}
    \includegraphics[width=0.33\linewidth, height = 5cm]{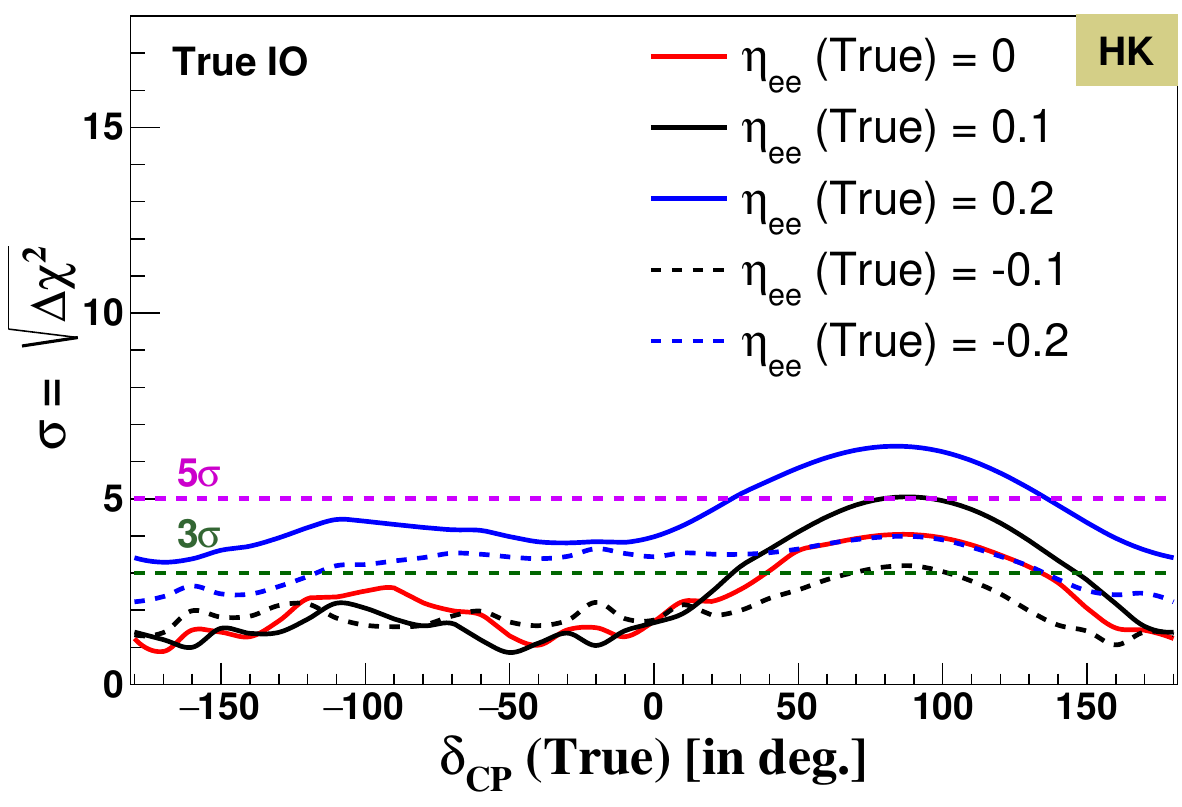} 
    \includegraphics[width=0.33\linewidth, height = 5cm]{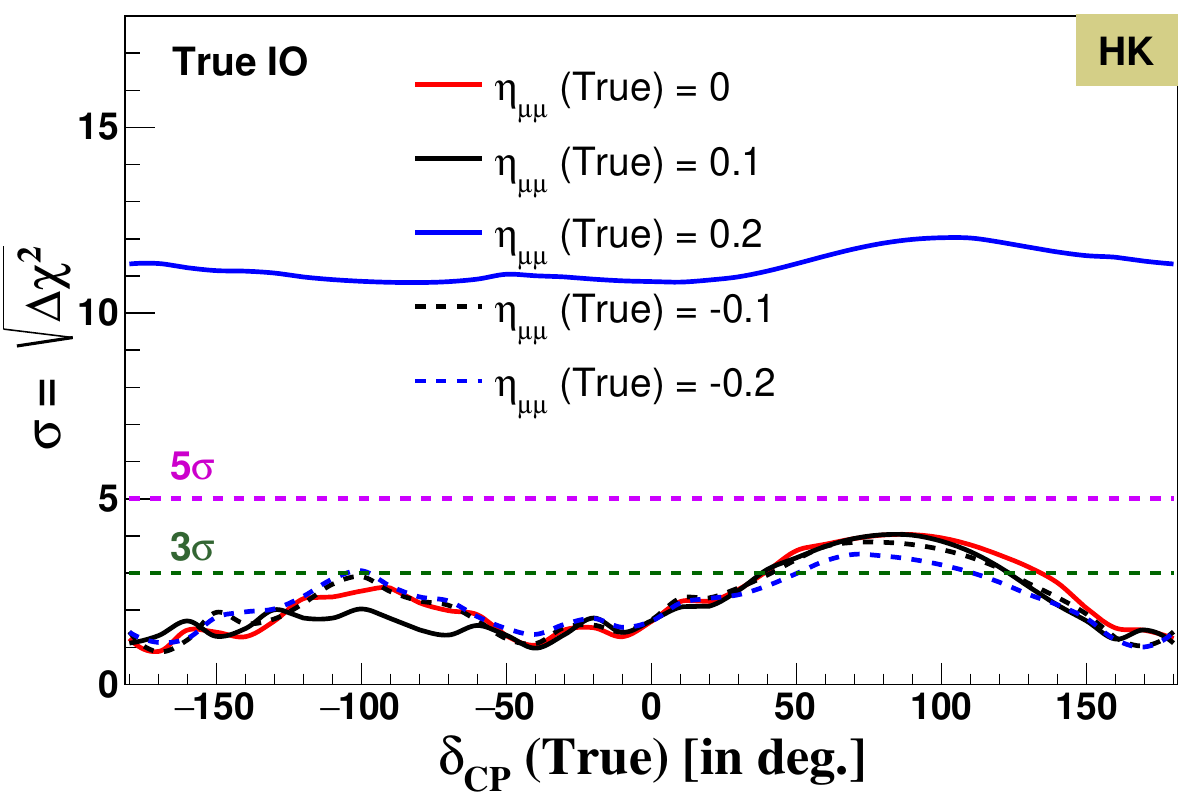} 
    \includegraphics[width=0.32\linewidth, height = 5cm]{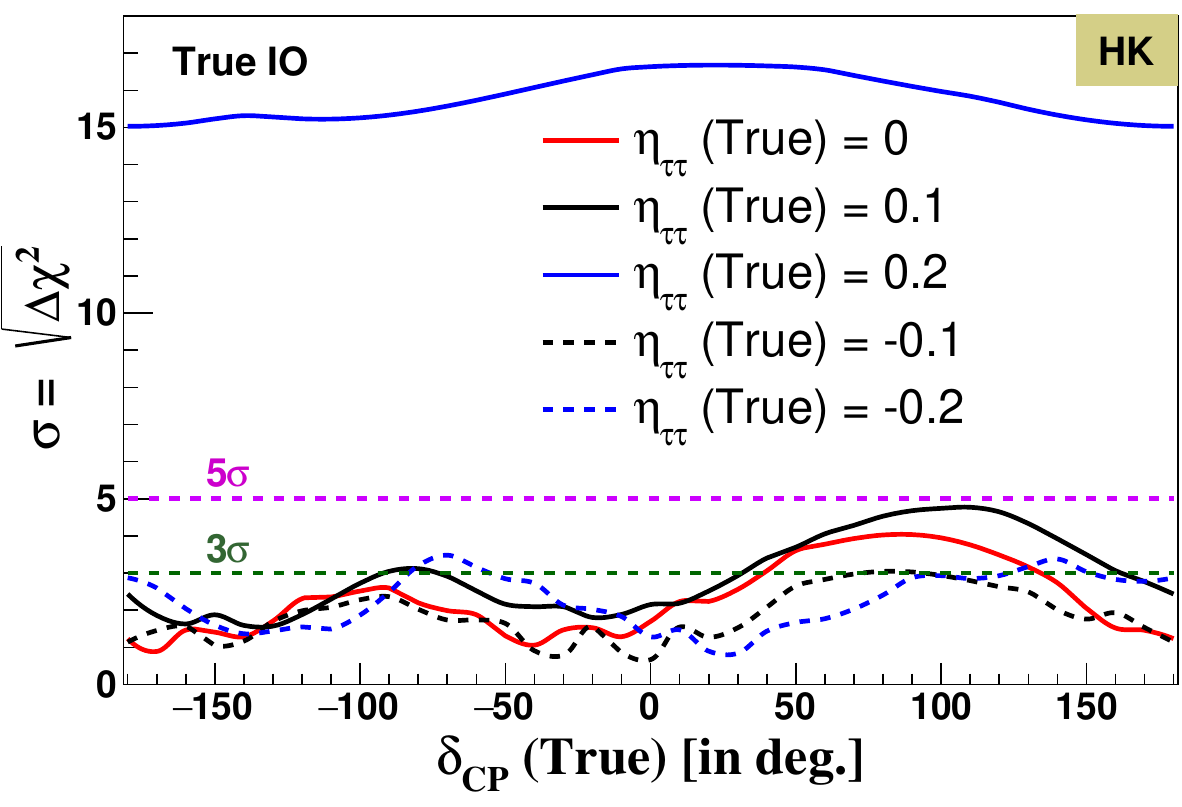} 
    \caption{The MO sensitivity of HK, in the presence of scalar NSI, for true NO (top--panel) and true IO (bottom--panel). The effects of diagonal scalar NSI parameters $\eta_{ee}$, $\eta_{\mu \mu}$ and $\eta_{\tau \tau}$ are shown in left, middle and right panels respectively. The solid red line is for no--scalar NSI case. The solid (dashed) lines are for non-zero positive (negative) scalar NSI parameters.}
    \label{fig:MH_2}
\end{figure}

\begin{figure}[!h]
	\centering
	\includegraphics[width=0.33\linewidth, height = 5cm]{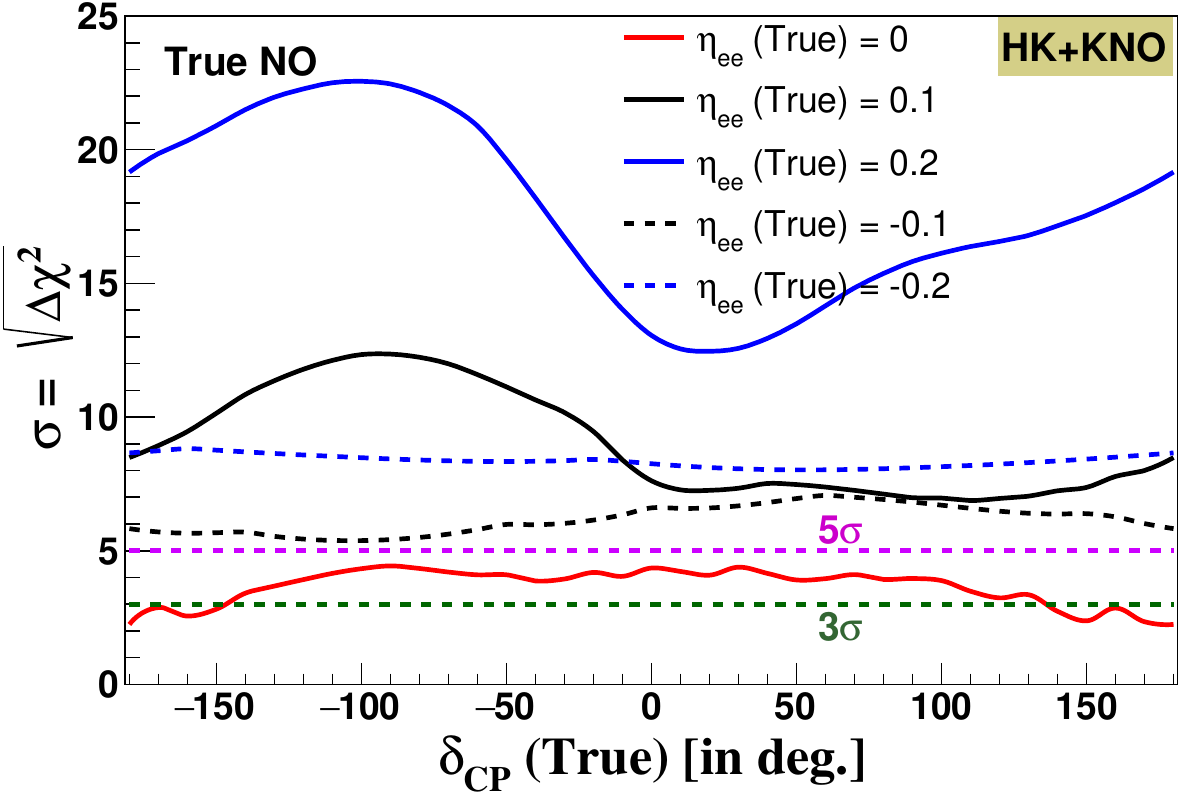} 
	\includegraphics[width=0.33\linewidth, height = 5cm]{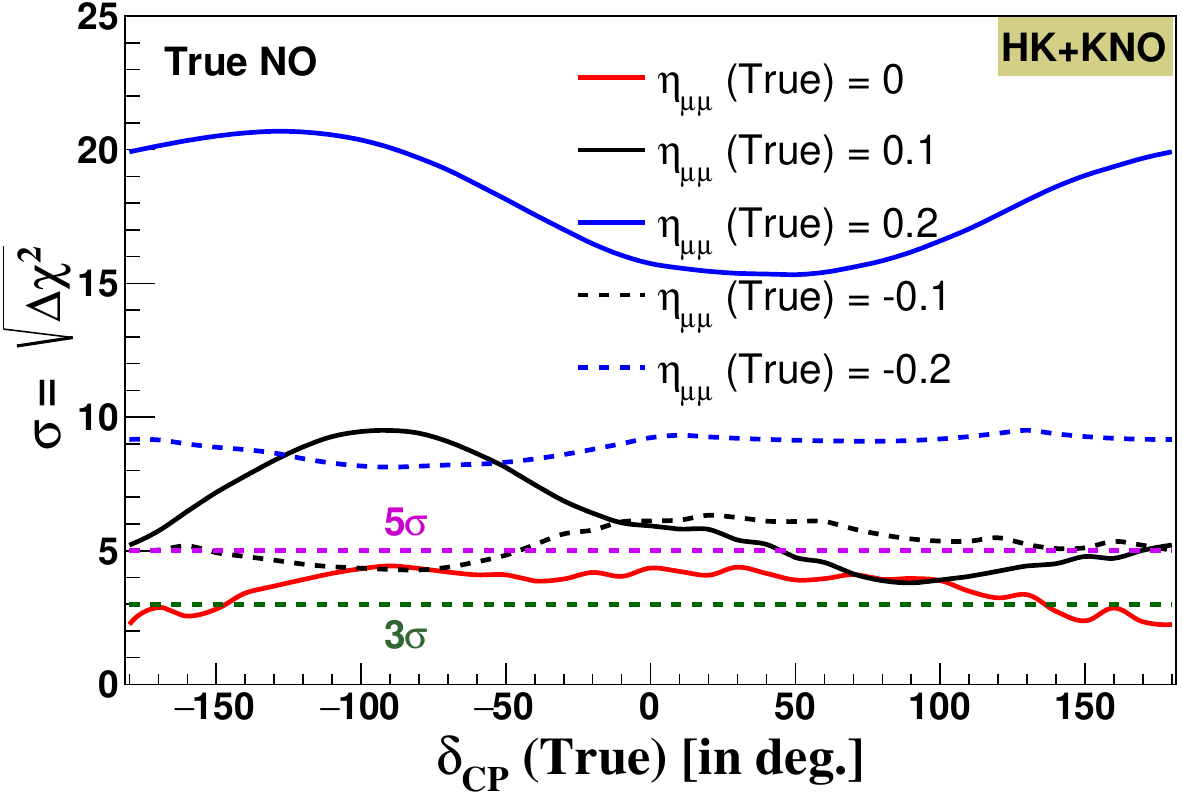} 
	\includegraphics[width=0.32\linewidth, height = 5cm]{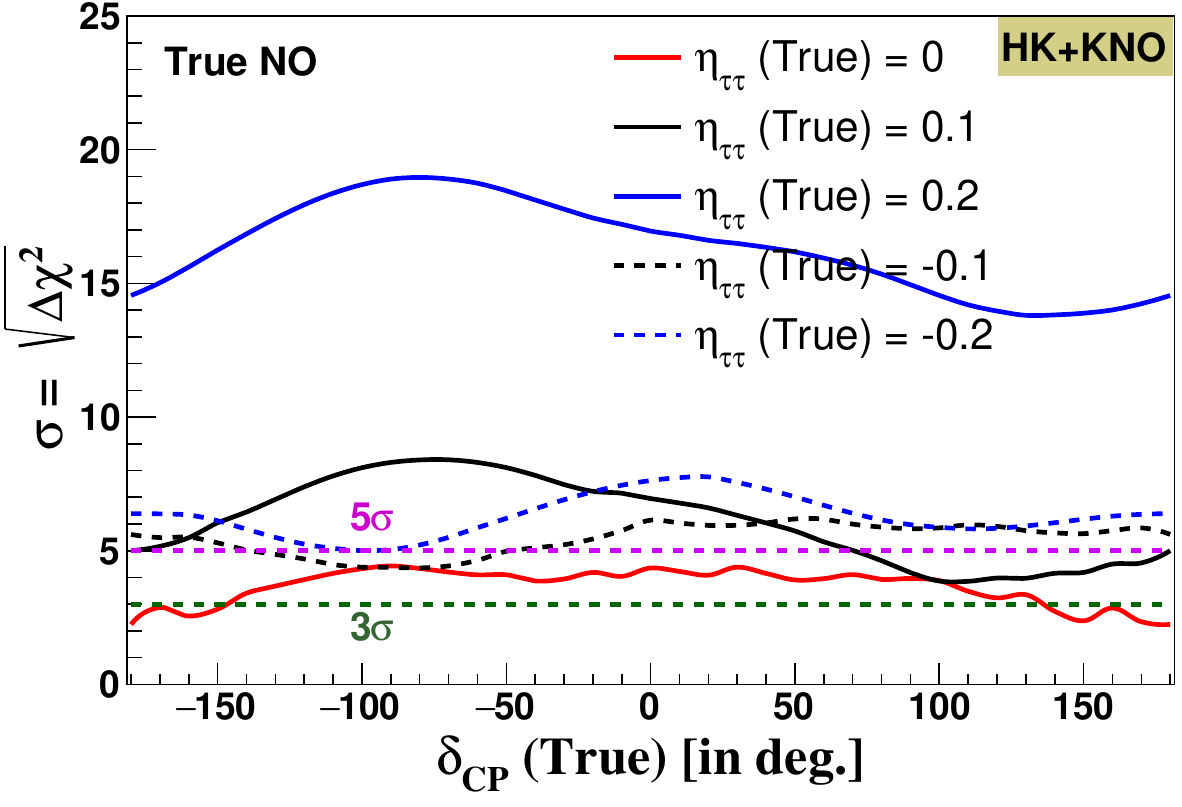} 
        \includegraphics[width=0.33\linewidth, height = 5cm]{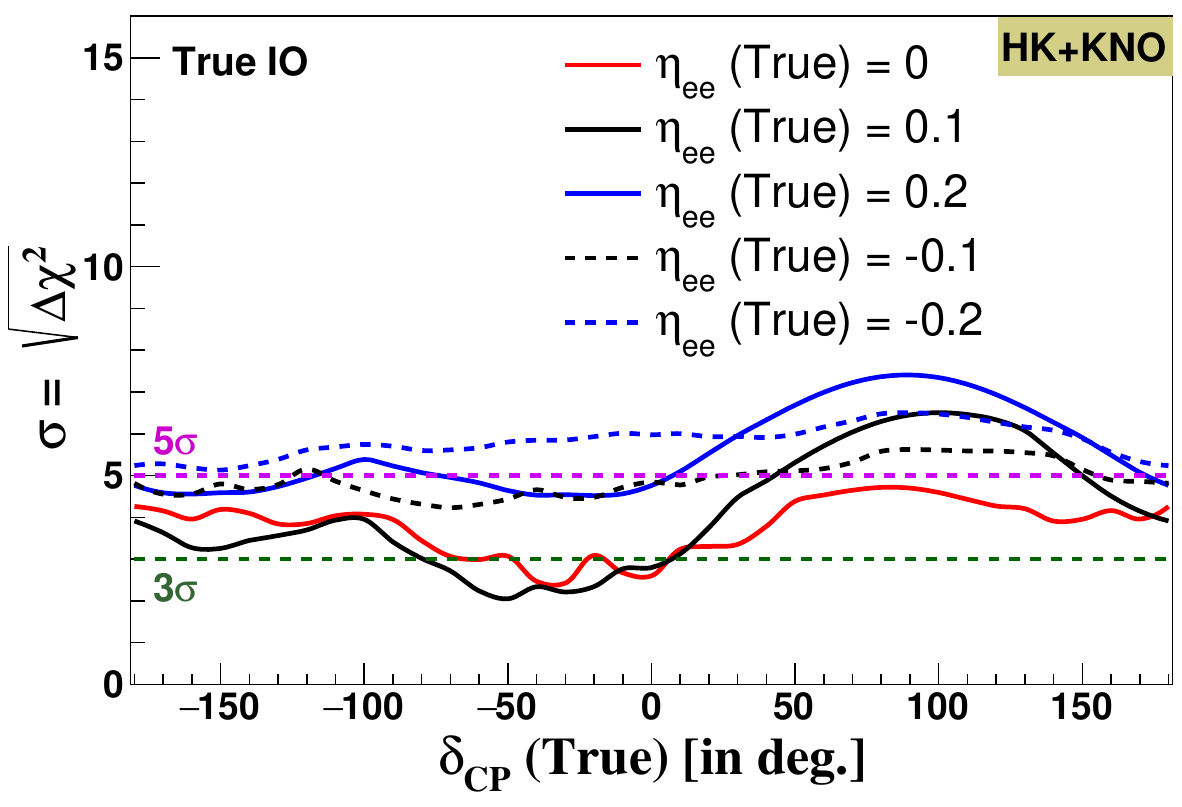} 
	\includegraphics[width=0.33\linewidth, height = 5cm]{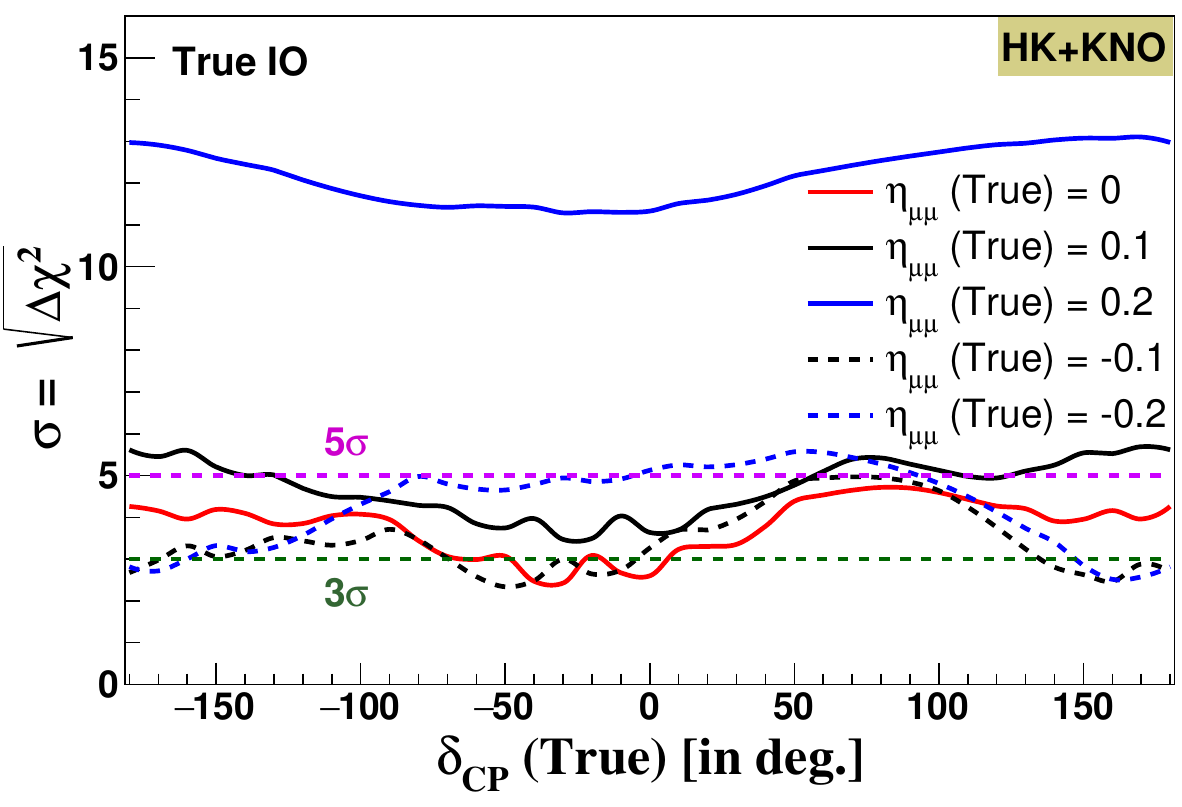} 
	\includegraphics[width=0.32\linewidth, height = 5cm]{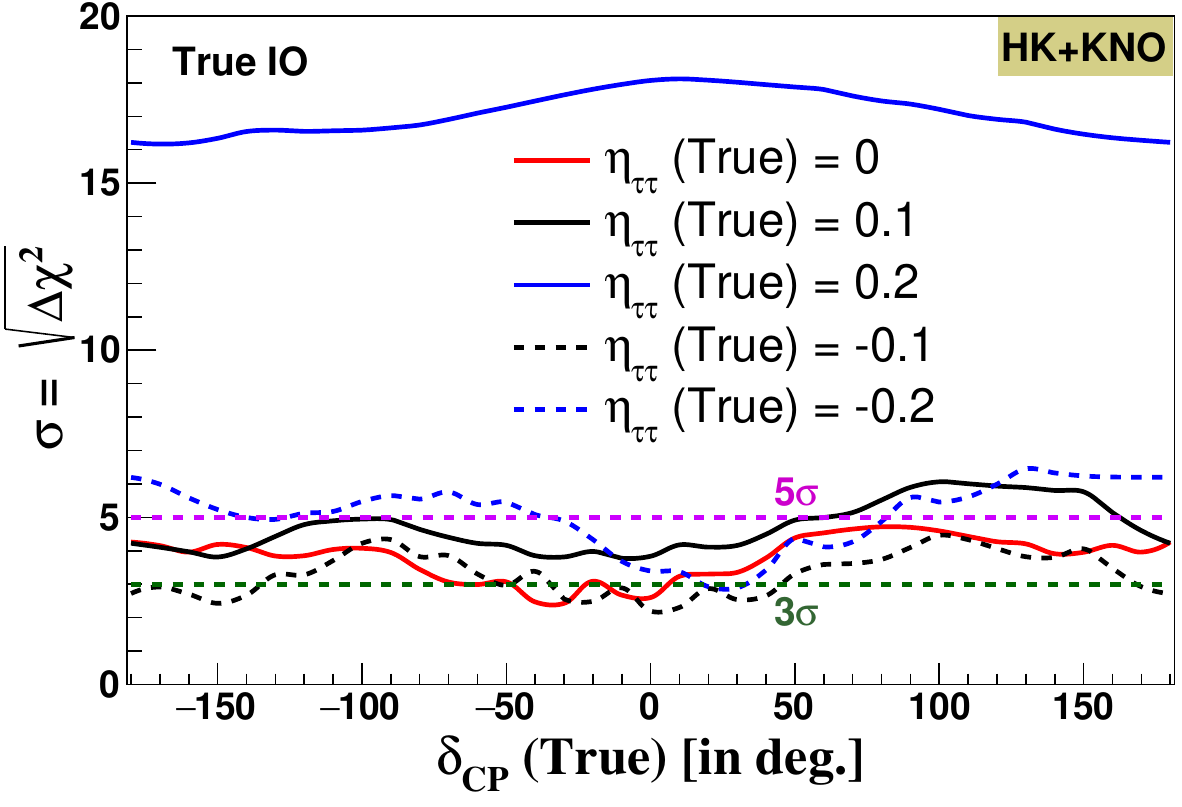} 
    \caption{The MO sensitivity of HK+KNO, in the presence of scalar NSI, for true NO (top--panel) and true IO (bottom--panel). The effects of diagonal scalar NSI parameters $\eta_{ee}$, $\eta_{\mu \mu}$ and $\eta_{\tau \tau}$ are shown in left, middle and right panels respectively. The solid red line is for no--scalar NSI case. The solid (dashed) lines are for non-zero positive (negative) scalar NSI parameters.}
\label{fig:MH_3}
\end{figure}

One of the major goals of the current and upcoming LBL $\nu$-oscillation experiments is the determination of neutrino mass ordering. The impact of scalar NSI on the MO sensitivities at three different LBL experiments i.e. DUNE, HK and HK+KNO is explored here. We have considered two different cases in our analysis i.e., 
\begin{itemize}
    \item Considering NO as the true mass ordering for which $\Delta \chi^{2}$ can be defined as, 
\begin{eqnarray}
\Delta\chi^2_\text{MO} &=& \chi^2_\text{NO} -\chi^2_\text{IO}~~~~(\text{for true NO}).
\end{eqnarray}
\item Considering IO as the true mass ordering for which $\Delta \chi^{2}$ can be defined as, 
\begin{eqnarray}
\Delta\chi^2_\text{MO} &=& \chi^2_\text{IO} -\chi^2_\text{NO}~~~~(\text{for true IO}).
\end{eqnarray}
\end{itemize}
To obtain the MO sensitivities, the statistical significance which is defined as $\sigma=\sqrt{\Delta \chi^{2}}$ is plotted for varying true $\delta_{CP}$ in [-$\pi$, $\pi$]. The other $\nu$-oscillation parameters are fixed at values as listed in table \ref{tab:mixing_parameters}. In figure \ref{fig:MH_1}-\ref{fig:MH_5}, we have marginalized over the test ordering as well as $\delta_{CP}$ in [-$\pi$, $\pi$] and $\theta_{23}$ in [$40^\circ$, $50^\circ$]. We have also additionally marginalized over the scalar NSI parameters in the range [-0.2,0.2].

In figure \ref{fig:MH_1}, we show the effects of scalar NSI on the MO sensitivity of DUNE considering NO (IO) as the true ordering in the top--panel (bottom--panel). The solid red line corresponds to the case with no scalar NSI effects. The solid (dashed) lines correspond to the positive (negative) values of scalar NSI parameters. The dashed magenta and green lines represent the 5$\sigma$ and 3$\sigma$ CL respectively. We observe that,
\begin{itemize}
    \item For true NO, positive (negative) values of $\eta_{ee}$ and $\eta_{\mu\mu}$ elements enhance (suppress) the MO sensitivities as compared to no scalar NSI case. For $\eta_{\tau\tau} = 0.2$, the sensitivities show an enhancement for most values of $\delta_{CP}$, while for $\eta_{\tau\tau}=0.1 $, the sensitivities lie below the no scalar NSI case for the whole $\delta_{CP}$ parameter space. For the chosen negative values of $\eta_{\tau\tau}$, the sensitivities lie mostly below the no scalar NSI case.
    \item For true IO, positive (negative) values of $\eta_{ee}$ and $\eta_{\tau\tau}$, enhance (suppress) the MO sensitivities as compared to when there is no scalar NSI present. However, for a positive (negative) $\eta_{\mu\mu}$ value the sensitivities get mostly suppressed (enhanced).
\end{itemize}

\noindent In figure \ref{fig:MH_2}, we present the MO sensitivities of HK in presence of scalar NSI parameters $\eta_{ee}$ (left--panel), $\eta_{\mu \mu}$ (middle--panel) and $\eta_{\tau \tau}$ (right--panel) assuming NO (IO) as true ordering in the top--panel (bottom--panel). The no scalar NSI case is depicted by the red line, whereas the positive (negative) scalar NSI parameters are shown by the solid (dashed) lines. The 5$\sigma$ (3$\sigma$) CL are represented by the horizontal magenta (green) lines. The observations made are listed below.
\begin{itemize}
    \item For true NO, a positive value of the $\eta_{\alpha\beta}$ mostly enhances the sensitivity towards MO determination. Even with $\eta_{\alpha \beta}=0.1$, above $5 \sigma$ sensitivity is observed in the negative half-plane of $\delta_{CP}$. For most values of $\delta_{CP}$, negative values of scalar NSI parameters also enhance the sensitivities as compared to no scalar NSI case. A suppression in the sensitivity can be observed for some combinations of $\delta_{CP}$ and negative $\eta_{\alpha \beta}$.
    \item For true IO, a significant enhancement in the ordering sensitivity is observed only for $\eta_{\alpha \beta} = 0.2$, particularly for $\eta_{\mu\mu}$ and $\eta_{\tau\tau}$. For $\eta_{\mu\mu}=0.1$, the sensitivity curves with and without scalar NSI cases are seen to nearly overlap for the entire range of $\delta_{CP}$ values.
\end{itemize}

\noindent In figure \ref{fig:MH_3}, we have shown the effects of diagonal scalar NSI parameters on the MO sensitivities at HK+KNO. The top (bottom) panel corresponds to MO sensitivity for true NO (IO) with non-zero $\eta_{ee}$ (left--panel), $\eta_{\mu \mu}$ (middle--panel) and $\eta_{\tau \tau}$ (right--panel). The red solid line is in the absence of scalar NSI, whereas, the black and blue lines correspond to non-zero scalar NSI parameters. The solid (dashed) lines correspond to positive (negative) values of scalar NSI parameters. The $5\sigma$ (dashed magenta) and $3 \sigma$ (dashed green) CL are drawn as a reference. The following are the observations from figure \ref{fig:MH_3}.
\begin{itemize}
    \item For NO as the true ordering, the sensitivity is enhanced for both positive and negative values of scalar NSI parameters. In general, with scalar NSI we get above $5 \sigma$ MO sensitivity except for the case when $\eta_{\mu\mu}$ and $\eta_{\tau\tau}$ has the value of $-0.1$ and $\delta_{CP}$ falls within the range of $[-120^\circ,-60^\circ]$.

    \item For IO as the true ordering, a substantial enhancement in sensitivity is observed only when $\eta_{\mu \mu}$ and $\eta_{\tau \tau}$ is equal to 0.2. The effect of scalar NSI, for all other cases, is nominal. Although, certain values of $\eta_{ee}$ can improve the sensitivity beyond 5$\sigma$ for some range of $\delta_{CP}$.
\end{itemize}

\subsection{The MO sensitivity for the Synergy of the LBL experiments}\label{sec:Combined_study}
We here show the effects of scalar NSI on the MO sensitivities after combining the data from two LBL experiments viz. DUNE+HK (in section \ref{sec:syn_1}) and DUNE+HK+KNO (in section \ref{sec:syn_2}) respectively.

\subsubsection{Combining DUNE with HK} \label{sec:syn_1}
\begin{figure}[!h]
	\centering
	\includegraphics[width=0.33\linewidth, height = 5cm]{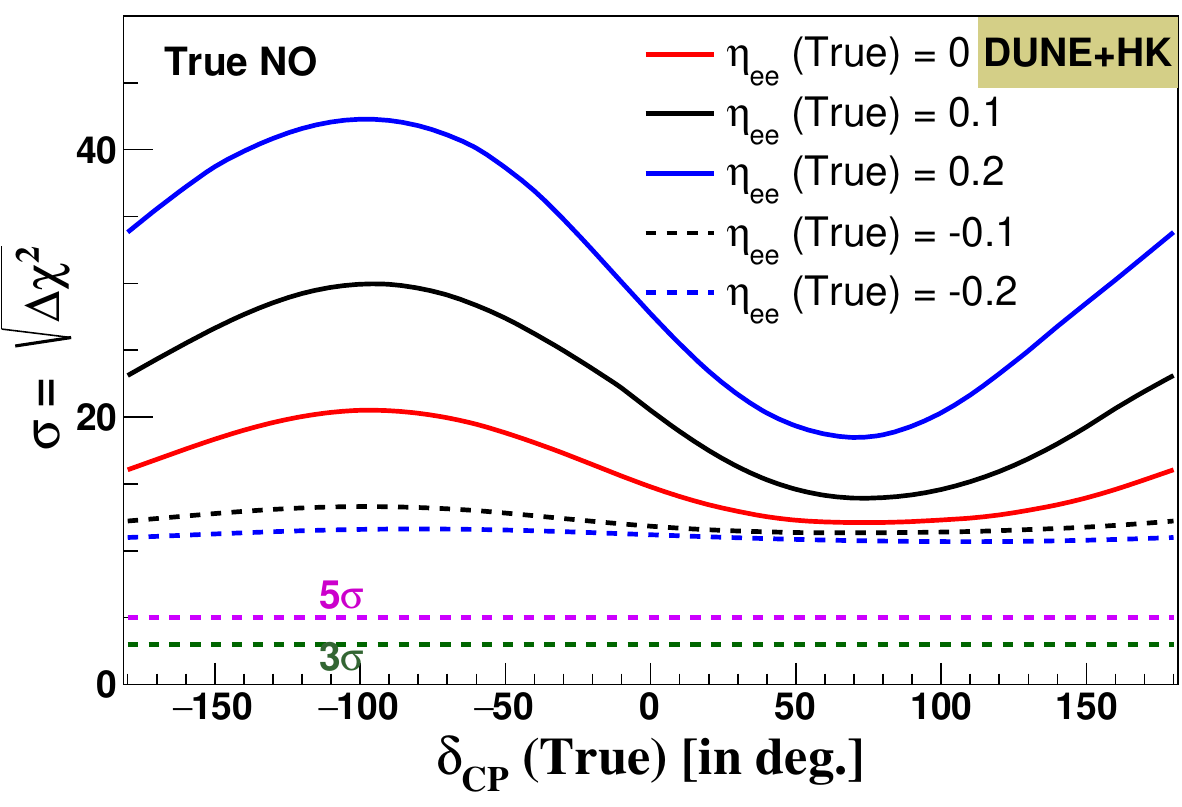}
	\includegraphics[width=0.33\linewidth, height = 5cm]{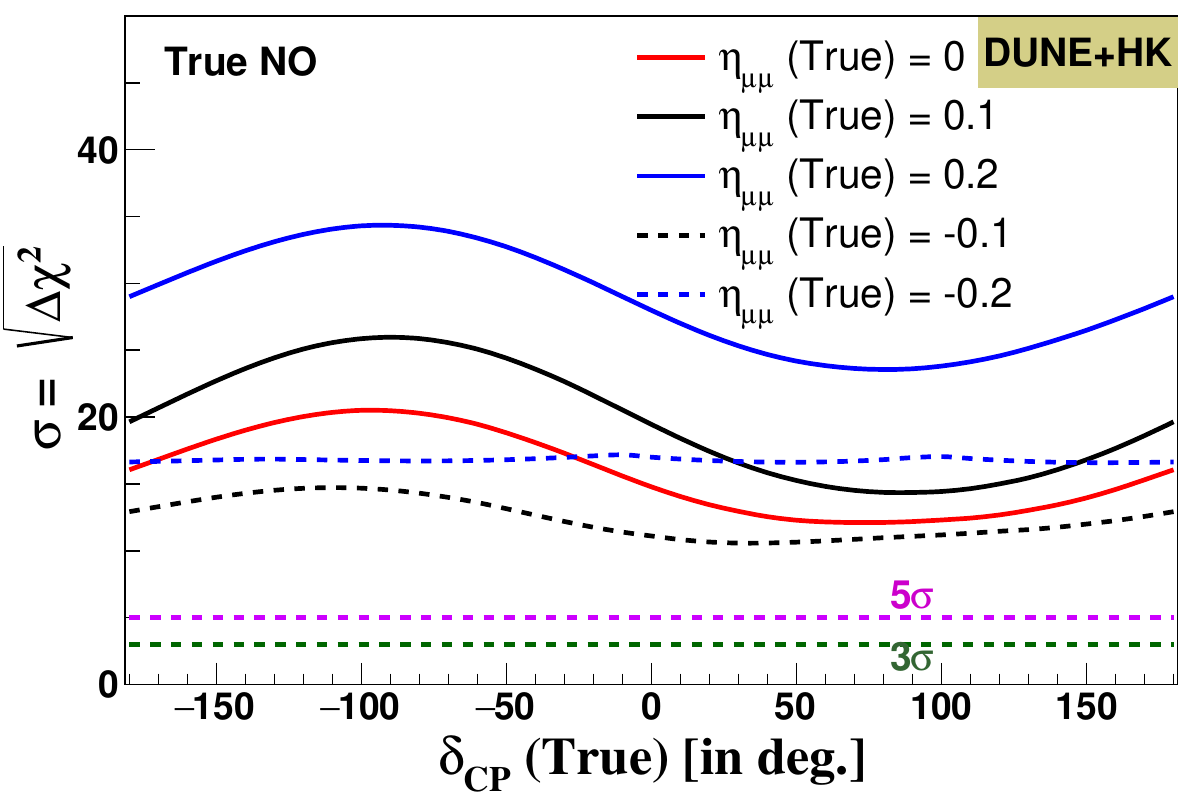}
	\includegraphics[width=0.32\linewidth, height = 5cm]{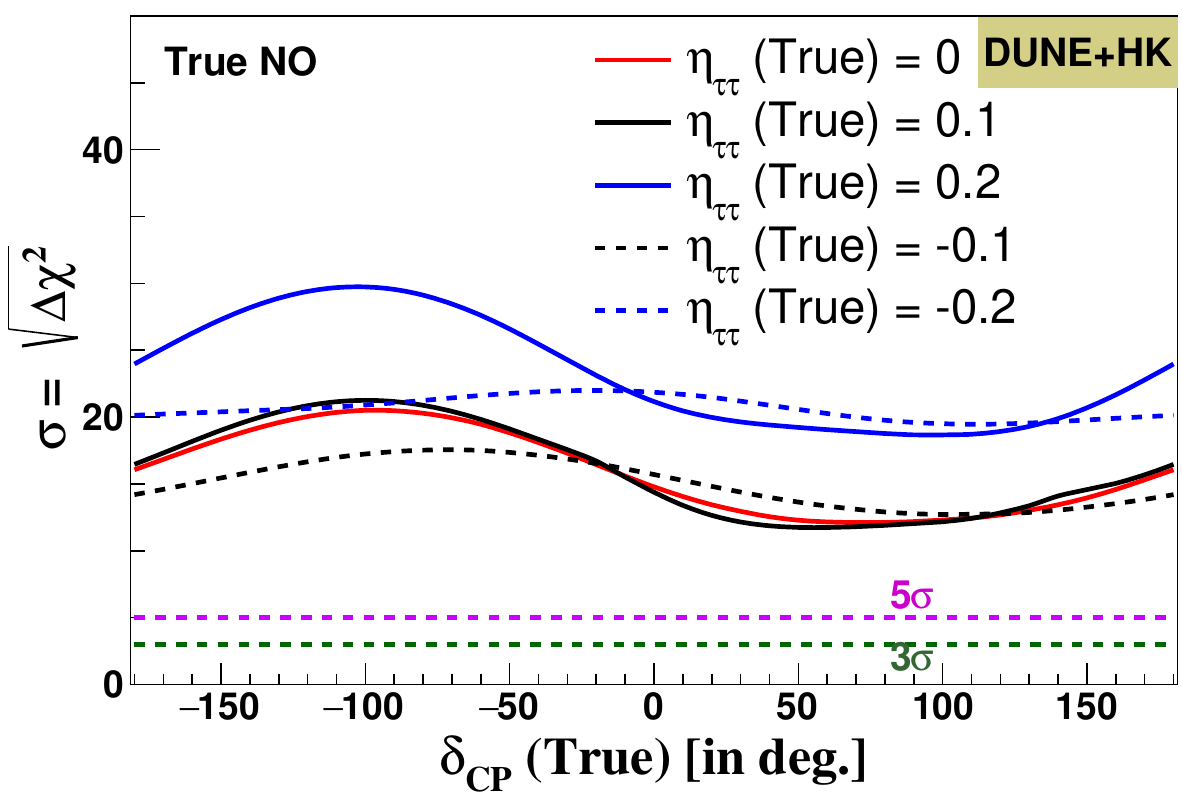}
        \includegraphics[width=0.33\linewidth, height = 5cm]{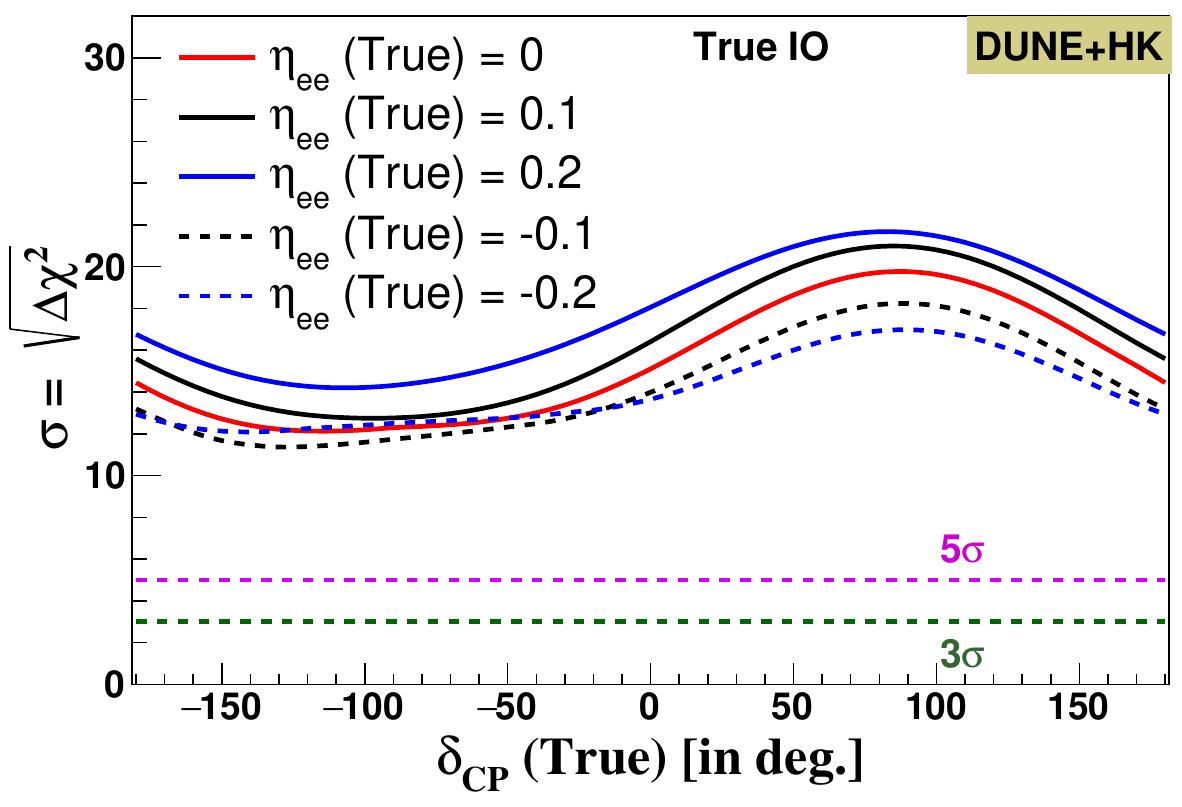}
	\includegraphics[width=0.33\linewidth, height = 5cm]{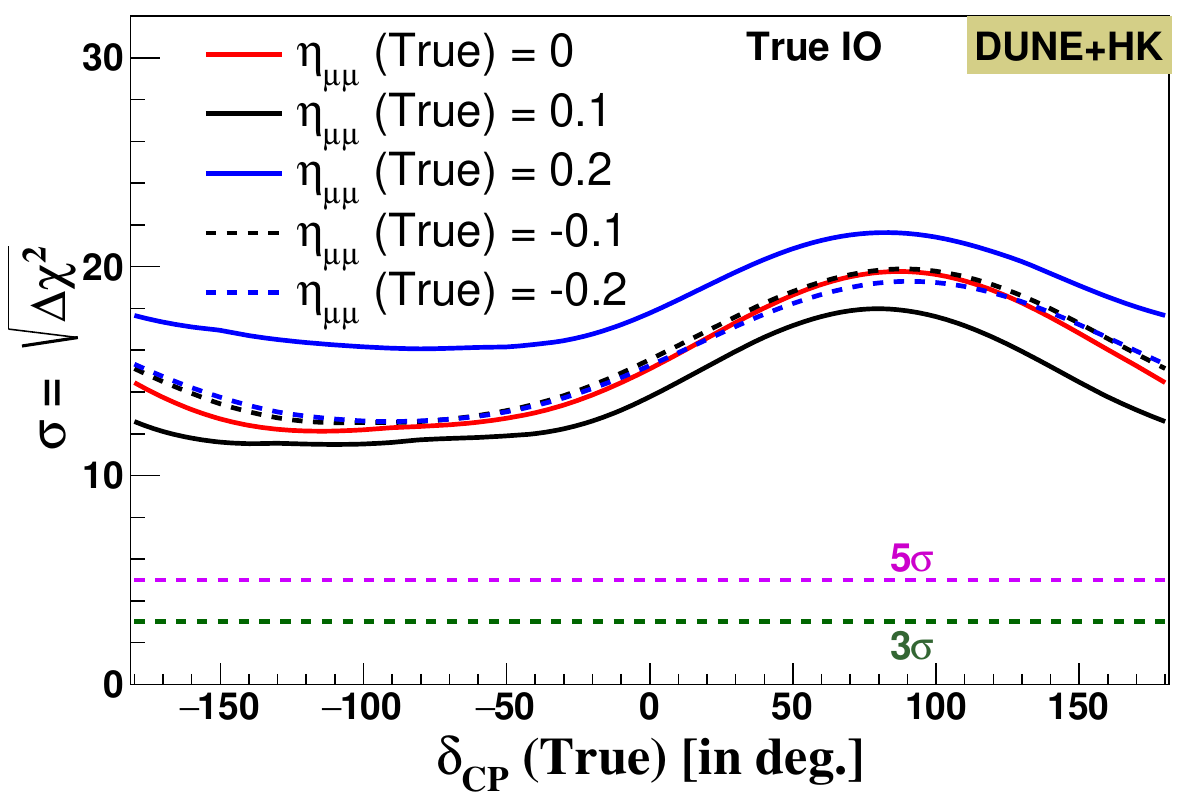}
	\includegraphics[width=0.32\linewidth, height = 5cm]{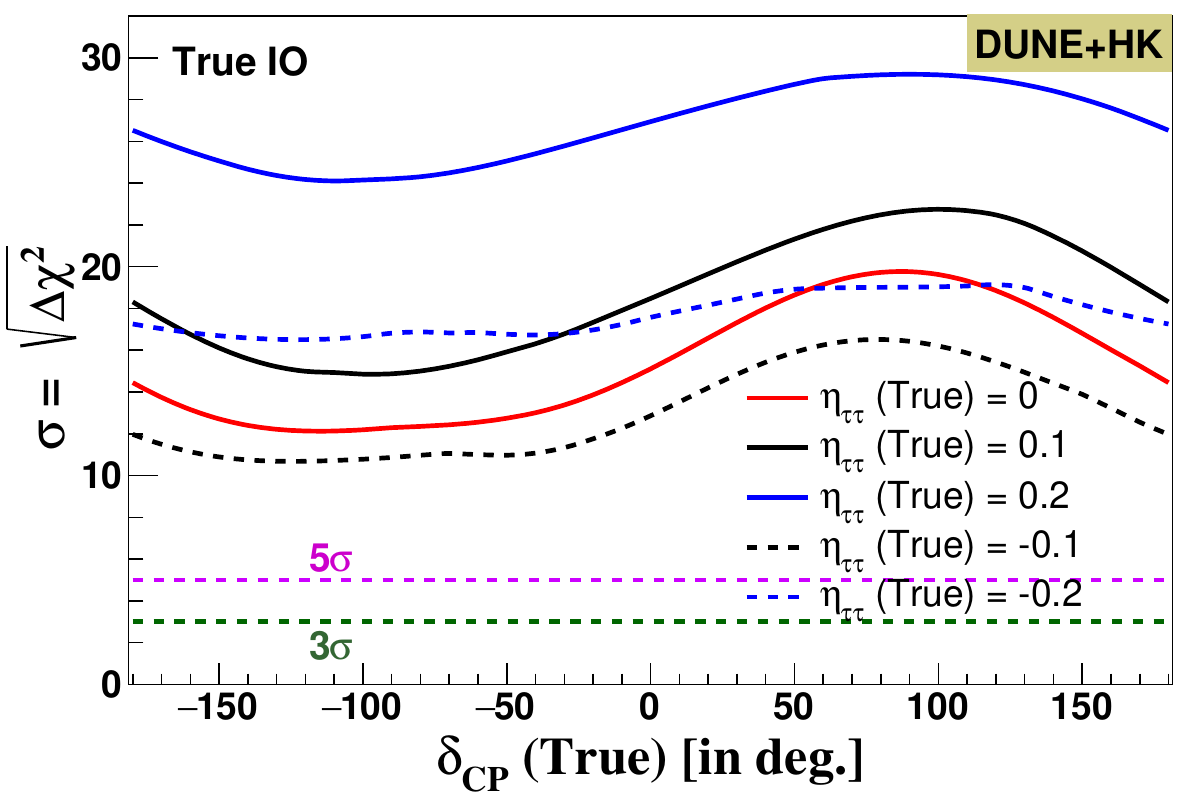}
	\caption{The impact of diagonal scalar NSI parameters on the MO sensitivities for DUNE+HK for true NO (top--panel) and IO (bottom--panel) in the presence of diagonal scalar NSI parameters: $\eta_{ee}$ (left--panel), $\eta_{\mu\mu}$ (middle--panel) and $\eta_{\tau\tau}$ (right--panel). The solid red line is for no--scalar NSI case. The solid (dashed) lines are for non-zero positive (negative) scalar NSI parameters.}
	\label{fig:MH_4}
\end{figure}

In figure \ref{fig:MH_4}, the neutrino MO sensitivity of the combination of DUNE and HK in the presence of scalar NSI is shown. The top panel represents the true NO, whereas the bottom panel represents the true IO case. The non-zero positive (negative) values of diagonal scalar NSI parameters are depicted by the blue and black solid (dashed) curves. The red solid curve corresponds to the no scalar NSI case, with $5\sigma$ and $3\sigma$ CL represented by dashed magenta and green lines respectively. The observations from figure \ref{fig:MH_4} are,

\begin{itemize}
    \item In case of DUNE+HK, the MO sensitivity is above $5\sigma$ CL for all values of $\delta_{CP}$. In the presence of scalar NSI, for the true NO case, positive (negative) $\eta_{ee}$ and $\eta_{\mu \mu}$ mostly improve (deteriorate) the MO sensitivity. However, for $\eta_{\mu \mu}$ and $\eta_{\tau \tau}$ fixed at -0.2, an enhancement in the sensitivity towards MO determination can be seen. 
    \item For IO as true ordering, the positive (negative) $\eta_{ee}$ and $\eta_{\tau \tau}$ enhances (suppresses) the MO sensitivity for DUNE+HK. The effect is more prominent for $\eta_{\tau\tau}$ as compared to $\eta_{ee}$. Although for $\eta_{\tau\tau}$ = -0.2, the MO sensitivity is enhanced for most values of $\delta_{CP}$.  For $\eta_{\mu\mu} = 0.2$, the sensitivity is enhanced, whereas $\eta_{\mu\mu} =0.1$ suppresses the MO sensitivity. In the case of negative $\eta_{\mu\mu}$, the sensitivity curves largely overlap with the no scalar NSI case across various values of $\delta_{CP}$.
\end{itemize}
\begin{figure}[!h]
	\centering
	\includegraphics[width=0.33\linewidth, height = 5cm]{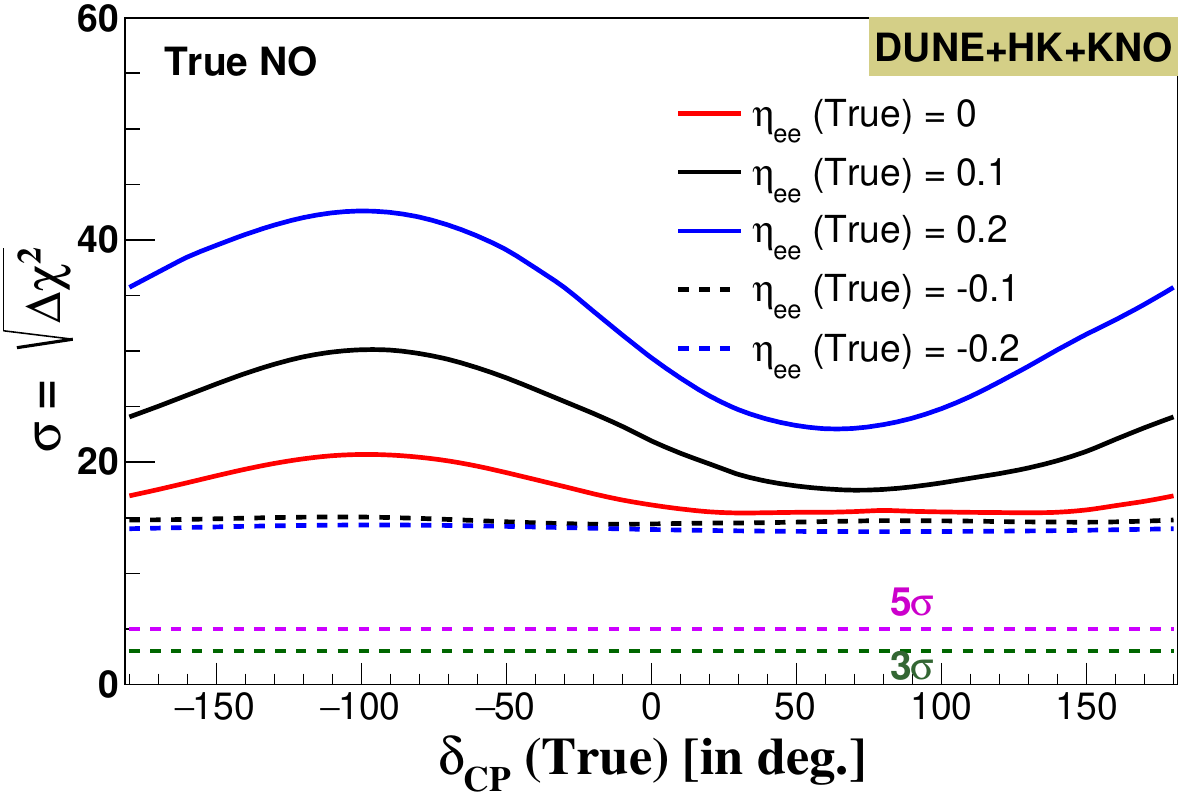}
	\includegraphics[width=0.33\linewidth, height = 5cm]{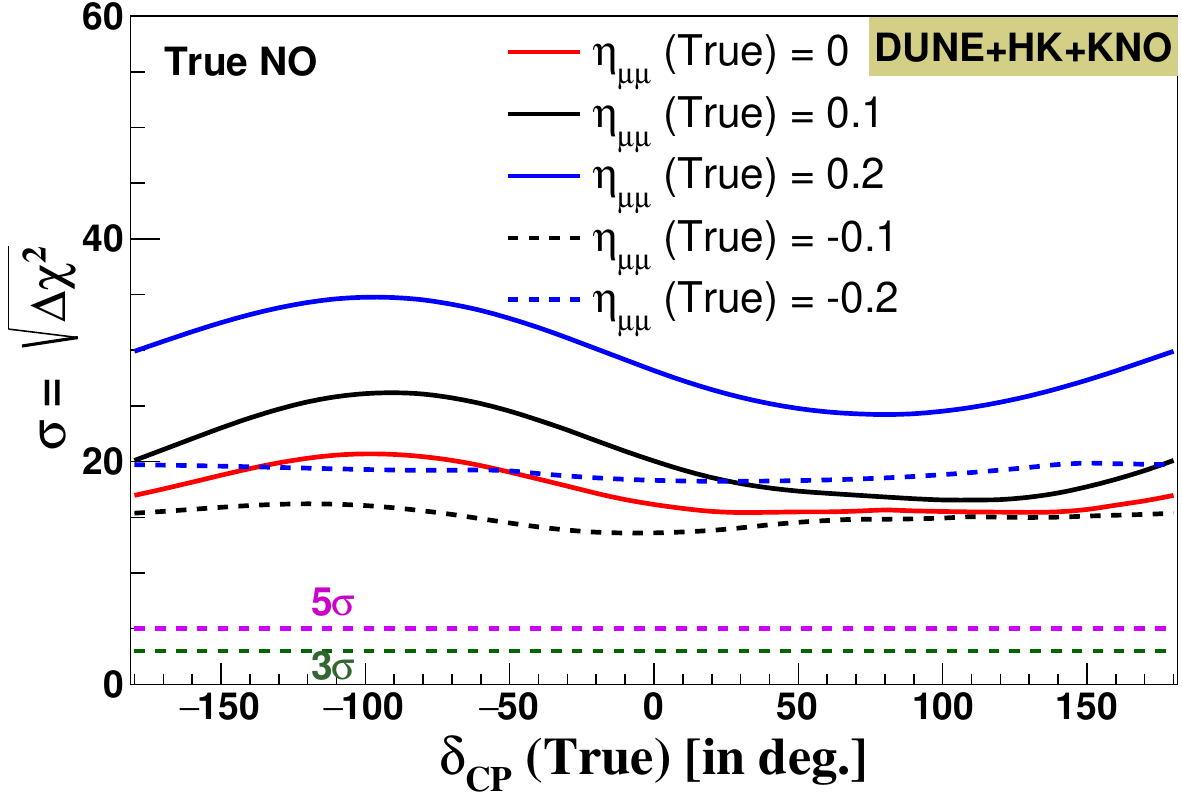} 
	\includegraphics[width=0.32\linewidth, height = 5cm]{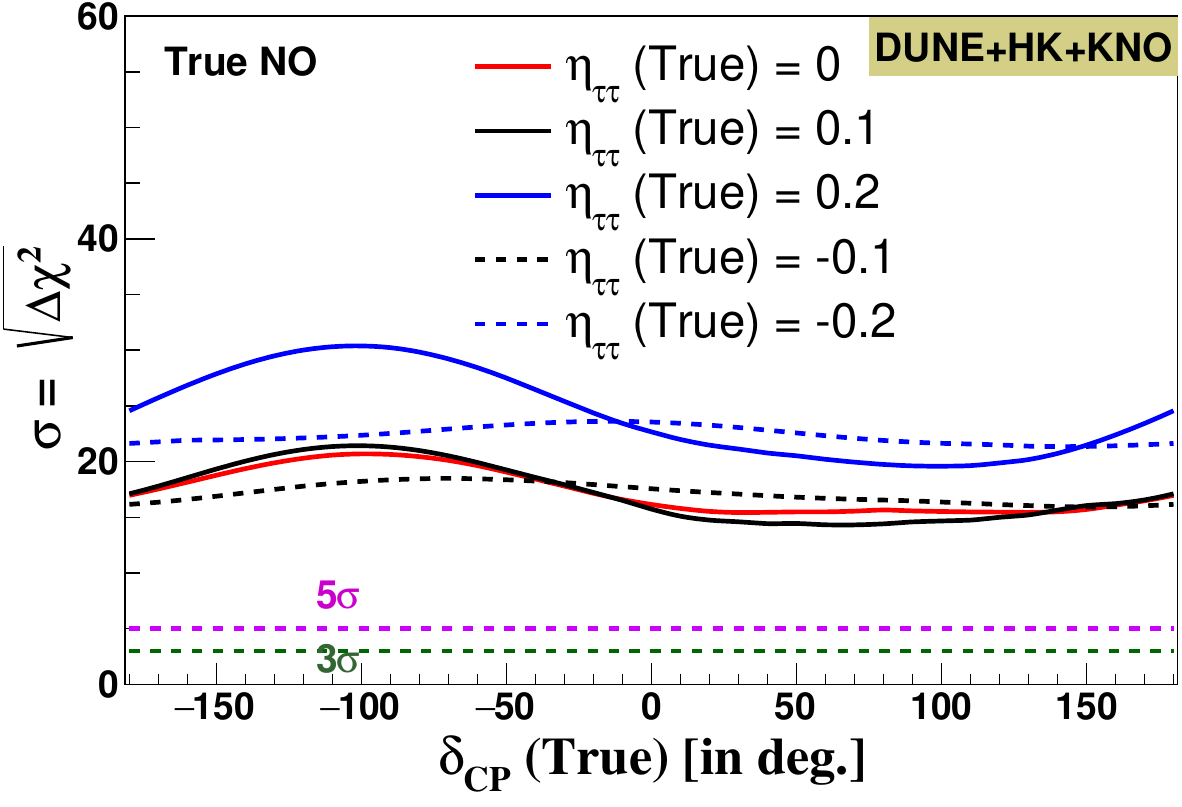}  

        \includegraphics[width=0.33\linewidth, height = 5cm]{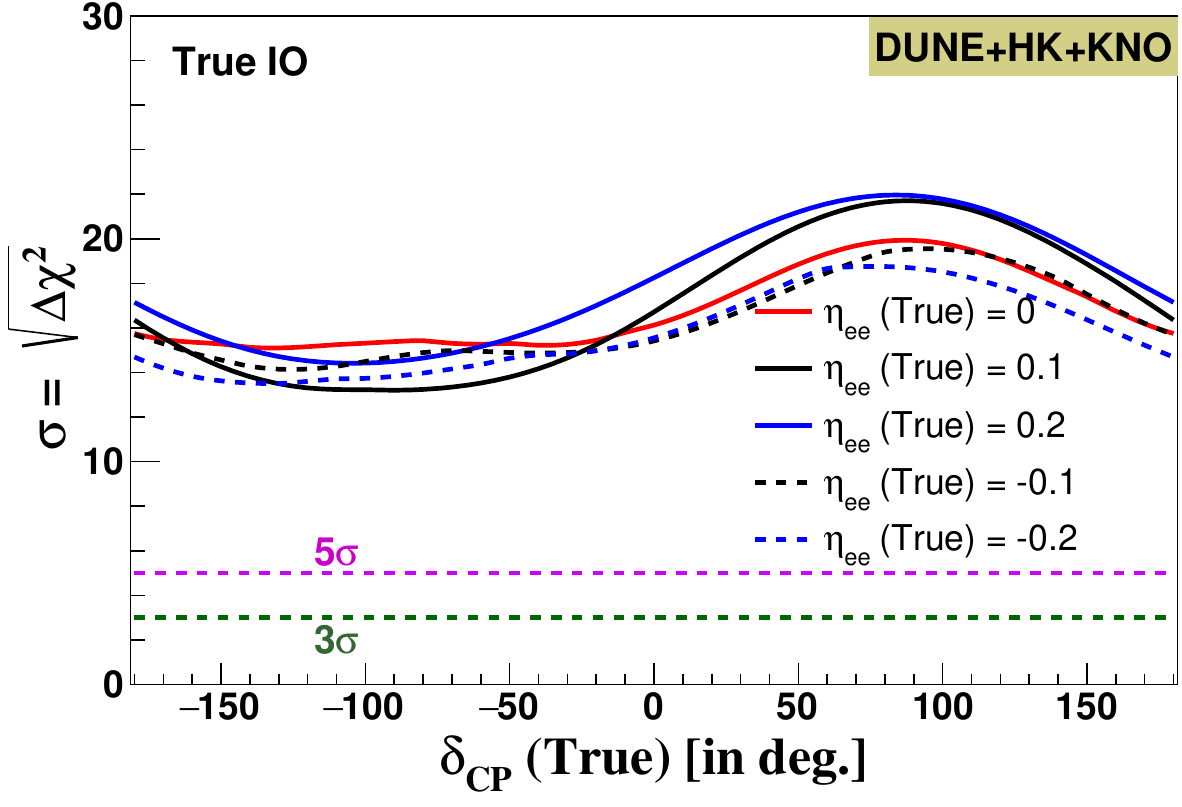} 
	\includegraphics[width=0.33\linewidth, height = 5cm]{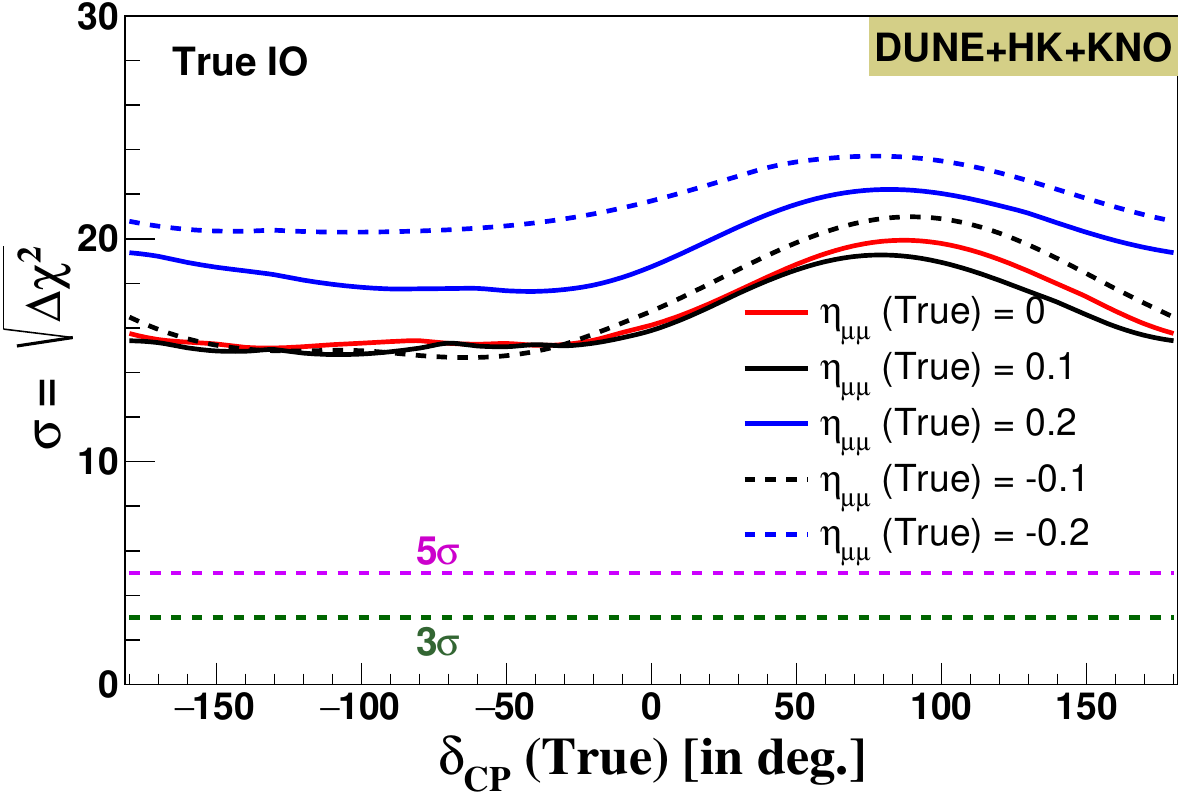} 
	\includegraphics[width=0.32\linewidth, height = 5cm]{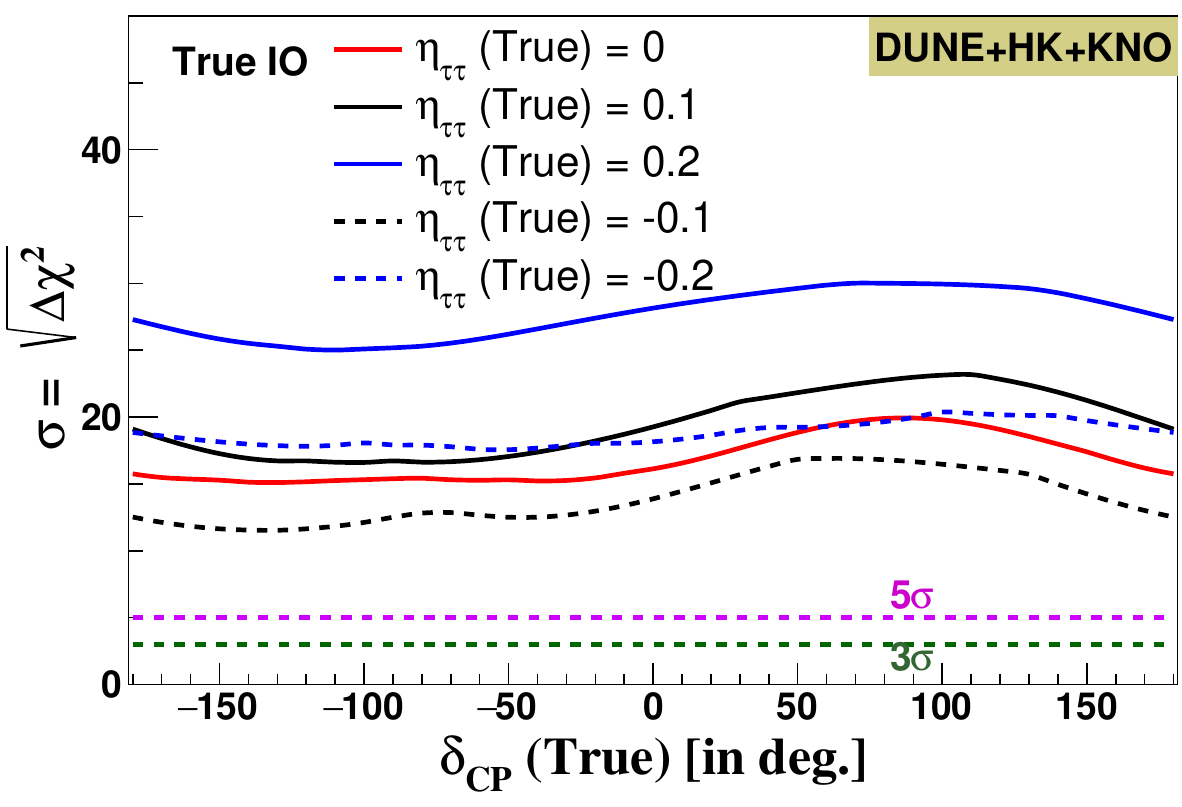} 
	\caption{The impact of scalar NSI on MO sensitivities for DUNE+HK+KNO in the presence of diagonal scalar NSI parameters: $\eta_{ee}$ (left--panel), $\eta_{\mu\mu}$ (middle--panel) and $\eta_{\tau\tau}$ (right--panel). The top--panel corresponds to true NO and the bottom--panel corresponds to true IO. The solid red line is for no--scalar NSI case. The solid (dashed) lines are for non-zero positive (negative) scalar NSI parameters.}
    \label{fig:MH_5}
\end{figure}
\subsubsection{Combining DUNE with HK+KNO} \label{sec:syn_2}
The MO sensitivity of DUNE$+$HK+KNO for true NO (top-panel) and true IO (bottom-panel) is shown in figure \ref{fig:MH_5}. The red solid line represents the case without scalar NSI and other coloured lines represent non-zero values of $\eta_{ee}$ (left--panel), $\eta_{\mu \mu}$ (middle--panel) and $\eta_{\tau \tau}$ (right--panel) as indicated in the legends. The $5\sigma$ and $3\sigma$ CL are drawn in magenta and green colours. We observe that,

\begin{itemize}
    \item For true NO, a positive (negative) $\eta_{ee}$ and $\eta_{\mu\mu}$ enhances (suppresses) the MO sensitivities for most values of $\delta_{CP}$ except for $\eta_{\mu\mu}$=-0.2. For $\eta_{\tau\tau}=\pm$0.2, MO sensitivities are enhanced whereas $\eta_{\tau\tau}=$0.1 mostly overlaps with the standard MO sensitivities and $\eta_{\tau\tau}=$-0.1 suppresses in the region of $\delta_{CP}$ $\in$ [$-40^\circ, -180^\circ$].

    \item For true IO, the effect of $\eta_{\tau\tau}$ is prominent as compared to $\eta_{ee}$ and $\eta_{\mu\mu}$. The positive values of $\eta_{\alpha\beta}$ enhances the sensitivities except for $\eta_{ee}$ in the negative $\delta_{CP}$-plane. A negative value of $\eta_{ee}$ mostly suppresses the MO sensitivities. For negative $\eta_{\mu\mu}$, we see an enhancement in the MO sensitivities. For $\eta_{\tau\tau}$ fixed at -0.2 (-0.1), the MO sensitivities are enhanced (suppressed) for the complete $\delta_{CP}$ space.
    
    \item In both the scenarios of true NO and true IO, we observe a similar trend to those of the DUNE+HK configuration. This similarity is expected since the detectors used in both setups are identical. However, in the case of DUNE+HK+KNO, the effects are more prominent due to its longer baseline, as scalar NSI varies linearly with the environmental matter density. 
\end{itemize}

\subsection{Precision measurement of $\Delta m_{31}^2$}\label{sec:prec}
\begin{figure}[!h]
\centering
\includegraphics[width=0.32\linewidth, height = 5cm]{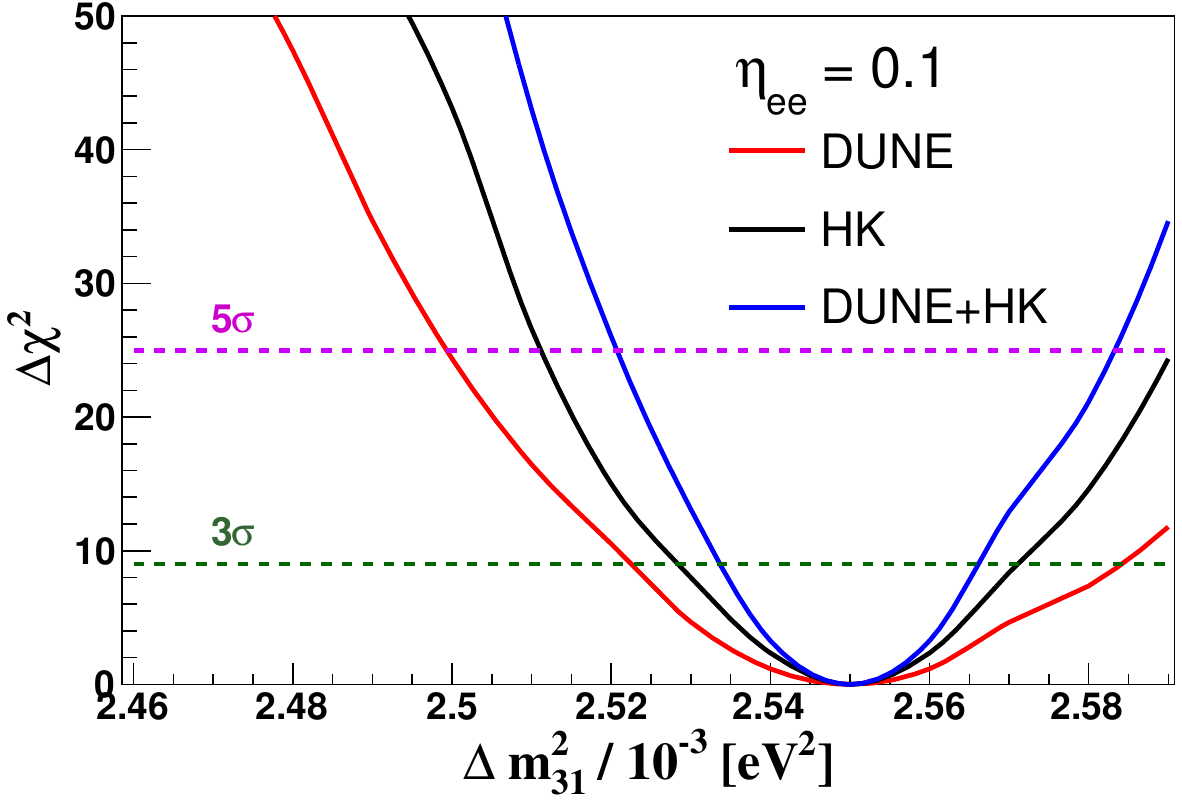} 
\includegraphics[width=0.32\linewidth, height = 5cm]{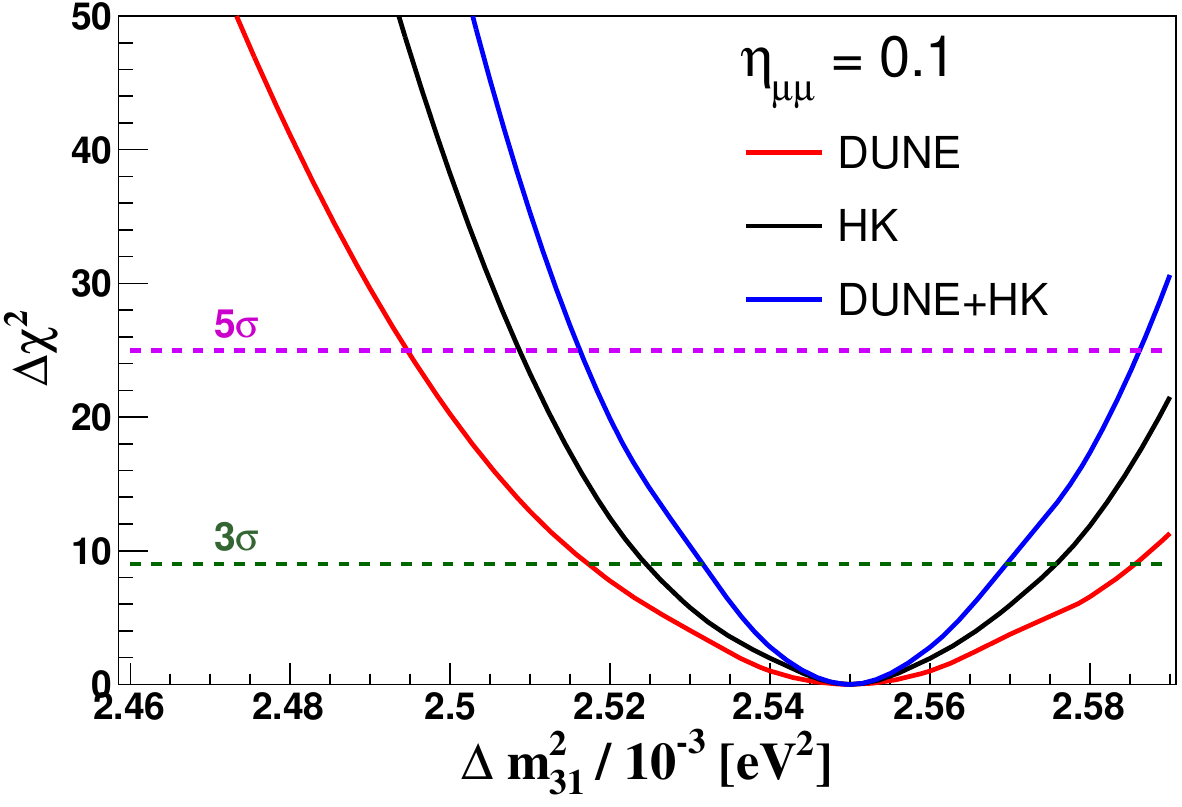}
\includegraphics[width=0.32\linewidth, height = 5cm]{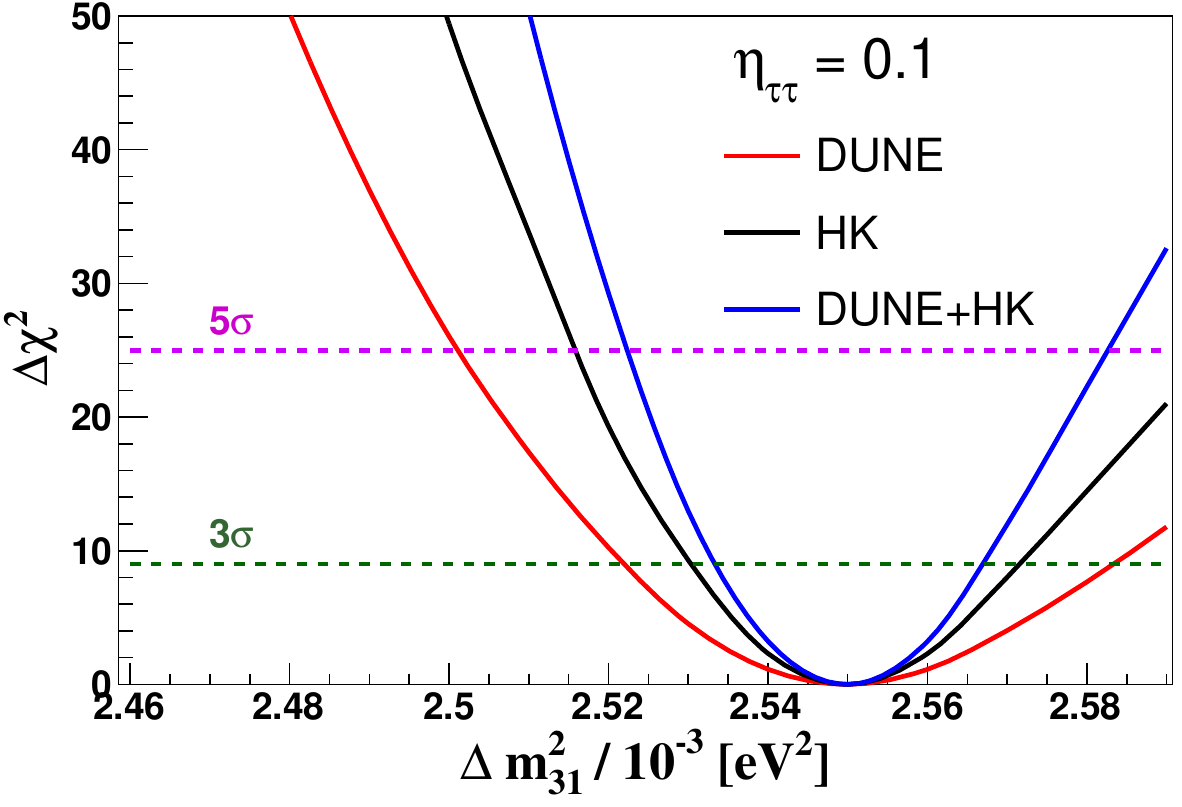}
\includegraphics[width=0.32\linewidth, height = 5cm]{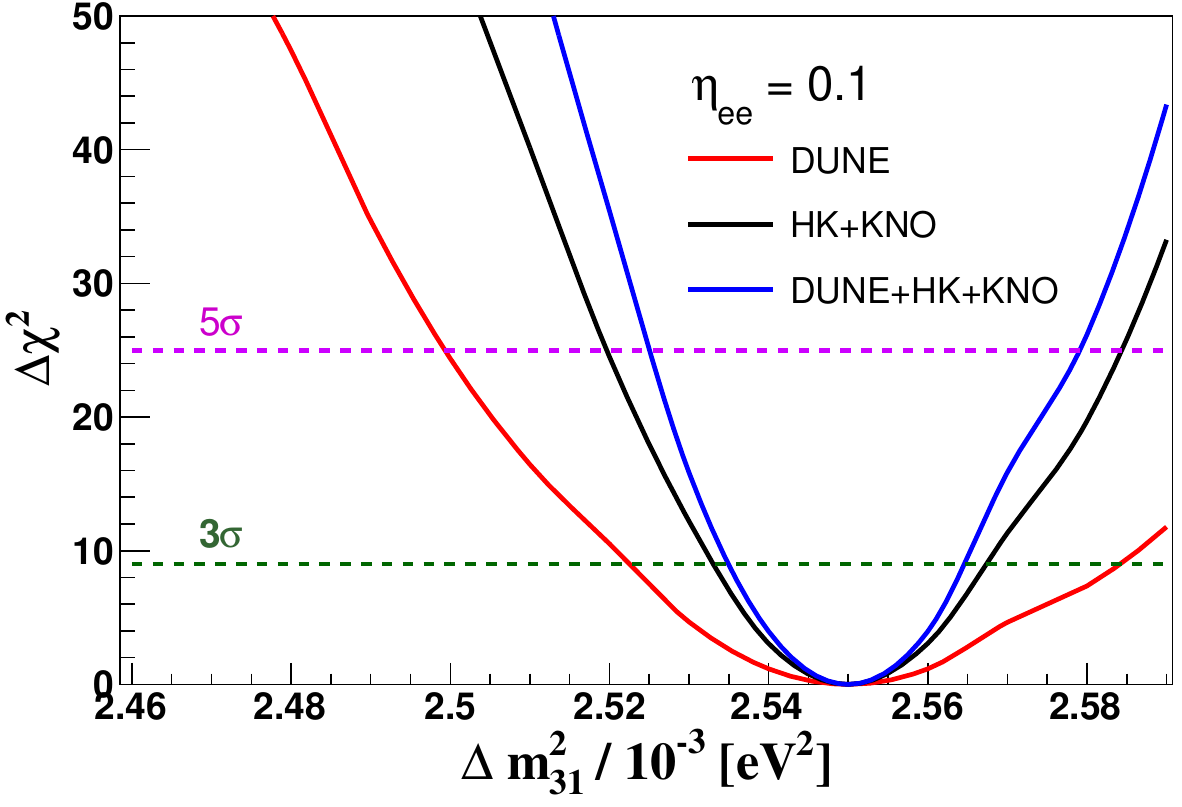}
\includegraphics[width=0.32\linewidth, height = 5cm]{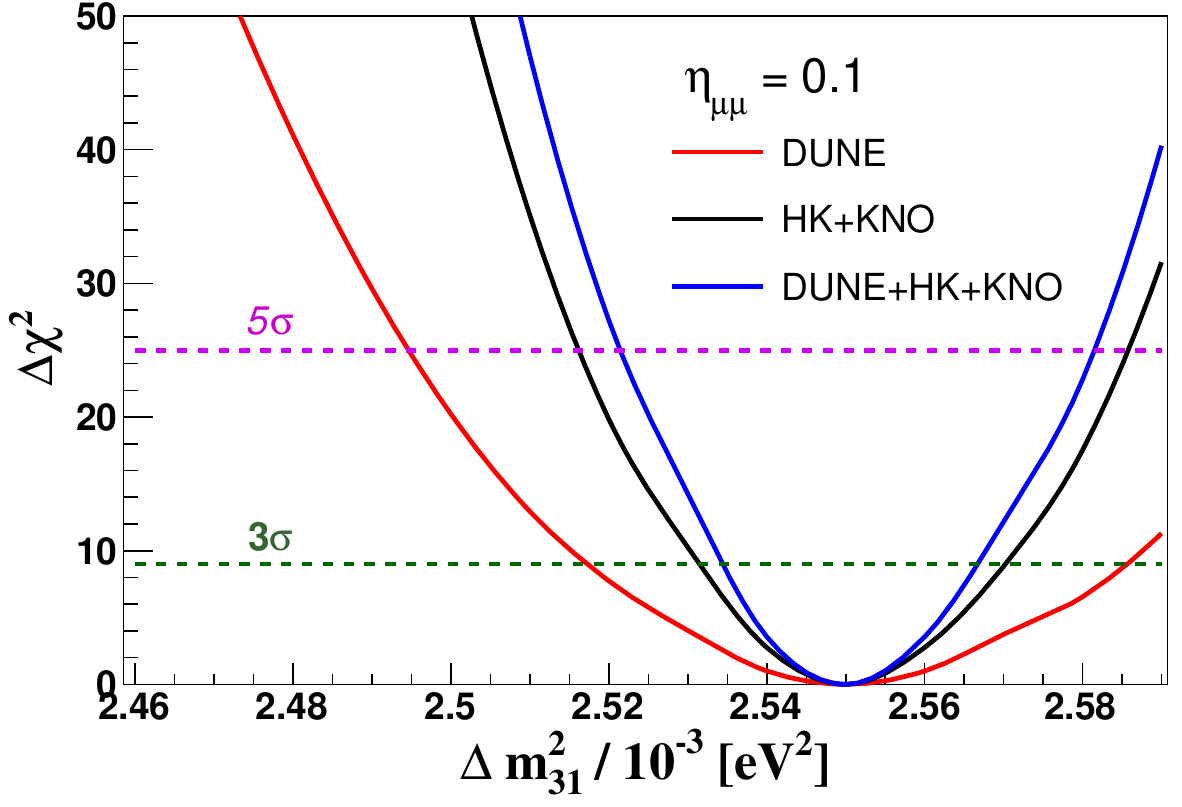}
\includegraphics[width=0.32\linewidth, height = 5cm]{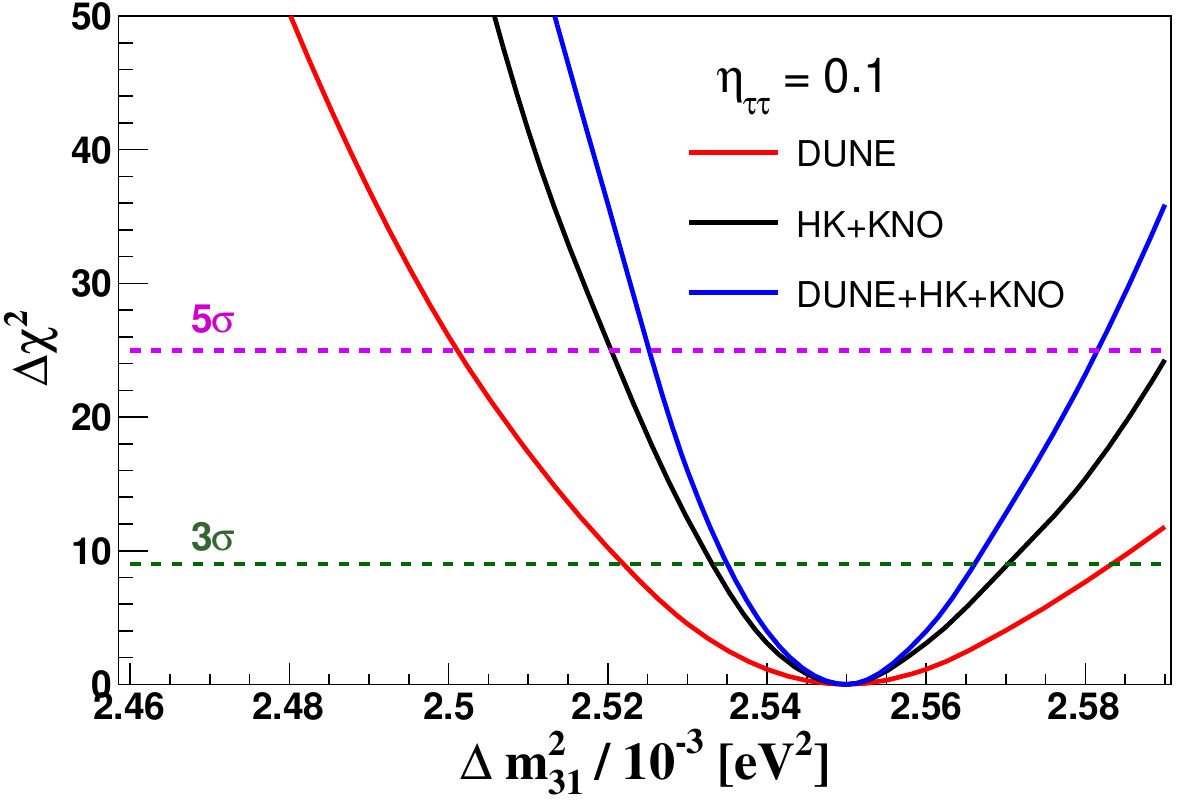}
\caption{The impact of scalar NSI on $\Delta m_{31}^{2}$ constraining capability in the presence of diagonal scalar NSI parameters $\eta_{ee}$ (left--panel), $\eta_{\mu\mu}$ (middle--panel) and $\eta_{\tau\tau}$ (right--panel) for true NO. The top--panel corresponds to DUNE (red), HK (black) and the combined analysis of DUNE+HK (blue), whereas the bottom--panel corresponds to DUNE (red), HK+KNO (black) and DUNE+HK+KNO (blue).}
\label{fig:MH_6}
\end{figure}

We look into the impact of diagonal scalar NSI elements $\eta_{ee}$, $\eta_{\mu\mu}$ and $\eta_{\tau\tau}$ on the precision measurements of $\Delta m_{31}^{2}$ at DUNE, HK and HK+KNO. We have also performed combined analyses for DUNE+HK and DUNE+HK+KNO. We present the results in figure \ref{fig:MH_6}. The magenta and green dashed lines represent the $5\sigma$ and $3\sigma$ CL respectively. The red solid line represents the DUNE case. In the top (bottom) panel, the black and blue solid lines represent the case HK (HK+KNO) and DUNE+HK (DUNE+HK+KNO) respectively. We have considered the higher $\theta_{23}$ octant as the true octant and NO as the true mass ordering. The true value of $\Delta m_{31}^{2}$ is fixed at 2.55$\times10^{-3}eV^{2}$ and the test value is varied in $[2.46,2.59]\times10^{-3}eV^{2}$. We have additionally marginalized over $\delta_{CP}$, $\theta_{23}$ and $\eta_{\alpha\beta}$. The true value of $\eta_{\alpha\beta}$ is fixed at 0.1 for all the experimental configurations. The findings obtained are stated below.

\begin{itemize}
    \item The constraining capability of $\Delta m_{31}^{2}$ for DUNE+HK and DUNE+HK+KNO configuration is better compared to DUNE, HK and HK+KNO for all the cases of scalar NSI parameters. The synergies of experiments lead to a better constraining of $\Delta m_{31}^{2}$ in each case.

    \item In presence of $\eta_{ee}$ and $\eta_{\tau\tau}$, the constraining of $\Delta m_{31}^{2}$ is nominally better in comparison to $\eta_{\mu\mu}$. This can be seen for both the synergy cases i.e. DUNE+HK and DUNE+HK+KNO. 
\end{itemize}

\noindent To quantify the precision measurement of $\Delta m_{31}^2$, we can define a quantity $\mathbb{P}$ as,
\begin{equation}
    \mathbb{P}= \frac{p_{max}-p_{min}}{p_{max}+p_{min}},
\end{equation}

\noindent where, $p_{max}$ and $p_{min}$ are the maximum value and minimum value of $\Delta m_{31}^2$ respectively, at a fixed CL. For example, if $\eta_{ee}$ is determined elsewhere to be  0.1, DUNE will be able to measure $\Delta m_{31}^2$ at a precision of $\sim$ 1.17\% at 3$\sigma$ CL, whereas, HK will be able to measure at a precision of $\sim$ 0.78\%. HK+KNO is expected to measure the same at a precision of $\sim$ 0.6\%. The synergy between DUNE and HK may significantly improve the precision to $\sim$ 0.58\%. In addition, the synergy with DUNE, HK and KNO may further improve it to $\sim$ 0.5\%. In presence of $\eta_{\mu\mu}$, DUNE, HK and HK+KNO will be able to measure at a precision of $\sim$ 1.27 \%, $\sim$ 0.98 \% and  $\sim$ 0.78 \% respectively at 3$\sigma$ CL.. A synergy of DUNE and HK (HK+KNO) improves the number to $\sim$ 0.68 \% ($\sim$ 0.58\%). Similarly, in presence of $\eta_{\tau\tau}$, DUNE, HK and HK+KNO will be able to measure at a precision of $\sim$ 1.17 \%, $\sim$ 0.78 \% and $\sim$ 0.68 \% respectively at 3$\sigma$ CL. And a synergy of DUNE+HK and DUNE+HK+KNO may improve the precision to $\sim$ 0.58\%.

\section{Summary and concluding remarks }\label{sec:summary}
The upcoming $\nu$-oscillation experiments are centred on achieving precise measurements of $\nu$-mixing parameters and addressing the unknowns in the field of neutrino physics like the determination of neutrino MO, octant of $\theta_{23}$ and the leptonic CP phase. The determination of the true neutrino MO is crucial as it holds the potential to significantly enhance our understanding of the fundamental nature of neutrinos. The existence of NSIs of neutrinos has the potential to influence the physics sensitivities of these LBL $\nu$-experiments. The potential scalar coupling of neutrinos is one such non-standard physics scenario which may affect the neutrino MO sensitivities. Scalar NSIs of neutrinos bring a direct contribution to the $\nu$-mass matrix which is sub-dominant in nature. 

In this work, we have explored the effects of diagonal scalar NSI parameters $(\eta_{ee},\eta_{\mu\mu},\eta_{\tau\tau})$ on the neutrino MO sensitivities at three upcoming LBL experiments i.e. DUNE, HK and HK+KNO. We have first explored the impact of scalar NSI at the probability level for both the neutrino MO, where if the value of $\eta_{\alpha \beta}$ is exactly known from other experiments, a larger separation between the two probability bands would indicate a better measurement of the MO. We observe that the discrimination between the mass orderings is enhanced in the presence of $\eta_{ee}$ for all the chosen experiments. We then explored the MO sensitivities for the individual experiments in the presence of diagonal scalar NSI parameters. We further probed the potential improvement due to the two synergy cases i.e. DUNE+HK and DUNE+HK+KNO. We have mainly focused on the impact of the presence of scalar coupling of neutrinos on the MO sensitivities. We observe that the presence of scalar NSI can significantly alter the $\nu$-oscillation probabilities which in turn affects the sensitivities of LBL experiments towards MO determination. In MO sensitivity studies, we mostly observed an enhancement in the sensitivities for different choices of positive $\eta_{\alpha \beta}$ parameters considering true NO. However, for negative values of scalar NSI parameters, we observed suppression/enhancement depending on the combination of $\eta_{\alpha \beta}$ and $\delta_{CP}$ values. For true IO, we observe a significant overlapping of sensitivities for the standard case with the scalar NSI case for negative $\eta_{\alpha \beta}$. However, the combined analysis of DUNE+HK and/or DUNE+HK+KNO can remove the overlapping for $\eta_{\tau\tau}$ parameter due to the increased statistics over an enhanced parameter region. In presence of scalar NSI, we observe that HK and HK+KNO show better constraining capability of $\Delta m_{31}^{2}$ in comparison to DUNE. The combined analyses for DUNE+HK and DUNE+HK+KNO can enhance the overall constraining capability for all non-zero scalar NSI parameters. 

The neutrino mass ordering stands as one of the fundamental mysteries in particle physics, the determination of which would mark a significant breakthrough in our understanding of the universe. Therefore, understanding the impact of subdominant effects arising from scalar NSIs on the MO sensitivities of LBL experiments is imperative. It is also crucial to impose constraints on the scalar NSI parameters to ensure accurate interpretation of experimental data. 

\section*{Acknowledgments}
 The authors acknowledge the DST SERB grant CRG/2021/002961. AS would also acknowledge the fellowship received from CSIR-HRDG (09/0796(12409)/2021-EMR-I). AM thanks the support of the Research and Innovation Grant 2021 (DoRD/RIG/10-73/ 1592-A) funded by Tezpur University. DB acknowledges DST for the financial support provided through the INSPIRE fellowship. 

\section*{A  Scalar NSI features involving probability and sensitivity} 
\subsection*{A.1  Dependence of oscillation probabilities on $\nu$-mass}\label{app:abs_mass_dep}

\begin{figure}[!h]
    \centering
    \includegraphics[width=0.48\textwidth, height = 5.5cm]{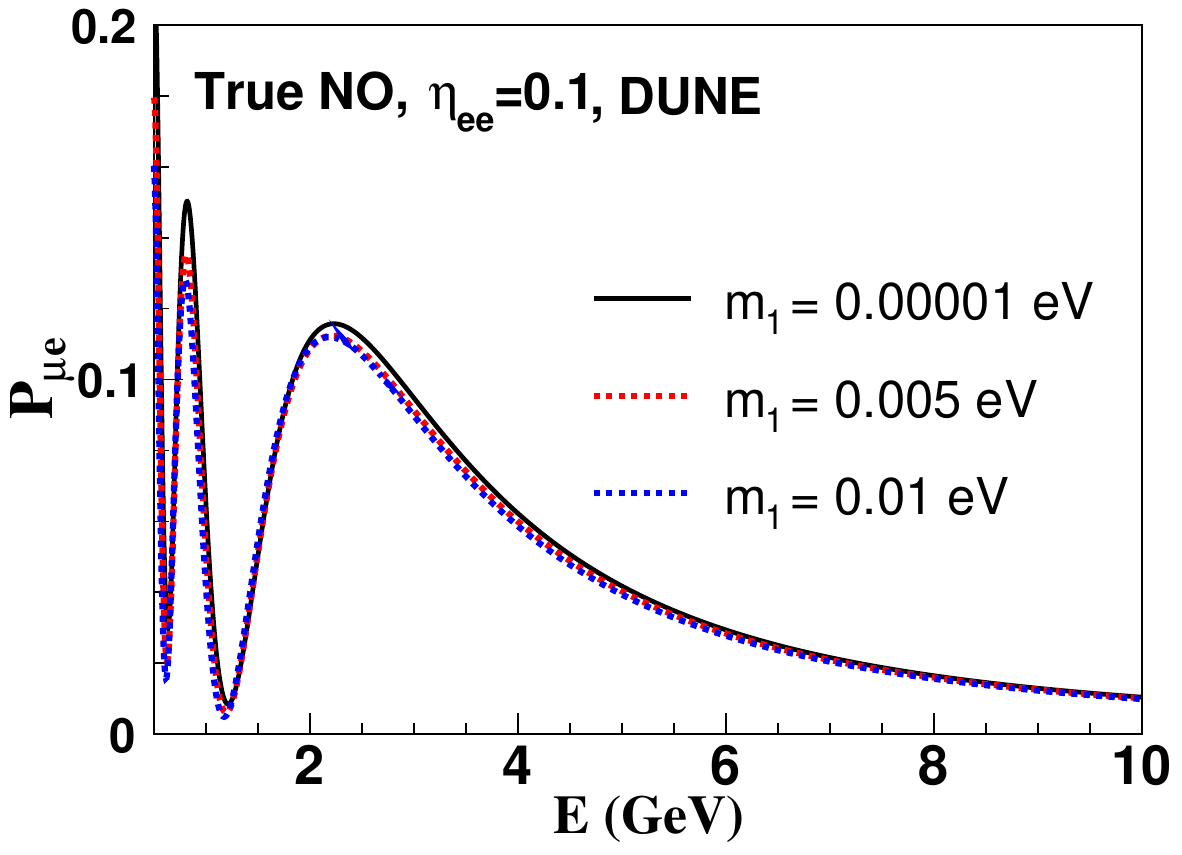}
    \includegraphics[width=0.48\textwidth, height = 5.5cm]{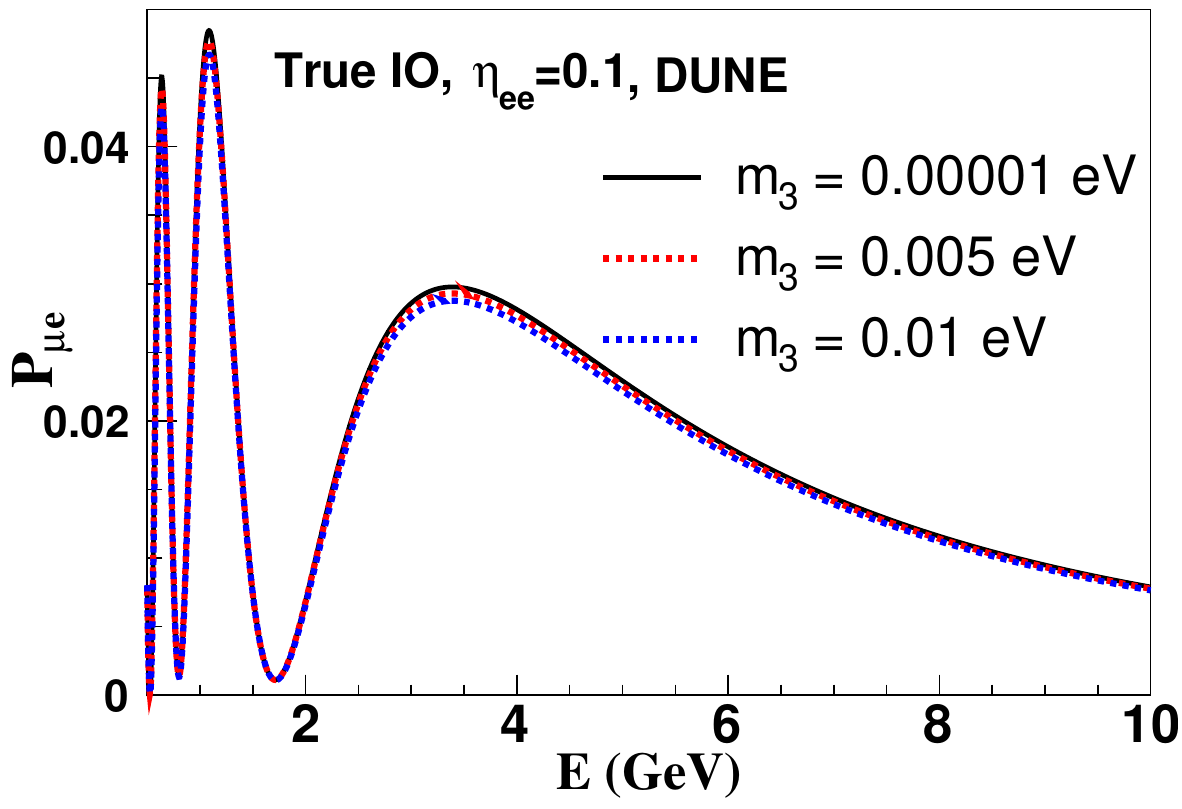}
    \caption{Impact of different choices of lightest $\nu$-mass on $P_{\mu e}$ for the baseline of DUNE in presence of scalar NSI parameter $\eta_{ee}$ fixed at 0.1. The left panel represents true NO and the right panel represents true IO.}
    \label{fig:prob}
\end{figure}

As scalar NSI contributes directly to the neutrino mass matrix in the effective Hamiltonian. The $\nu$-oscillation probabilities will depend not only on the mass-squared differences but also on the absolute neutrino masses. We have explored the impact of various choices of $m_{lightest}$ on the $\nu$-oscillation probabilities. In figure \ref{fig:prob}, the $P_{\mu e}$ vs Energy for such neutrino mass choices are shown for both of the mass orderings at DUNE baseline. The choices of the values of the lightest neutrino mass are such that the sum of neutrino masses follows the cosmological bound i.e. $\sum m_{i}<0.12$ eV. The other oscillation parameters are fixed at values as mentioned in table \ref{tab:mixing_parameters}. For both mass orderings, we observe that an increase in $m_{lightest}$ nominally suppresses $P_{\mu e}$.

\begin{figure}[!h]
    \centering
    \includegraphics[width=0.48\textwidth, height = 5.5cm]{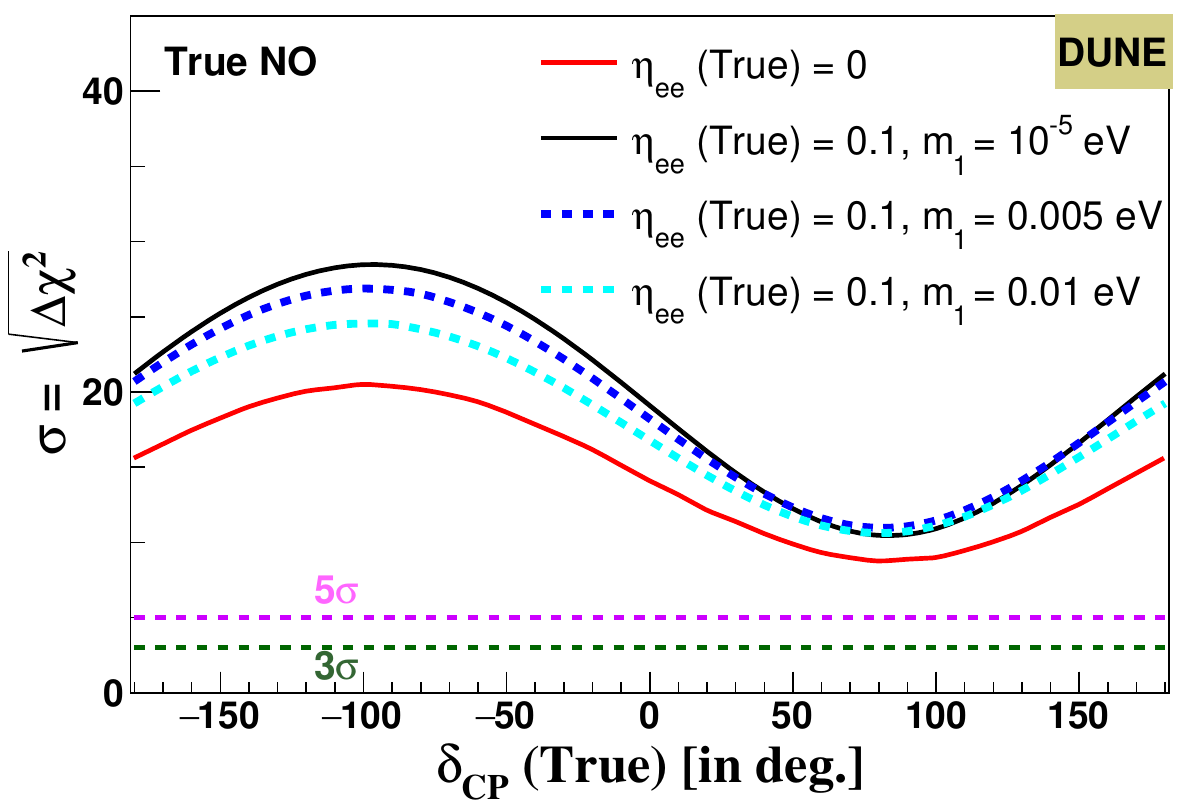}
    \includegraphics[width=0.48\textwidth, height = 5.5cm]{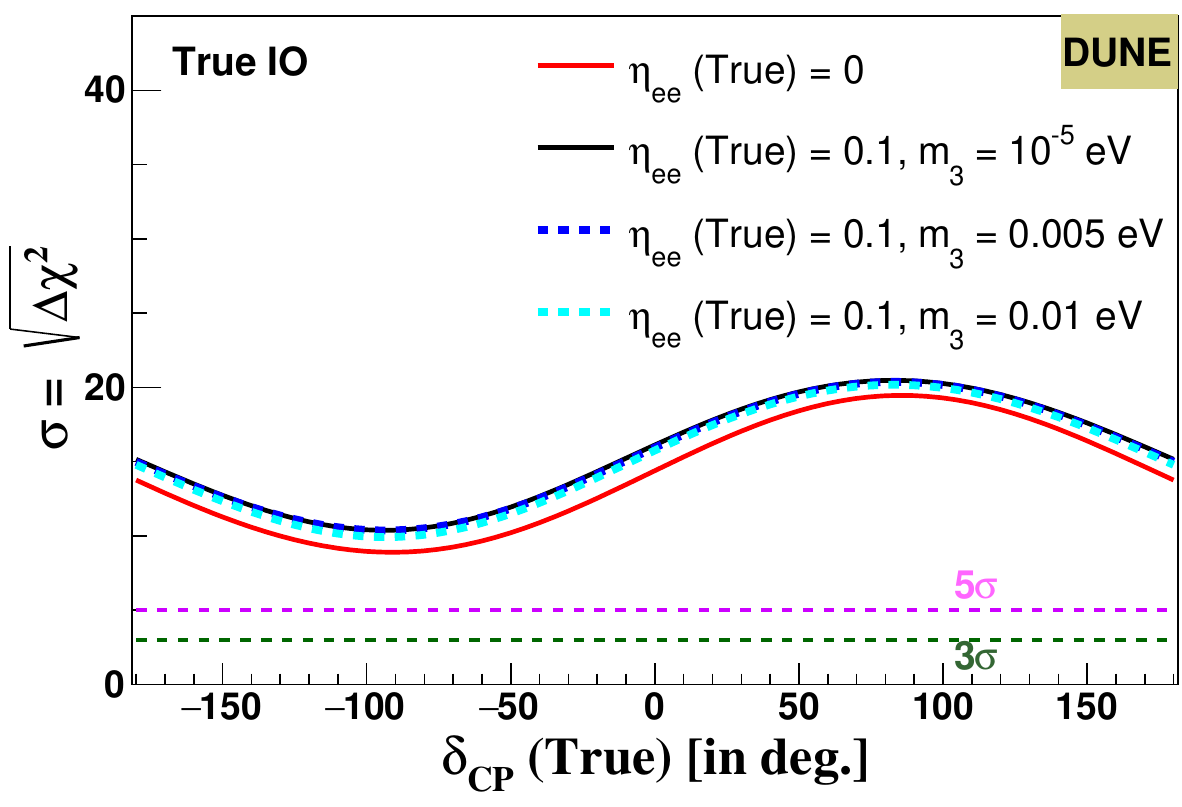}
    \caption{Impact of different choices of lightest $\nu$-mass on the MO sensitivity at DUNE in the presence of scalar NSI parameter $\eta_{ee}$ fixed at 0.1. The left (right) panel represents the true NO (IO) case. The red solid line represents the SI case. The solid black, dashed blue and dashed cyan line represents the sensitivity in the presence of $\eta_{ee}$ with the lightest $\nu$-mass fixed at $10^{-5}$, $0.005$ and $0.01$ eV respectively.}
    \label{fig:chi2_mass}
\end{figure}

In figure \ref{fig:chi2_mass}, the impact of the choices of $m_{lightest}$ on the MO sensitivity of DUNE is shown. The red solid line represents the SI case. The solid black, dashed blue and dashed cyan lines represent the sensitivity in the presence of $\eta_{ee}$ with the lightest $\nu$-mass fixed at $10^{-5}$, $0.005$ and $0.01$ eV respectively. For both of the mass orderings, the overall trend of the MO sensitivity curve remains similar. For NO, the impact of $\nu$-mass is relatively nominal in the positive $\delta_{CP}$-plane, whereas, a suppression with increasing $\nu$-mass is observed in the negative $\delta_{CP}$-plane. For IO, the MO sensitivity is marginally affected by different choices of lightest $\nu$-mass.

\subsection*{A.2  Variation of $\Delta P_{\mu e}$ in $\Delta m_{31}^{2}$-E parameter space} \label{app:deg}
In order to quantify the impact of scalar NSI, we define a quantity $\Delta P_{\mu e}$ as $$\Delta P_{\mu e} = P_{\mu e}^{NSI}- P_{\mu e}^{SI}$$ where, $P_{\mu e}^{NSI}$ and $P_{\mu e}^{SI}$ represents the appearance probabilities in the presence of scalar NSI and no scalar NSI cases respectively. 
\begin{figure}[!h]
\centering
\includegraphics[width=0.33\linewidth, height = 5.5cm]{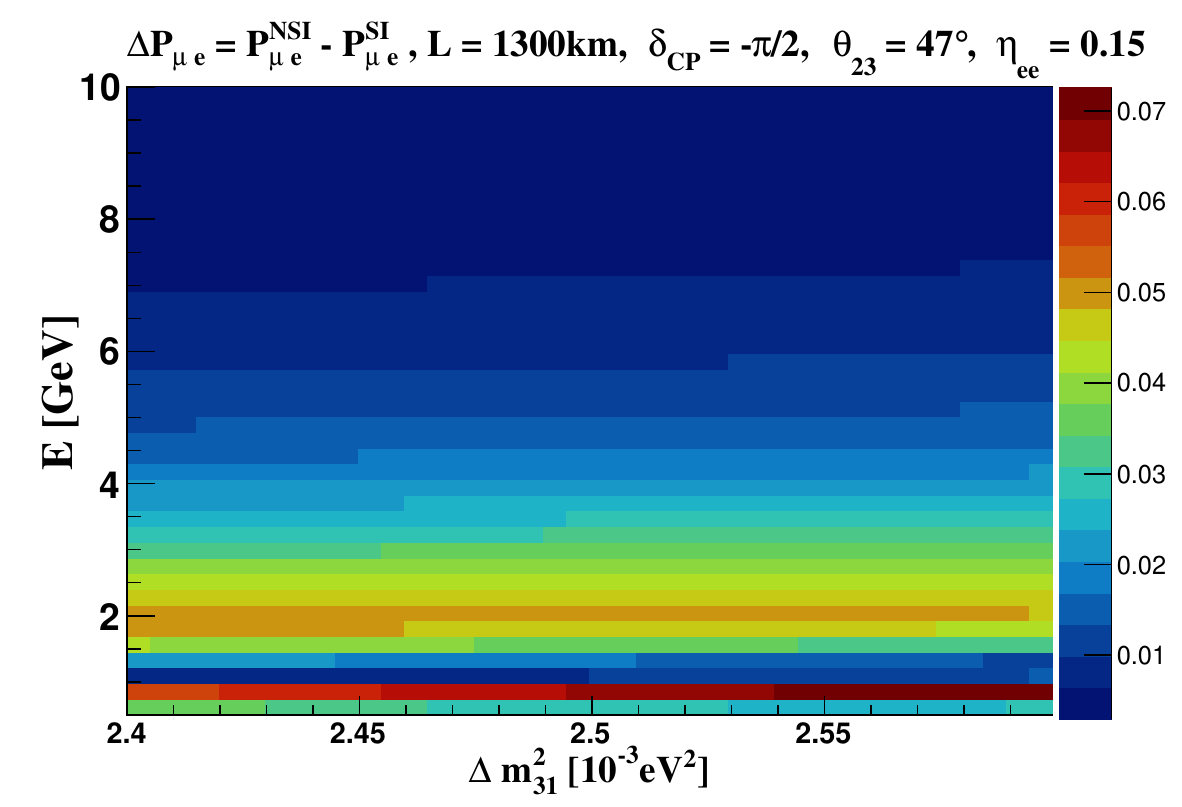} 
\includegraphics[width=0.33\linewidth, height = 5.5cm]{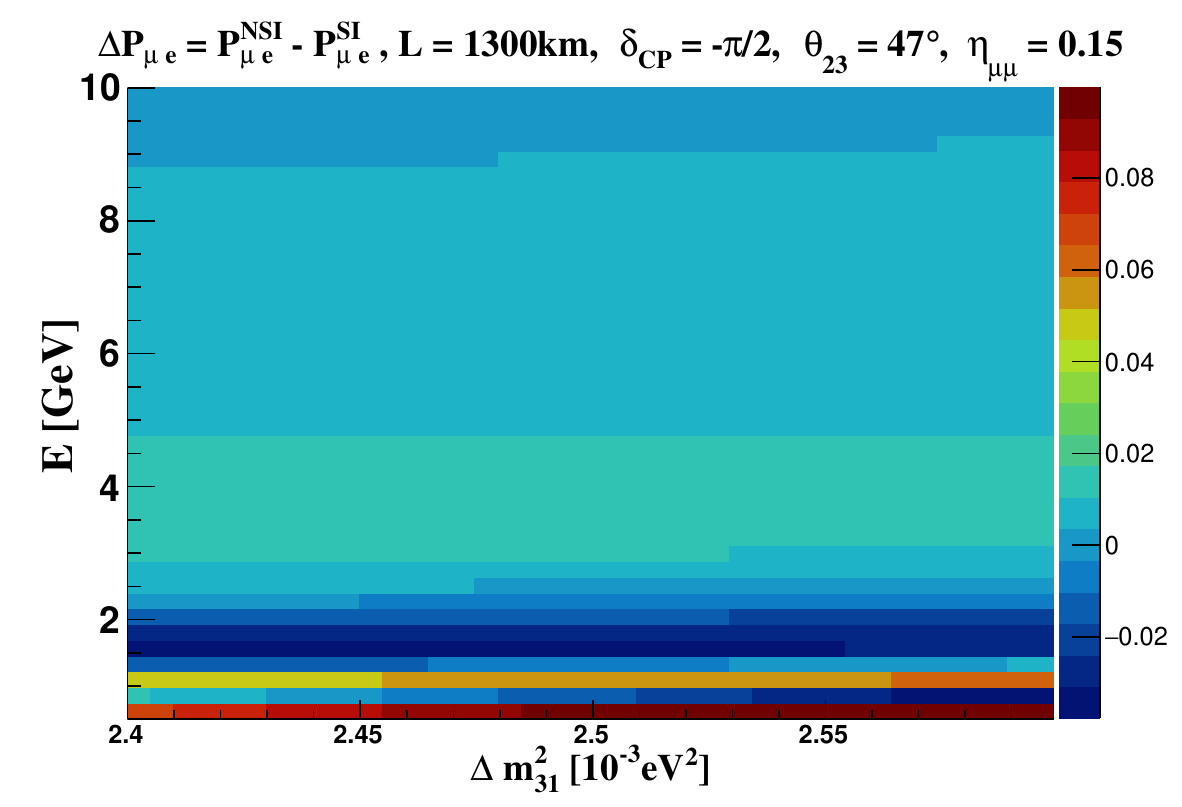} 
\includegraphics[width=0.32\linewidth, height = 5.5cm]{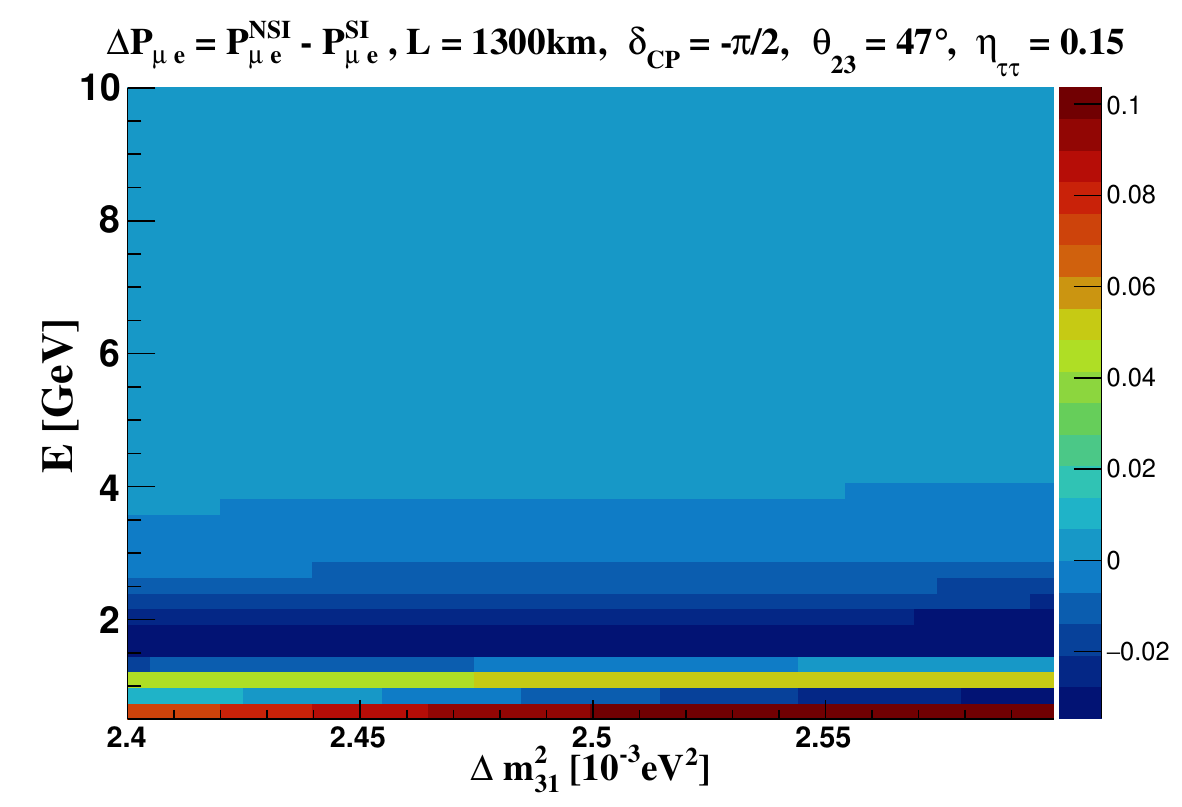} 
\caption{Variation of $\Delta P_{\mu e}$ in $\Delta m_{31}^{2}$-E parameter space in the presence of scalar NSI elements $\eta_{ee}$ (left--panel), $\eta_{\mu\mu}$ (middle--panel) and $\eta_{\tau\tau}$ (right--panel) for DUNE.}
\label{fig:2d_E_M_DUNE}
\end{figure}
\begin{figure}[!h]
\centering
\includegraphics[width=0.33\linewidth, height = 5.5cm]{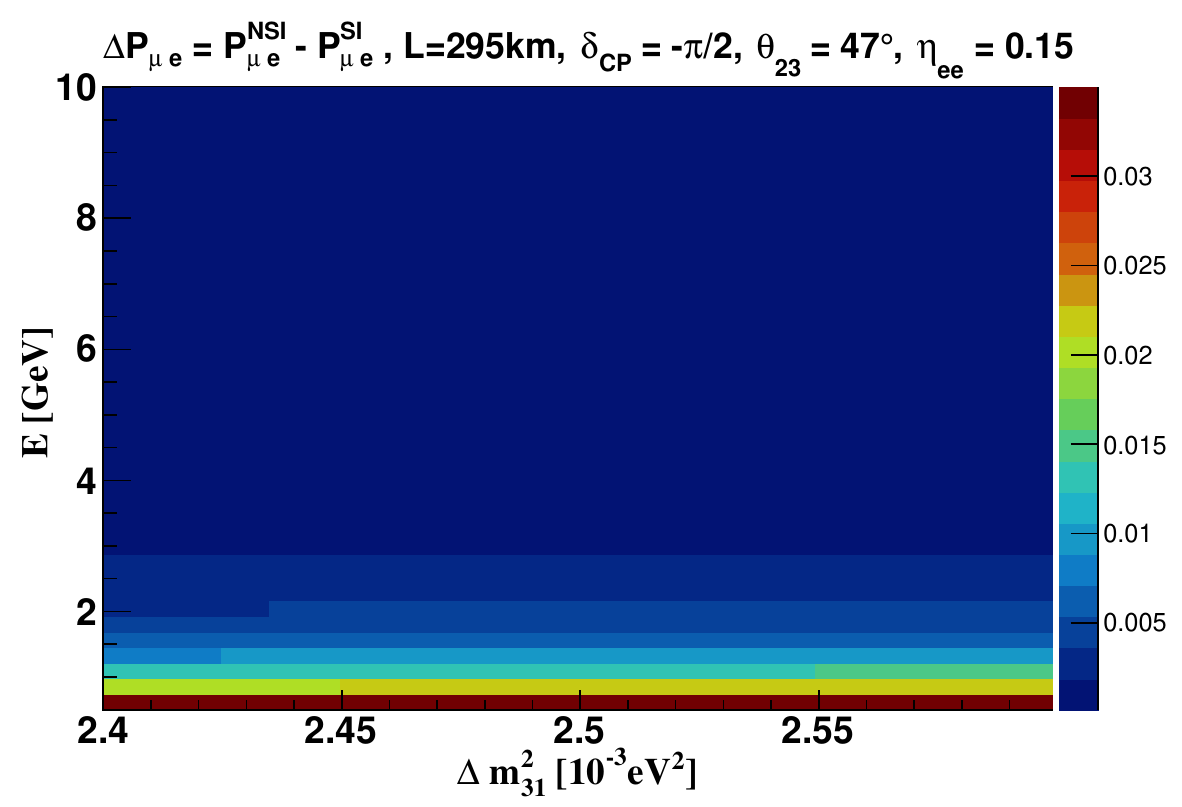} 
\includegraphics[width=0.33\linewidth, height = 5.5cm]{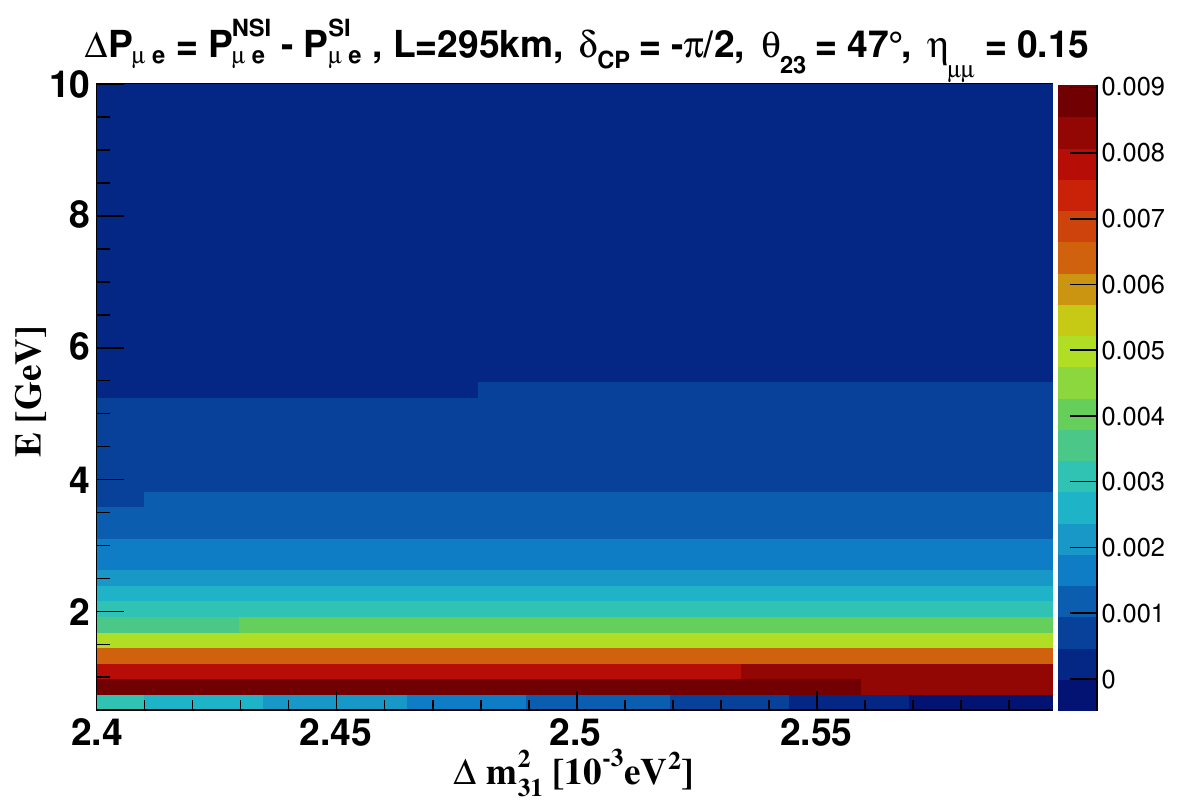}
\includegraphics[width=0.32\linewidth, height = 5.5cm]{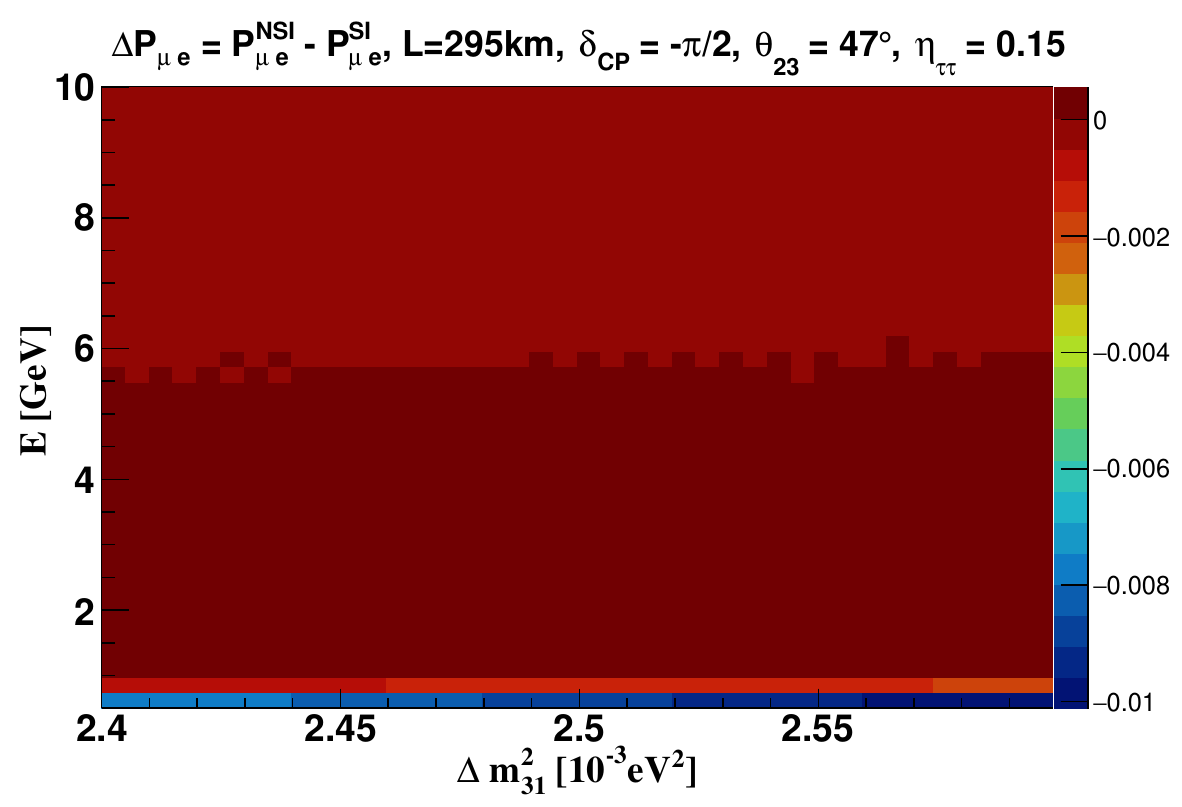} 
\caption{Variation of $\Delta P_{\mu e}$ in $\Delta m_{31}^{2}$-E parameter space in the presence of scalar NSI elements $\eta_{ee}$ (left--panel), $\eta_{\mu\mu}$ (middle--panel) and $\eta_{\tau\tau}$ (right--panel) for HK.}
\label{fig:2d_E_M_T2HK}
\end{figure}
We have plotted $\Delta P_{\mu e}$ for varying neutrino energy and $\Delta m_{31}^{2}$ for true NO. We have shown the variation in the presence of scalar NSI for all the considered experiments i.e. DUNE, HK and HK+KNO. We consider the diagonal scalar NSI parameters i.e. $\eta_{ee}$ (left--panel), $\eta_{\mu\mu}$ (middle--panel) and $\eta_{\tau\tau}$ (right--panel), one element a time. In figure \ref{fig:2d_E_M_DUNE}, \ref{fig:2d_E_M_T2HK} and \ref{fig:2d_E_M_T2HKK}, we have shown the variation of $\Delta P_{\mu e}$ at DUNE, HK and HK+KNO respectively. In all the figures, we have fixed the value of $\eta_{\alpha\beta}$ at 0.15. The other oscillation parameters are fixed at values shown in table \ref{tab:mixing_parameters}. 

We observe that for higher values of neutrino energies, the value of $\Delta P_{\mu e}$ changes in the complete $\Delta m_{31}^{2}$ range for DUNE, HK and HK+KNO. In the case of DUNE, the same trend is seen in the presence of all scalar NSI parameters. In case of HK \& HK+KNO, the impact on $\Delta P_{\mu e}$ is more prominent in the presence of $\eta_{ee}$ in comparison to $\eta_{\mu\mu}$ and $\eta_{\tau\tau}$. For a fixed value of energy, $\Delta P_{\mu e}$ remains almost the same for the whole $\Delta m_{31}^{2}$ range.
\begin{figure}[!h]
\centering
\includegraphics[width=0.33\linewidth, height = 5.5cm]{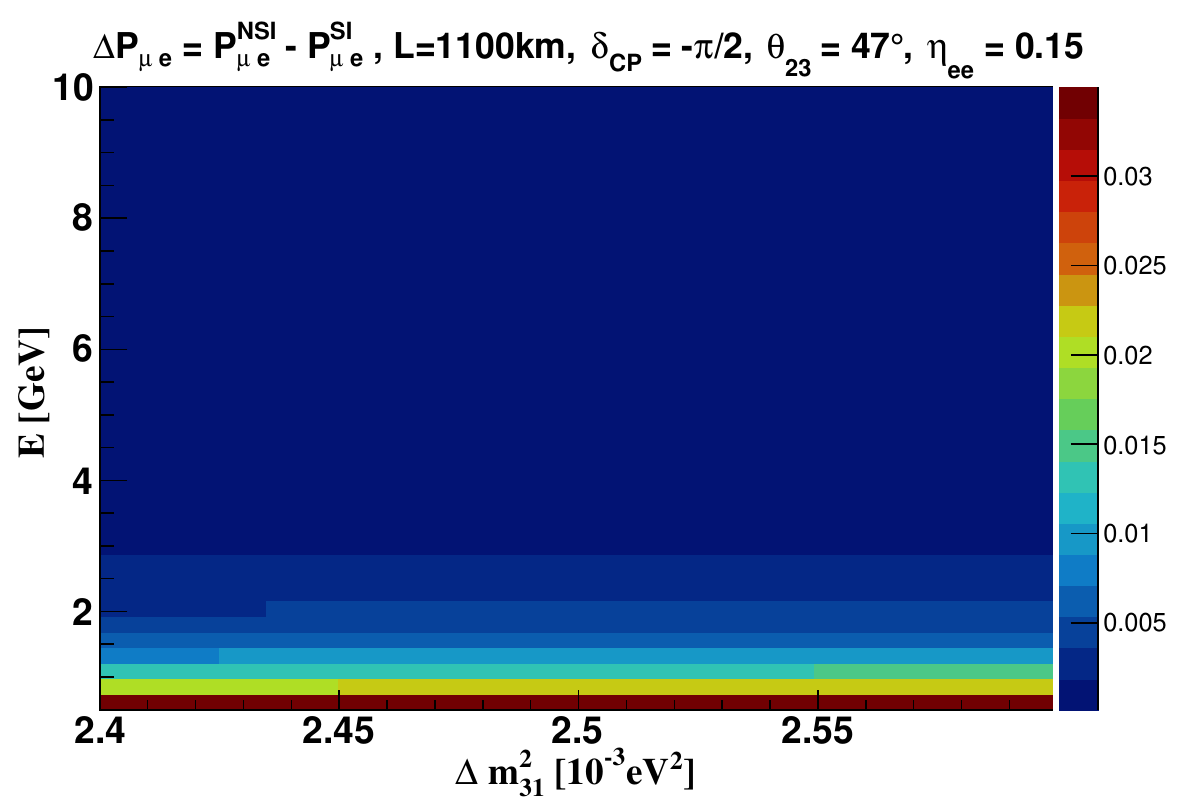}
\includegraphics[width=0.33\linewidth, height = 5.5cm]{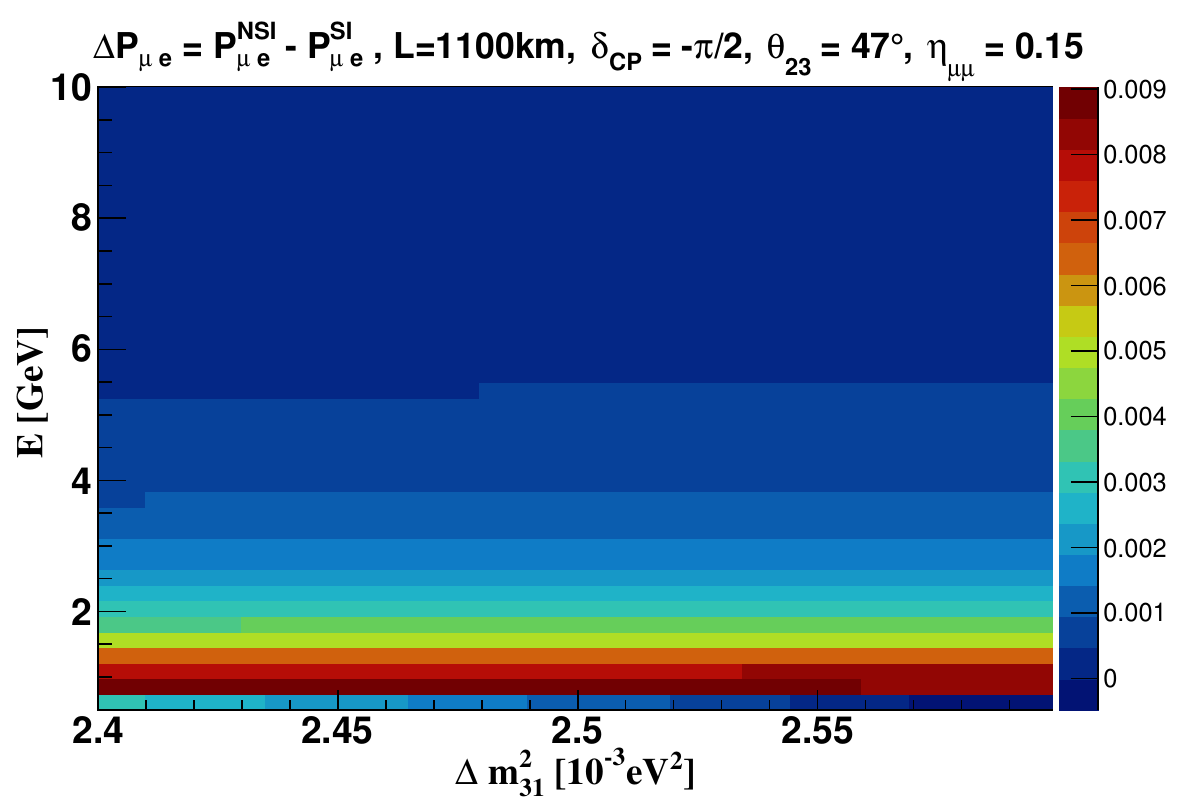}	
\includegraphics[width=0.32\linewidth, height = 5.5cm]{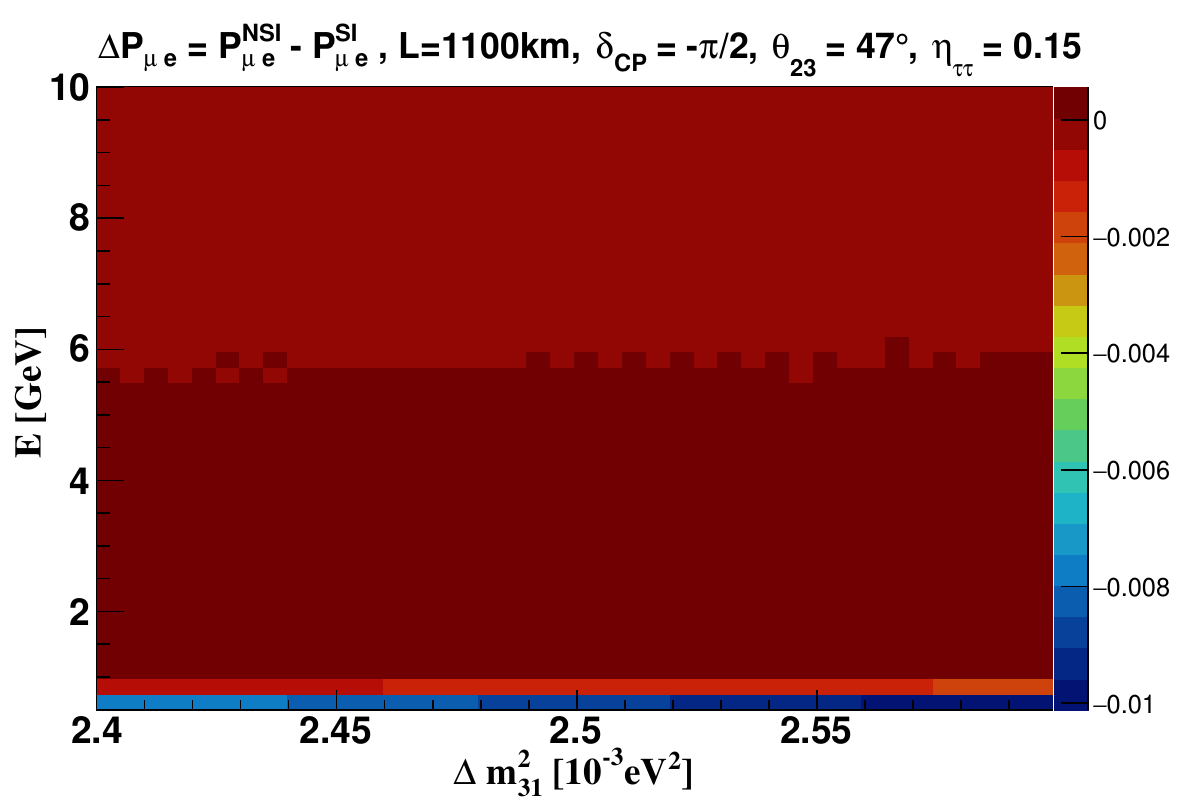}
\caption{Variation of $\Delta P_{\mu e}$ in $\Delta m_{31}^{2}$-E parameter space in the presence of scalar NSI elements $\eta_{ee}$ (left--panel), $\eta_{\mu\mu}$ (middle--panel) and $\eta_{\tau\tau}$ (right--panel) for HK+KNO.}
\label{fig:2d_E_M_T2HKK}
\end{figure}

\subsection*{A.3  Impact of different fixed values of $\Delta m_{31}^{2}$ on the MO sensitivities}
We have also explored the impact of different fixed values of $\Delta m_{31}^{2}$ on the MO sensitivities.
\begin{figure}[h]
    \centering
    \includegraphics[width=0.48\textwidth, height = 5.5cm]{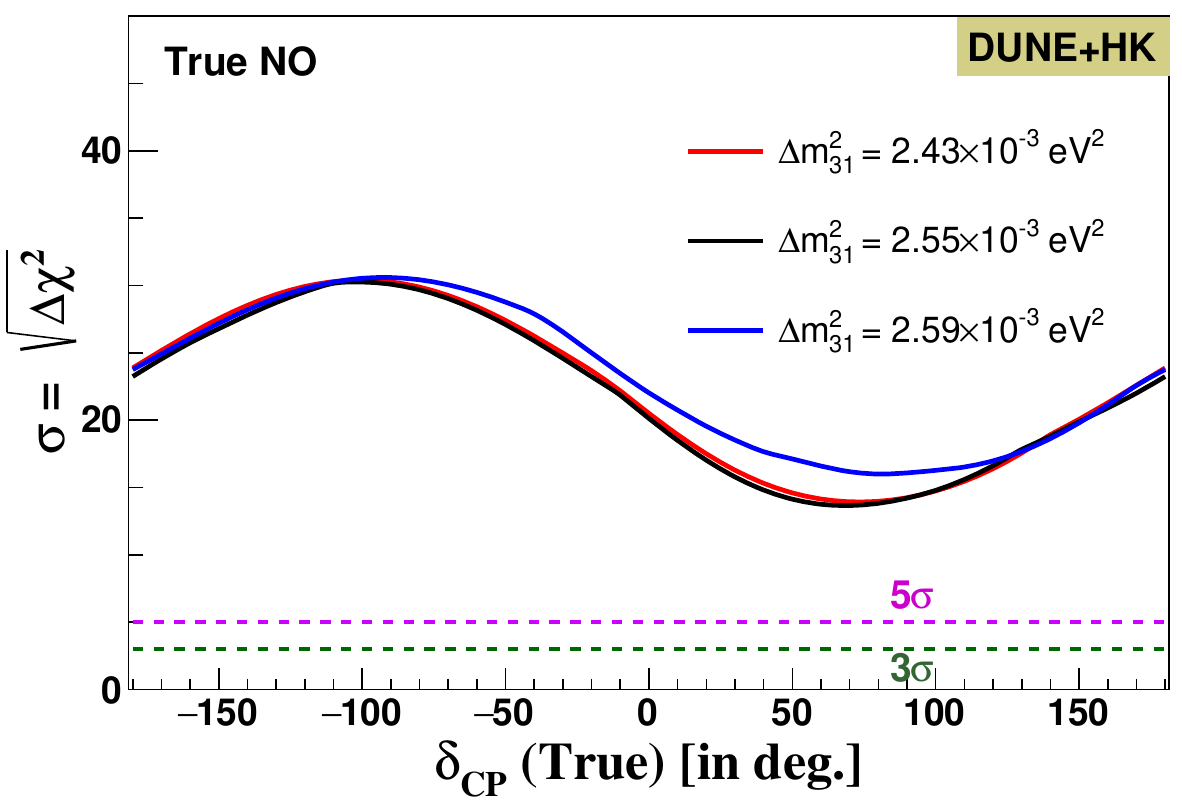}
    \includegraphics[width=0.48\textwidth, height = 5.5cm]{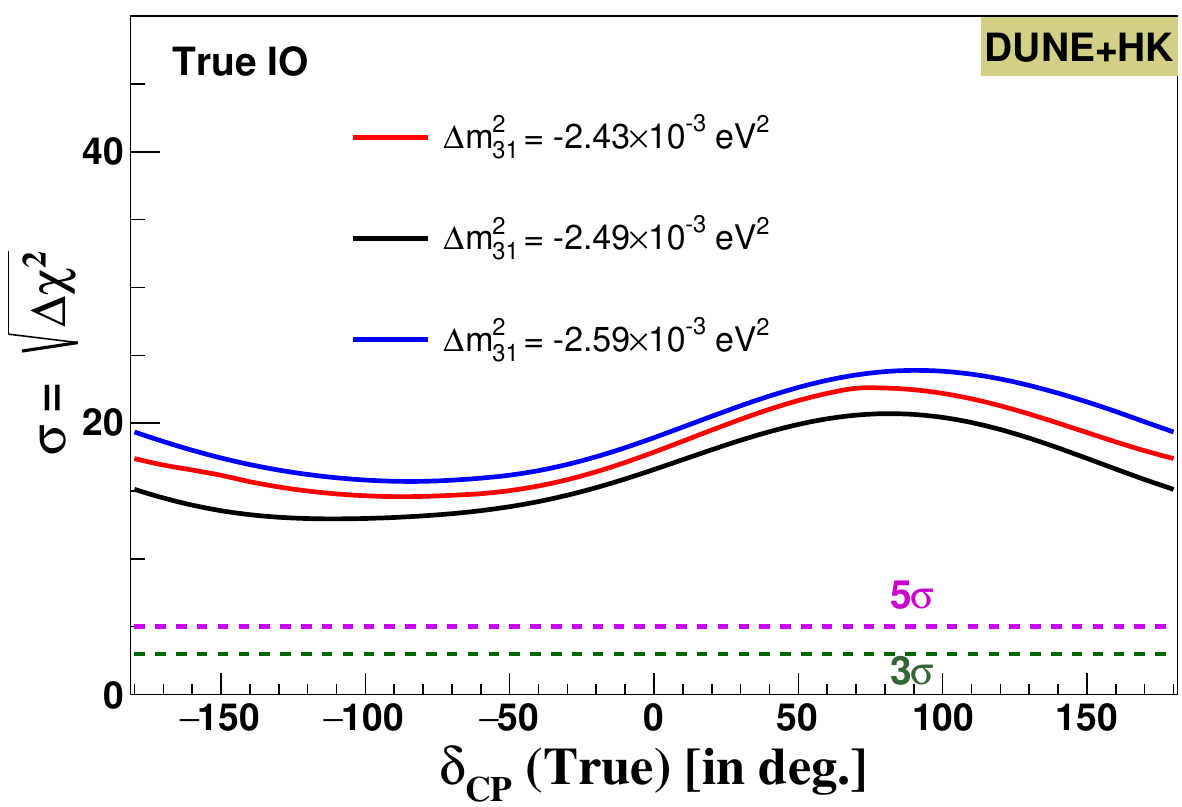}
    \caption{Impact of $\Delta m_{31}^{2}$ on the MO sensitivities for DUNE+HK configuration in the presence of $\eta_{ee}$ fixed at 0.1. The left panel represents true NO and the right panel represents true IO.}
    \label{fig1}
\end{figure}
In figure \ref{fig1}, we have examined how different true values of $\Delta m^{2}_{31}$ can affect the MO sensitivities for DUNE+HK configuration in the presence of $\eta_{ee}$. We have fixed the value of $\eta_{ee}$ at 0.1 and the oscillation parameters as described in the manuscript. The left panel represents true NO whereas, the right panel represents true IO. In our analysis, we have kept the value of $|\Delta m^{2}_{31}|$ fixed for given mass ordering.

\bibliographystyle{JHEP}
\bibliography{scalar_NSI_MH}

\end{document}